\documentstyle[rmp,aps,harvard]{revtex}

\input{epsf}

\newcommand{\BA}{\begin{eqnarray}}
\newcommand{\EA}{\end{eqnarray}}
\newcommand{\beqa}{\begin{eqnarray}}
\newcommand{\eeqa}{\end{eqnarray}}
\newcommand{\beq}{\begin{equation}}
\newcommand{\eeq}{\end{equation}}
\newcommand{\eq}[1]{Eq.~(\ref{#1})}
\newcommand{\fig}[1]{Fig.~\ref{#1}}
\renewcommand{\d}{{\rm d}}
\newcommand{\hlf}{{\frac{1}{2}}}
\newcommand{\drd}{{\frac{1}{3}}}
\newcommand{\n}{{\bf n}}
\newcommand{\p}{{\bf p}}
\newcommand{\q}{{\bf q}}
\newcommand{\qt}{{\bf q_\perp}}
\newcommand{\qtl}{{ q_\perp}} 
\newcommand{\roo}{\rho}
\newcommand{\rhoo}{\rho}
\newcommand{\alfa}{\alpha}
\newcommand{\nn}{\nonumber \\}

\renewcommand{\L}{{\cal L}}
\newcommand{\I}{{\cal I}}
\newcommand{\J}{{\cal J}}
\newcommand{\T}{{\cal T}}
\newcommand{\R}{{\bf{R}}}

\newcommand{\tOmega}{{\tilde{\Omega}}}
\newcommand{\Omegamark}{{\tilde{\Omega}}} 
\def\expect#1{\langle{#1}\rangle}
\newcommand{\eexp}[1]{{\rm e}^{#1}}
\newcommand{\tb}{\overline{t}}

\begin{document}
\bibliographystyle{rmp}
\citationstyle{dcu}

\title{Multiple scattering of classical waves: from microscopy to
mesoscopy and diffusion}
\author{M.\ C.\ W.\ van Rossum$^1$ and Th.\ M.\ Nieuwenhuizen$^2$}
\address{$^1$
Room 123, Anat.-Chem. Bldg.,
Dep. Neuroscience, University of Pennsylvania, Philadelphia, PA 19104-6058,
U.S.A. \\
$^2$ Van der Waals-Zeeman Instituut,
Valckenierstraat 65,1018 XE Amsterdam, The Netherlands}
\date{\today}
\maketitle

\begin{abstract}
A tutorial discussion of the propagation of waves in random media is presented.
In first approximation the transport of the multiple scattered waves is given
by diffusion theory, but important corrections are present. These corrections
are calculated with the  radiative transfer or Schwarzschild-Milne equation,
which describes intensity transport at the ``mesoscopic'' level and is derived
from the ``microscopic'' wave equation. A precise treatment of the diffuse
intensity is derived which automatically includes the effects of boundary
layers. Effects such as the enhanced backscatter cone and imaging of objects in
opaque media are also discussed within this framework. In the second part the
approach is extended to mesoscopic correlations between multiple scattered
intensities which arise when scattering is strong.  These correlations  arise
from the underlying wave character. The derivation of correlation functions and
intensity distribution functions is given and experimental data are discussed.
Although the focus is on light scattering, the theory is also applicable to
micro waves, sound waves and non-interacting electrons.

\end{abstract}

\newpage

\tableofcontents

\section{Introduction}
\label{chintro}

Transport of waves through opaque media is a subject of interest in daily life.
Examples are light transported through fog, clouds, milky liquids, white paint,
paper, and porcelain, but also electromagnetic waves transported through
stellar atmospheres and interstellar clouds.  Nowadays transport of visible
light through human tissue is being used as a noninvasive technique to detect,
for instance,  breast cancer.

Some basic properties of diffuse light are well known. On a cloud-less day we
see the immediate radiation from the sun. When a  cloud passes in front of the
sun, the light first becomes weaker and diffuse. When the cloud has become
thick enough,  the sun becomes invisible; this happens when the cloud thickness
is of the order of a mean free path. On a cloudy day there is no direct view of
the sun, but there is still light: it is diffuse light coming from many
directions. It has propagated diffusely through the cloud and leaves it in
random directions, partly in the direction we are looking.

The study of diffuse wave transport was started by astrophysicists. They wanted
to understand how radiation created in the center of stars is affected when it
traverses an interstellar cloud. Well-known books in this field were written by
\citeasnoun{chandrasekhar} and \citeasnoun{vanderhulst2}. For more mundane
applications, such as the detection of a school of fish using acoustic waves,
see \citeasnoun{ishimaru}.

It is the purpose of this review to present a comprehensive, self-contained
text intended for laboratory applications of diffuse wave transport. We explain
how the transfer equation follows from the wave equation and how radiative
corrections can be calculated.  Though the approach can be applied to any
geometry,  we shall focus mainly on slab geometries. 

In Sec.~\ref{theomacroscopy}  we discuss some general aspects of wave
scattering, such as diffusion and Anderson localization. In
Sec.~\ref{theomesoscopy} we  recall the basic concept of the radiative transfer
equation.  Next we begin a detailed analysis in section ~\ref{theomicroscopy}
with the underlying wave equation  in the scalar approximation. We explain the
notions of $t$ matrix and  cross section. In Sec.~\ref{theogreensfies}  we
consider the amplitude or dressed Green's function. We discuss how extinction
is related to the self-energy. Next we consider propagation of intensity via
the Bethe-Salpeter equation. We use the simplest case, the ladder
approximation, to derive a transport equation equivalent to the radiative
transfer equation. This equation then describes not only  diffusion in the
bulk, but also the precise behavior at the boundaries. Various experimentally
relevant situations are considered in detail: transport in a bulk
medium,through  very thick (semi-infinite), and through finite slabs; the
enhanced backscatter cone; exact solutions of the Schwarzschild-Milne equation;
the regime of large index-mismatch; semiballistic transport; and imaging of
objects hidden in opaque media (sections VI-XIII).

With these ingredients we consider in the second part correlations of 
intensities. This leads to three kinds of observable correlation functions, 
the angular correlation function $C_{1}$, the correlation of total transmission
$C_{2}$, and the correlation of the conductance $C_{3}$. We explain how these
functions are calculated and next we consider correlations between three
intensities (third cumulants) and the full distribution of intensities.

\subsection{Length scales}

The diffuse regime is formulated in the three inequalities
\cite{kavehvanhaeringen},
\begin{equation} \lambda \ll \ell\ll L \ll L_{\rm abs}, L_{\rm inc},
\end{equation}
where $\lambda$ is the wavelength, $\ell$ the mean free path, $L$ the sample
size, $L_{\rm abs}$ the absorption length (for optical systems), and $L_{\rm
inc}$ the incoherence length (for electronic systems).
The first inequality ensures that localization effects (see
below) are small; the second inequality implies that many scatterings
occur if the wave transverses the system; the third inequality ensures that
not all radiation is absorbed.

The description of radiation transport can occur on roughly three length
scales:

\indent -{\it Macroscopic:} On scales much larger than the mean free
path the average intensity satisfies a diffusion equation. The diffusion 
coefficient $D$ enters as a system parameter that has to be calculated
on mesoscopic length scales.

\indent -{\it Mesoscopic:} On length scales of the mean free path $\ell$, the
problem is described by the radiative transfer equation or Schwarzschild-Milne
equation. This is the Boltzmann-type equation for the system. At this level one
needs as input the mean free path $\ell$ and the speed of transport $v$, which
should be derived from microscopics. In the diffusive regime this approach 
leads to the diffusion coefficient $D=v\ell/3$.

\indent -{\it Microscopic:} The appropriate wave equation, such as Maxwell's
equations, the Schr\"odinger equation, or an acoustic wave equation, is used on
this length scale. The precise locations and shapes of scatterers are assumed
to be known. Together with the wave nature they determine the interference
effects of scattered waves. In light-scattering systems the scatterers often
have a size in the micron regime, comparable to the wavelength $\lambda$, which
could lead to important resonance effects. The mesoscopic or Boltzmann
description follows by considering the so-called ladder diagrams.

The microscopic approach will be the starting point for this review.
Fundamental quantities, such as the self-energy and the Hikami box, can only be
calculated on this level. The drawback of the microscopic approach is that it
is too detailed. In practice the precise shape and positions of the scatterers
often are not known and a mesoscopic or macroscopic description is necessary. 

\subsection{Weak localization, closed paths and the backscatter cone.}

The diffusion equation is a classical equation that fully neglects interference
effects inherent to wave propagation. At this level of description there is no
 difference between diffusion of particles and of wave intensity. Whereas a
transmission pattern of monochromatic (laser) light through an opaque medium is
known to consist of speckles (bright spots on a dark background), the
diffusion equation and the radiative transfer equation only describe the
average intensity.

The wave nature of light immediately leads to a reduction in transmission due
to interference effects. Following \citeasnoun{rayleigh} we suppose that the 
transmission amplitude $E$ for a certain experiment is composed of many terms
$C_p$ arising from physically different interference paths $p$.

Some paths have closed loops, i.e. loops that return to the same scatterer.
When such loops contain 
two or more common intermediate scatterers, they can be 
traversed in two directions. Let us consider one of such loops $C_p$, and
denote the contribution of the second, reversed loop by $D_p$. Summing over
all paths we have \begin{equation} E=\sum_p(C_p+D_p). \end{equation} 
The intensity is \begin{equation}
E^2=\sum_p(C^\ast_pC_p+D^\ast_pD_p) +\sum_p(C^\ast_pD_p+D^\ast_pC_p)
+\sum_{p\neq p'}(C^\ast_p+D^\ast_p)(C_{p'}+D_{p'}). \end{equation} 
When applied to electrons, quantum mechanics tells us that the $C^\ast C$ and
$D^\ast D$ terms are probabilities, while $C^\ast D$ and $D^\ast C$ terms are
interference contributions. Naively, one expects the second and third term to
be small. Thus in Boltzmann theory only probabilities are taken into account,
that is to say, the first term. 

However, if there is time-reversal invariance, the second term $\sum_p(
C^\ast_pD_p+ D^\ast_pC_p)$ will be equally large: {\it there is a factor of $2$
for each closed loop}. As a result, for the wave intensity there is a larger
probability of return. 

In optics there is a simply measurable effect due to this, namely the enhanced
backscatter cone. When the incoming and outgoing light are in exactly the 
same direction, the light path may be considered closed. As predicted by
\citeasnoun{barbaranenkov} and detected by several groups
\cite{kuga,albada3,wolf} the {\it average} intensity in the backscatter
direction has a small cone of height almost one and angular width $\delta
\theta\sim \lambda/\ell$. This observation has given an enormous
push to the research of weak localization phenomena in optics.

This effect is a so-called weak localization effect. These effects occur  if
$\lambda/\ell \ll 1$ and are precursor of the so-called strong localization
effects which occur if $\lambda/\ell \sim 1$. These effects are also known as
``mesoscopic,'' indicating length scales between macroscopic (the diffusion
equation) and microscopic (individual scattering events). Indeed, in electron
systems these effects only show up in rather small samples due to inelastic
scattering. For optical systems there is no such size restriction. The study of
mesoscopic effects started in the field of electronic systems, and we shall use
many results derived there. 

For electronic systems weak localization effects were first analyzed by
\citeasnoun{altshuler5} in order to explain the Sharvin-Sharvin fluctuations of
a resistance as function of the applied magnetic field \cite{sharvin}. These
so-called ``magneto-fingerprints'' show a seemingly chaotic conductance as the
applied field is varied, but are perfectly reproducible as they are solely
determined by the location of the scatterers. (Only at very low
temperatures electron scattering is dominated by impurity scattering;
at higher temperature scattering mainly arises from phonons.)

\subsection{Anderson localization} 

If scattering is very strong, the ``weak localization'' effects become of order
unity, i.e., $\ell\sim\lambda$ ( $\lambda/\ell$ is of the order of  0.001-0.01 
for visible light in standard laboratory situations). According to the
criterion of \citeasnoun{ioffe} the diffusion constant will tend to zero at
that point. The scattering process that causes the enhanced backscatter also
reduces the diffusion constant. The amplitudes that make up the intensity,
split and one of the amplitudes visits some of the scatters in reverse
sequence, thus forming a loop. The diffusion constant is lowered by these
processes. (The loop is somewhat similar to the renormalization of the
effective mass in field theories.) If scattering is increased, the contribution
of these loops becomes more and more important. It is not simply that the
diffusion constant tends linearly to zero as scattering increases. Rather, the
return probability of the intensity becomes higher and higher, reducing the
diffusion constant in a stronger fashion. The diffusion constant can thus
become zero at finite scatterer strength, meaning that the wave can no longer
escape from its original region in space. This is the transition to the
well-known Anderson localization \cite{anderson}. 

In the Anderson localization regime there are only localized states. In the
delocalized regime there exist extended states, responsible for diffusion. In
the localized regime the intensity typically decays exponentially over one
localization length, whereas in the diffuse regime the wave function extends up
to infinity. There are thus two different phases, the diffuse, metallic regime
and the localized, insulating regime. In three dimensions a phase transition
from the extended to the localized state can occur. In one and two dimensions
the states are always localized, provided the waves are non-interacting (for
electrons  also spin-orbit scattering has to be absent). Yet for a finite
sample the localization length can be much larger than the system size, in
which case the states appear to be extended and the conductance does not
vanish. Note that the localization is solely the result of the interference of
the waves scattered by the disorder. (This is not the only scenario for a
metal-insulator transition in electron systems. Due to their fermionic nature
and interactions, electron systems allow for a whole range of possible
transitions between the conducting and insulating regimes; see
\citeasnoun{mott}). Anderson localization is also called strong localization.

The precise behavior near and at the transition is not fully understood. The
standard diagrammatic perturbation technique, which we use, works well for the
description of diffusion and low order corrections, but it is not
suited for the study of the transition. Therefore various techniques have been
developed to study behavior near the transition and the phase transition
itself.

An important step was the scaling theory of localization put forward by
\citeasnoun{abrahams}, which states that near the localization the only
parameter of importance is the dimensionless conductance $g$ (the
conductance measured in units of $e^2/h$). The scaling of $g$ as a function of
sample size was studied earlier \cite{thouless1,thouless2,wegner2}. 
Abrahams {\it et al.} extended those ideas and derived
renormalization-group equations. They concluded that in one and two
dimensions there is no real phase transition; the states are always localized.
In three dimensions a phase transition can occur. 

Classical diffusive transport is described by the ladder diagrams. The
so-called {\em maximally crossed diagrams} are the next most
 important diagrams in
the diffuse regime. They describe the leading
interference terms responsible for the
backscatter cone and reduction of the diffusion constant. 
Vollhardt and W\"olfle
summed self-consistently combinations of the ladder diagrams and the maximally
crossed ones in the diffuse regime \cite{vollhardt1,vollhardt2,vollhardt3}.
They found that the diffusion constant vanishes at strong enough scatterer
strength, thus providing a microscopic picture of the Anderson transition. As
the maximally crossed diagrams yield loops of intensity, the approach can be seen as
a self-consistent one-loop summation. Although a self-consistent approach will
certainly not include all diagrams of higher order, the method works fine even
close the transition \cite{kroha1,kroha2}, because the first few higher order
correction terms vanish.

Another approach is to perform an exact average over the disorder within a
field-theoretic approach. Next, one integrates over the fast fluctuations and
is left with the slow variables of the system. This technique yields the
so-called {\em nonlinear sigma model}
\cite{wegner,hikami,efetov1,altshulerboek}. With the resulting action it is
possible to generate systematically all corrections to the diffusion process
(in $2+\epsilon$ dimensions), allowing for an explicit foundation of the
scaling theory. 

A recent approach is that of {\em random matrix theory}. The basic assumption
here is that the total scattering matrix of the system, although very
complicated, can be described just by random matrix elements respecting
the symmetries of the system. Surprisingly, this method works well, and just
from the assumption of the random ensemble it predicts many features of the
systems correctly and in a simple way. Its is, unfortunately, only applicable
to quasi one-dimensional situations. An overview of the theory and applications
of random matrix theory is given by \citeasnoun{stoneboek} and
\citeasnoun{beenakker2}.

Although some years ago there was hope that the Anderson transition might soon
be reached for light scattered on disordered samples, 
it was not observed when this review was written.
One did observe that in time resolved
transmission measurements the average transmission time became very long. This
was interpreted as an indication that the mean free path was very small, which
would mean that the light was close to localization. It turned out, however,
that the long transmission times arise since the light spends much time inside
the scatterers. This meant that the Anderson localization was still out of
reach~\cite{albada2}.

\subsection{Correlation of different diffusons}

Another interaction effect is the interference of a diffuse intensity with
another that has, for instance, a different frequency or position.
 In this work we
shall concentrate on such processes which, lead to correlations in,
for instance, the transmitted beams. There are advantages to studying the
correlations above the loop-effects. The correlations can be measured more
accurately and easily in experiments than can renormalization effects.
 Secondly,
it is an interesting feature of optical systems that there are three different
transmission measurements: \\

\indent - Coming in with a monochromatic beam in one
direction (this we call ``channel $a$'') one can measure the intensity in
the outgoing direction $b$ and define the 
{\it angular transmission coefficient} $T_{ab}$. 
Its correlation function is called the $C_1$ correlation and is of
order unity, describing the large intensity difference between dark and bright
transmission spots.

\indent - One can also measure all outgoing light. In practice one
uses an integrating
sphere. This leads to the measurement of the {\it total transmission},
 \begin{equation}
T_a=\sum_{b}T_{ab}. \end{equation} Its correlation function is called 
the $C_2$ correlation
function.

\indent - Finally, one can also add the results of coming in from all possible
directions, either by repeating the experiment under many different incoming
angles or by using diffuse incoming light. This leads to a quantity 
\begin{equation}
T=\sum_aT_a=\sum_{ab}T_{ab}.\end{equation} In analogy with electronic
 systems it is called the {\it conductance}.
Its fluctuations are called $C_3$ fluctuations in optics.
In electronics (where, for instance, the magnetic field is varied),
they are called universal conductance fluctuations (UCF). 
Notice that these fluctuations are not
temporal but static, since the scatterers are fixed. 
Both $C_2$ and $C_3$ correlations are
low order corrections in $1/g$, which acts as the small parameter giving the
relative strength of the interference effects.

There are various books and proceedings on localization and mesoscopics in
electron systems 
\cite{nagaoka,ando,physicaa,vanhaeringen,vanhaeringen2,altshulerboekgeheel,hankeboek}
and classical waves \cite{kirkpatrick,sheng,kretaboek}. Many aspects of the
Anderson transition have been discussed by ~\citeasnoun{lee4}.

\section{Macroscopics: the diffusion approximation}
\label{theomacroscopy}

\subsection{Transmission through a slab and Ohm's law}

Consider the propagation of light through a slab of thickness $L$ (``plane
parallel geometry''). As indicated in Fig.~\ref{skin.eps} a plane wave
impinges on the surface of an opaque medium at an angle $\theta_a$.
The index of refraction of the medium is $n_0$, and that 
of the surroundings is $n_1$. The ratio of the indices of refraction,
\begin{equation} m=\frac{n_0}{n_1}, \end{equation}
is larger than unity if a dry substance is placed in air ($n_1\approx 1$).
If the substance is placed in a liquid, one may have $m<1$.
If $m\neq 1$ the refracted beam inside the medium will have an angle 
$\theta_a'$ different from $\theta_a$. 

Due to multiple scattering the incoming beam decays exponentially,
\begin{equation} I_{in}(z)=I_0e^{-z/\ell\cos\theta_a} ,\end{equation}
where the cosine is a geometrical factor expressing the total path length in
terms of $z$. This exponential decay is known as Lambert-Beer law. It describes
the decay of the direct sunlight in clouds or of headlights of cars in fog.
The light source becomes invisible when the thickness of the diffuse medium is
greater than one mean free path. The decay of direct light intensity occurs
because it is transformed into diffuse light.

The scattering of the incoming beam into diffuse light occurs in a skin 
layer with characteristic thickness of one mean free path. Later on this will
be discussed in greater length when we solve the Schwarzschild-Milne equation.
In the diffusion approach one simply assumes that the incoming beam is 
partly reflected and partly converted into a diffuse beam in the skin layer.
Next one assumes that effectively 
diffuse intensity enters the system in a trapping plane
located at distance $z_0\sim \ell$ outside the scattering medium.
The value of $z_0$ is phenomenological, but a precise analysis of
the Schwarzschild-Milne equation reveals that this picture is valid.
One usually takes $z_0=0.7104\ell$, the exact value for isotropic point
scattering of scalar waves. \citeasnoun{vanderhulst2}
investigated many possible
cross sections and numerically always found values very close to this.
\citeasnoun{amic} considered the limit of very strong forward
scattering and found that $z_0$ lies only
$1.1\%$ above the quoted value for isotropic scattering.

Once the light has entered the bulk, the diffuse intensity obeys the diffusion
equation
\begin{equation}
\frac{\partial}{\partial t} I({\bf r},t)=D\nabla^2I({\bf r},t).
\end{equation}
In the steady state the time-derivative vanishes and the diffusion coefficient
plays no role. For a slab geometry with plane waves
 there is no $x,y$ dependence, and we have
to solve $I''(z)=0$. The trapping plane
boundary conditions are $I(-z_0)=I_0$, $I(L+z_0)=0$. The solution is
a linear function of $z$,
\begin{equation} I(z)=I_0\frac{L+z_0-z}{L+2z_0}.
\end{equation} 
The transmitted intensity is essentially equal to this expression at
$z=L$ (at a distance $z_0\sim\ell$ from the trapping plane at $L+z_0$)
and equal to $z_0$ times the derivative at that point;
this estimate will turn out to be qualitatively correct
\begin{eqnarray} \label{Tequiv}
T&=&\frac{I(z=L)}{I_0}=-\frac{z_0}{I_0} \frac{dI(z)}{dz}
\left|_{z=L+z_0}. \right. \nonumber\\ 
&=&\frac{z_0}{L+2z_0}\approx \frac{z_0}{L}. \end{eqnarray}
Without having paid attention to details we find
the behavior $T\sim \ell/L$, equivalent to Ohm's law for conductors.
Generally a conductor has conductance $e^2/h$ per channel.
Here there are $A/\lambda^2$ channels, of which a fraction $T$ is
open. This yields the conductance 
\begin{equation} \Sigma\sim\frac{e^2}{h} \frac{A}{\lambda^2}T\sim
\frac{e^2\ell A}{h\lambda^2L} .
\end{equation}

\subsection{Diffusion propagator for slabs}
In the presence of absorption the diffusion equation reads
\begin{equation}\label{D1}
 \partial_t I({\bf r},t)=D\nabla^2 I({\bf r},t)-D\kappa^2I({\bf r},t)
+S({\bf r},t), \end{equation}
where $S$ is a source term, 
the second term on the right-hand side describes absorption, 
and the absorption length is $L_{abs}=1/\kappa$.
In a bulk system the solution with initial conditions
$I({\bf r},0)=\delta ({\bf r})$ and $S=0$ reads
\begin{equation}\label{D2}
 I({\bf r},t)=(4\pi Dt)^{-3/2}\eexp{-\frac{r^2}{4Dt}-D\kappa^2 t}.
 \end{equation}
Its Fourier-Laplace transform 
\begin{equation} \label{D4}
I(\q,\Omega)= \int \d^3r \eexp{-i\q\cdot{\bf r}} \int_0^\infty \d t 
\eexp{-i\Omega t}
 I({\bf r},t), \end{equation}
 takes the form
\begin{equation} \label{D5}
I(\q,\Omega)=\frac{ I(\q ,{\bf t} =0) } { D\q^2+D\kappa^2+i\Omega }.
\end{equation}
For $\kappa=0$ these expressions diverge in the limit $\q,\Omega\to
0$. This divergency is called the ``diffusion pole'' or ``diffuson''.

For a slab geometry there will only be translational invariance in the
$\rho=(x,y)$ plane. Suppose we have a source $S({\bf r})= 
\tilde S(x,y)\delta (z-z')$.
Denoting the perpendicular wave vector by 
$\qt$ we have for diffusion from $z'$ to $z$
\begin{equation} \label{D7}
 I''(z,z')=M^2I(z,z')-\frac{\tilde S(\qt)}{D}\delta(z-z'), \end{equation}
with the ``mass'' $M$ defined by
\begin{equation} \label{D8} 
M^2=\qt^2+\kappa^2+i\frac{\Omega}{D}.\label{eqdefM} \end{equation}
$M$ is the inverse depth at which a given intensity contribution
$\psi_a^\ast\psi_b$ of an
incoming beam $\psi=\sum_a\psi_a$ has decayed by a factor $1/e$, 
due to spatial de-phasing of the amplitudes (encoded in $\qt$), temporal
de-phasing (expressed by $\Omega$) or absorption (expressed by
$\kappa$.) 

By realizing that the diffusion equation Eq.~(\ref{D7}) is just a wave
equation with complex frequency, one sees that the solutions to the diffusion
equation are a linear combination of hyperbolic sines and cosines. 
(The solution can also be obtained using the
method of ``image charges'' known from electrostatics.)
The diffuse intensity propagator has the form \cite{zhu,lisyansky}
\begin{eqnarray} \label{D9b} \label{Lint1} 
I(z,z';\qt;\Omega)=\frac{\tilde S(\qt)}{D}\,\,
\frac{[\sinh M\! z_<+M \!z_0 \cosh M \!z_< ] [ \sinh M(L\!-\!z_>)+ M\!z_0
\cosh M(L\!-\!z>) ]}{(M+M^3 z_0^2)\sinh M L+2M^2 z_0 \cosh ML }  ,
\end{eqnarray} 
where
\begin{eqnarray} z_<={\rm min}(z,z'),\qquad z_>={\rm max}(z,z').\end{eqnarray}
In the stationary limit in the absence of absorption ($\Omega=\kappa=0$,
$\qt={\bf 0}$ $\to$ $M=0$), Eq.~(\ref{D9}) reduces to a tent-shaped function,
\begin{equation} \label{D10} I(z,z';\qt={\bf 0};\Omega=0)
=\frac{\tilde S(\qt)}{D}\,\,
\frac{[z_<+ z_0 ] [ L\!-\!z_>+ \!z_0]}{L+2 z_0}.
\end{equation}

The propagator describes the diffuse propagation from one point in the slab  to
another. With roughly equal indices of refraction inside and outside the
sample, the extrapolation length $z_0$ is a few mean free paths and thus the
terms involving $z_0$ yield contributions of  order $\ell/L$. For optically
thick samples ($L\gg \ell$), this is negligible and one has 
\begin{eqnarray} \label{D9} I(z,z';\qt;\Omega)=\frac{\tilde S(\qt)}{D}\,\,
\frac{\sinh(Mz_<)\sinh(ML-Mz_>)} {M\sinh(ML)}.
\end{eqnarray}.

The diffusion equation does not hold if the intensity gradient is steep, i.e.,
if $q \ell \sim 1$. It can therefore not describe properly the diffuse
intensity near the surface of the medium. Nevertheless,  in a heuristic manner
one often makes the following  assumptions in the diffusion approach
\cite{ishimaru}: (1)The diffuse intensity from an outside plane-wave source is
assumed to be given by substituting $z=\ell $ in \eq{Lint1}, as if all diffuse
intensity originated from this $z=\ell$ plane. Likewise, the coupling outwards
is obtained by taking $z'= L-\ell$. In this way two extra propagators, the
incoming ($z< 0$,  $0<z'<L$) and the outgoing ($0<z<L$, $z'>L$) are constructed
from the internal diffuson ($0<z,z'<L$).  (2) The boundary conditions are
\cite{zhu} \begin{equation} \left. {I(z,z';M)}\right|_{z=0+,L-} =z_0\left.
{\partial_z I(z,z';M)}\right|_{z=0+,L-} \label{eqrvwdif} . \end{equation}  The
form Eq.~(\ref{Lint1}) fulfills these conditions.  The extrapolation length
$z_0$ is determined by the reflectance at the surface. In the following we
calculate it precisely. 

\section{Mesoscopics: the radiative transfer equation}
\label{theomesoscopy}

The study of multiple light scattering was initiated in astrophysics with
the goal to derive, on the basis of energy conservation, the radiative
transfer equation. This is the ``Boltzmann'' transport equation of the problem
(i.e., the mesoscopic balance equation that neglects all memory effects). It
has been solved in particular for slabs (plane-parallel geometries). This
approach has been described in the books by \citeasnoun{chandrasekhar} and
\citeasnoun{vanderhulst2}. It will be shown below that the radiative transfer
equation can also be derived from the ladder approximation to the
Bethe-Salpeter equation. In other words: there exists a microscopic derivation
of the radiative transfer equation. Once this is shown, more details
can be incorporated in the microscopics and more subtle effects, such as 
backscatter and correlations, can be derived microscopically in a way
closely related to the radiative transfer equation.

\subsection{Specific intensity}

The specific intensity $\I({\bf r},\n)=\I({\bf r},\theta,\phi)$ is
defined as the radiation density emitted at
position ${\bf r}$ in direction $\n=(\sin\theta\cos\phi,\sin\theta\sin\phi,
\cos\theta)$ in a system with density $n$ of scatterers.
Let $\d U$ be the radiation energy in a given frequency interval 
 $(\omega-\frac{1}{2}\Delta\omega,\omega+\frac{1}{2}\Delta\omega)$,
transported through a surface 
$\d\sigma$ in directions lying within a solid angle $\d\n$
centered around $\n$, during a time interval $\d t$.
This energy is related to the specific intensity as \cite{chandrasekhar}
\begin{equation} \d U=\I\cos{\Theta}\,\Delta \omega\,\, 
\d\sigma\,\,\d\n\,\,\d t,  \end{equation}
where $\Theta$ is the angle between the director $\n$ 
of the emitted radiation and the normal of $\d\sigma$. 
The energy depends in general on the position ${\bf r}=(x,y,z)$,
 the direction $\n$, the frequency $\omega$, and the time $t$.

We are mainly interested in stationary, monochromatic situations for a slab
with axial symmetry. Then $\I$ depends only on $z$ and $\mu=\cos\theta$, where
$\theta$ is the angle between the $z$ axis and the direction of emitted
radiation.  We consider propagation in a medium with density $n$ of 
rotationally symmetric scatterers with extinction cross section $\sigma_{ex}$
and phase function $p(\n,\n')=p(\n\cdot\n')=p(\cos\Theta)$. Thus
$p(\n\cdot\n')\,\,\d\n'\,\,\d\n$ is the fraction of radiation entering inside a
narrow cone of width $\,\d\n'$  around incoming direction $\n'$ and leaving
inside a narrow cone $\d\n$ around $\n$. For spherically symmetric scatterers
$p$ is a  function of $\n'\cdot\n=\cos\Theta$. 

If radiation propagates over a distance $\d s$, there is a loss of intensity
due to scattering into other directions and due to absorption (both
are incorporated in $\sigma_{ex}$),
\begin{equation} \d\I=-n\sigma_{ex} \I \d s . \end{equation}
There is also a gain term $n\sigma_{ex} \J \d s$
with the {\it source function} $\J$,
\begin{equation} \J({\bf r};\theta,\phi)=
\int_0^\pi\int_{-\pi}^\pi p(\theta,\phi;
\theta'\phi')\frac{\sin\theta'\d\theta' \d\phi'}{4\pi}
\I({\bf r};\theta',\phi').\label{bronfie} \end{equation}
$\J$ describes the radiation arriving at ${\bf r}$ in direction $\n'$
and scattered there in direction $\n$.
The {\it radiative transfer equation} expresses the 
net effect of the gain-loss mechanism
\begin{equation} \frac{1}{n\sigma_{ex}}\, \frac{\d\I}{ \d s}=\J-\I .
\end{equation}
Here it is derived from phenomenological considerations,
later we will provide a microscopic derivation.
 
The loss term $-\I$ leads to the Lambert-Beer law. Indeed, for a unidirectional
 beam
 $\I(0,\n)=I_0\delta(\n-\n_0)$, simple integration yield
 $\I({\bf r},\n)=I_0\delta(\n-\n_0) \exp(-r/\ell_{sc})+$scattered.
Which is again the Lambert-Beer law with scattering mean free path
\begin{equation} \ell_{sc}=\frac{1}{n{\sigma_{ex}}}. \end{equation}

\subsection{Slab geometry}

For homogeneous illumination of a slab $0\le z\le L$,
 physical quantities depend only on the depth 
$z$. It is useful to introduce the ``optical depth''
\begin{equation} \tau=\frac{z}{\ell_{sc}}.\end{equation} 
The optical thickness of the slab is then
\begin{equation} b=\frac{L}{\ell_{sc}}.\end{equation}

Let $\theta$ be the angle between the direction of radiation and the
positive $z$ axis, and $\phi$ its angle with respect to the positive
$x$ axis. 
This allows us to introduce the dimensionless form of
the radiative transfer equation,
\begin{equation}\label{RTE}
\mu \frac{\d\I(\tau,\mu,\phi)}{\d\tau}
=\J(\tau,\mu,\phi)-\I(\tau,\mu,\phi),
\end{equation}
where \begin{equation} \mu=\cos \theta. \end{equation}
For a plane wave with intensity $I_0$ incident under an angle
$\n_a(\theta_a,\phi_a)$ on the interface $z=0$, the 
boundary condition is $\I(0,\mu,\phi)$$=
I_0\delta(\mu-\mu_a)\delta(\phi-\phi_a)$, where $\mu_a=\cos\theta_a>0$. 

\subsubsection{Isotropic scattering}

The radiative transfer equation (\ref{RTE}) can be written as an integral
equation. For a slab this is usually called the 
Schwarzschild-Milne equation or, for short, the Milne equation. 
Let us consider a semi-infinite medium.
For $\mu<0$ Eq.~(\ref{RTE}) yields for the radiation in the $-z$ direction
(backward direction)
\begin{equation} \I(\tau,\mu,\phi)=\int_\tau^\infty {\cal J}(\tau',\mu,\phi)
\eexp{-(\tau'-\tau)/|\mu|}\frac {\d\tau'}{|\mu|}, \end{equation}
while for $\mu>0$ the specific intensity in the $+z$ 
direction satisfies
\begin{equation} \I(\tau,\mu,\phi)=\I(0,\mu,\phi)\eexp{-\tau/\mu}+\int_0^\tau 
{\cal J}(\tau',\mu,\phi)\eexp{-(\tau-\tau')/\mu}\frac{\d\tau'}{\mu} .
 \end{equation}
Of special interest is
the case of {\it isotropic scattering} or $s$-wave scattering for which
\begin{equation} p(\cos\Theta)=1 . \end{equation}
Isotropic scattering occurs for electron scattering from small impurities.
For {\it isotropic} scattering ${\cal J}$ does not depend on $\mu$ and
 $\phi$. It is common to introduce the dimensionless intensity
\begin{equation} \Gamma(\tau)=
\frac{1}{I_0}\int \d\mu' \d\phi'\I(\tau,\mu'\phi'),
\end{equation}
which here equals $4\pi\J/I_0$ .
Combining the last two equations yields for a plane wave incident 
in direction $(\mu_a,\phi_a)$ 
\begin{equation} \Gamma(\tau)= \eexp{-\tau/\mu_a}+\int_0^\infty d\tau'
\int_0^1 \frac{d\mu}{2\mu} \eexp{-|\tau'-\tau|/\mu}\Gamma(\tau')  .
\label{SMslab}
\end{equation}
This Boltzmann equation will also be found from the ladder approximation 
to the Bethe-Salpeter equation, which provides a microscopic foundation.
Corrections to the ladder approximation then yield the limit of validity
of the Boltzmann approach.

The precise solution of Eq.~(\ref{SMslab}) can be obtained numerically. See
\citeasnoun{vanderhulst2} and \citeasnoun{kagiwada} for details.
 We have plotted it in
Fig.~\ref{figdifsm} for a relatively thin slab ($L=4\ell$), using the data
from Table 17 in \citeasnoun{vanderhulst2}. The albedo is unity, and 
no index mismatch. The form of the solution near the incoming surface is quite
different for the three cases drawn: $\mu_a=1$, or perpendicular incidence;
$\mu_a=0.1$ (angle with the $z$ axis is 84 degrees); and diffuse incidence
uniformly distributed over all angles. At the outgoing surface all solutions
are alike: there is only a small deviation from the straight line
crossing zero at $L+z_0$.

\subsubsection{Anisotropic scattering and Rayleigh scattering}

For acoustic or electromagnetic waves scattered from particles much
smaller than the wavelength
there is no $s$-wave scattering. Instead one has to leading order
the dipole-dipole or $p$-wave scattering. It is expressed by the
phase function for {\it Rayleigh scattering},
\begin{equation} p(\cos\Theta)=\frac{3}{4}(1+\cos^2\Theta).\end{equation}

Scattering from spherically symmetric particles is usually 
anisotropic, but cylindrically symmetric with respect to the
incoming direction. For arbitrarily shaped scatterers this symmetry holds
only after averaging over their possible orientations. 
In such situations the phase function
 $p(\cos\Theta)$ depends only on $\phi-\phi'$, and
the average over $\phi$ can be carried out. This leads to the projected 
phase function, with $\mu=\cos\theta$,
\begin{equation} p_0(\mu,\mu')=\int\frac {\d\phi}{2\pi}
p(\theta\phi;\theta'\phi')
=\int\frac{\d\phi}{2\pi}\frac{\d\phi'}{2\pi}p(\theta\phi;\theta'\phi').
\end{equation}
For Rayleigh scattering one has, using the equality
$ \cos\Theta=\n\cdot\n'=\sin{\theta}\sin{\theta'}
\cos(\phi-\phi')+\cos{\theta}\cos{\theta'}$,
\begin{equation} p_0(\mu,\mu')=\frac{3}{8}(3-\mu^2-\mu'^2+3\mu^2\mu'^2).
\end{equation}

Because of the form of the $\J$ integral, it is useful to introduce
\begin{equation} \Gamma(\tau)=\frac{2\pi}{I_0}\int_{-1}^1 \d\mu
 \I(\tau,\mu)\qquad
 \Delta(\tau)=\frac{2\pi}{I_0}\int_{-1}^1 \d\mu \mu^2\I(\tau,\mu)
\label{GammaDelta}. \end{equation} 
The radiative transfer equation yields the coupled integral equations
\begin{eqnarray} \Gamma(\tau)&=&\eexp{-\tau/\mu_a} 
+(\frac{9}{16}E_1-\frac{3}{16}E_3)*\Gamma+
(\frac{9}{16}E_3-\frac{3}{16}E_1)*\Delta \nonumber
\\
\Delta(\tau)&=&\mu_a^2\eexp{-\tau/\mu_a}
+(\frac{9}{16}E_3-\frac{3}{16}E_5)*\Gamma+
(\frac{9}{16}E_5-\frac{3}{16}E_3)*\Delta \label{GammaDeltavgl}. \end{eqnarray}
Here we introduced the exponential integrals $E_k$,
 \begin{equation} E_k(\tau)=\int_0^1 \frac{\d\mu}{\mu}\mu^{k-1}
 \eexp{-\tau /\mu}=
\int_1^\infty \frac{\d y}{y^k}\eexp{-\tau y}, \end{equation} 
and the product
\begin{equation} (E*f)(\tau)=\int_0^\infty d\tau'
 E(|\tau-\tau'|)f(\tau').\end{equation}
Eqs.~(\ref{GammaDeltavgl}) involves two coupled functions and
represent the simplest extension of isotropic scattering. They 
have been analyzed by \citeasnoun{vanderhulst2}.

\subsubsection{The transport mean free path and the absorption length}

We have discussed how unscattered intensity decays exponentially as a function
of the distance from the source. This occurs because more and more light is
scattered out of the direction of the beam.  The characteristic  distance
between two scattering events is the scattering mean free path $\ell_{sc}$. Now
suppose that scattering is rather ineffective, so that at each scattering the
direction of radiation is not changed much.  The diffusion constant must then
be large. In other words, the factor $\ell$ in the identity
$D=\frac{1}{3}v\ell$  cannot be the scattering mean free path. Intuitively one
expects that  $\ell$ is the distance over which also the direction of radiation
gets lost. This length scale is called the transport mean free path. We show
now how it follows from the radiative transfer equation. 

The time dependent radiative transfer equation has the form
\begin{equation}\label{tSTV}
t_{sc} \frac{\partial}{\partial t}\I({\bf r},{\bf n},t)+
\ell_{sc}\n\cdot\nabla \I({\bf r},{\bf n},t)=
\int\frac{\d\n'}{4\pi} p(\n,\n')
\I({\bf r},{\bf n}',t)-\I({\bf r},{\bf n},t),
\end{equation}
where $t_{sc}$ is the mean time between two scatterings.
We can now introduce the local radiation density
$I$ and the local current density ${\bf J}$ as
 \begin{equation} I({\bf r},t)=\int \d\n \, \I({\bf r},\n,t), \qquad
{\bf J}({\bf r},t)=\frac{\ell_{sc}}{t_{sc}}\int \d\n \I({\bf r},\n,t)
\,\n.\end{equation}
Integration of Eq.~(\ref{tSTV}) yields the continuity equation
\begin{equation} \partial_t I({\bf r},t)+\nabla\cdot{\bf J}({\bf r},t)
=-\frac{1-a}{t_{sc}}I({\bf r},t).\end{equation}
Here we introduce the albedo (from the Latin {\it
albus}, white), the whiteness of the scatterer,
\begin{equation} a=\int\frac{\d\n}{4\pi}p(\n,\n').\end{equation}
For $a=1$ there is no absorption. If we multiply Eq.~(\ref{tSTV})
 by $\n$ and integrate over
 $\n$ we obtain
\begin{equation} \label{JJ}
\frac{t_{sc}^2}{\ell_{sc}}\partial_t {\bf J}+
\frac{t_{sc}}{\ell_{sc}}{\bf J}+\ell_{sc}\nabla\cdot 
\int \d\n \I({\bf r},\n,t)\n\n=
\int \d\n\, \d\n'\, p(\n,\n')\, \I({\bf r},\n',t)\n.\end{equation}
If scattering is (on the average) spherically symmetric, the integral 
 $\int \d\n \,p(\n,\n')\,\n$ can only be proportional to $\n'$.
If one takes the inner product with $\n'$,
one finds the average cosine of the scattering angle,
\begin{equation}\label{avcos}
\langle \cos\Theta\rangle= \langle\n\cdot\n'\rangle=
\int \frac{\d\n}{4\pi}
p(\n,\n')\,\n\cdot\n'=\int \frac{\d\n}{4\pi}p(\cos\Theta)\cos\Theta
=\int_{-1}^1\frac{\d\mu}{2}p(\mu)\mu.\end{equation}
Therefore Eq.~(\ref{JJ}) can be written as 
\begin{equation}
\frac{t_{sc}^2}{\ell_{sc}}\partial_t {\bf J}+
\frac{t_{sc}}{\ell_{sc}}(1-\langle\cos\Theta\rangle){\bf J}
=-\ell_{sc}\nabla\cdot \int \d\n \I({\bf r},\n,t)\,\n\,\n.
\end{equation}
The right-hand side depends on $I$ and ${\bf J}$.
If one assumes that the intensity distribution is almost 
isotropic, then the current is much smaller than the density. 
This allows one to make the approximation \cite{ishimaru}
\begin{equation}\label{IIJ} \I({\bf r},\n,t)\approx I({\bf r},t)
+\frac{3t_{sc}}{\ell_{sc}}
\n\cdot{\bf J}({\bf r},t)+\cdots,\end{equation}
since higher order terms are smaller.
Under these approximations it follows that
\begin{equation}\label{tdepdiffeq}
 \frac{t_{sc}^2}{\ell_{sc}}\partial_t {\bf J}+
(1-\langle \cos\Theta\rangle)\frac{t_{sc}}{\ell_{sc}}{\bf J}=-
\frac{\ell_{sc}}{3}\nabla I.\end{equation}
For processes that change slowly in time we thus find
\begin{equation}\label{hulp} {\bf J}({\bf r},t)
=-D\nabla I({\bf r},t). 
\label{eqextra} \end{equation}
When inserted in the continuity equation, Eq.~(\ref{eqextra})
 leads to the wanted 
diffusion equation for the density. Thus under the above assumptions the
radiative transfer equation leads to the diffusion equation,
\begin{eqnarray} \partial_t I({\bf r},t)&=&D\nabla^2 I({\bf r},t)
-\frac{1-a}{t_{sc}}I({\bf r},t)\nonumber\\
&\equiv &D\nabla^2 I({\bf r},t)-D\kappa^2 I({\bf r},t).\end{eqnarray}
The diffusion constant $D$ is given by
\begin{equation} D=\frac{\ell_{sc}^2}{3t_{sc}(1-\langle
 \cos\Theta\rangle)}\equiv
\frac{1}{3}\,v\,\ell_{tr}, \label{eqextra2} \end{equation}
where $v=\ell_{sc}/t_{sc}$ is the transport speed. We wish to stress that
in principle $t_{sc}$ has two contributions: the time to travel from one 
scatterer to the next and the dwell time spent in the neighborhood of 
one scatterer\cite{albada2}. 
In Eq.~(\ref{eqextra2}) the {\it transport mean free path} occurs
\begin{equation}
 \ell_{tr}=\frac{1}{1-\langle \cos\Theta\rangle } \ell_{sc}.\end{equation}
It is the mean distance after which the direction of radiation gets lost. For
strongly forward scattering  $\langle \cos\Theta\rangle $ will be close to
unity so that the transport mean free path becomes large. 
In this excercise we also found an
explicit expression for the  absorption length $L_{abs}$ and the inverse
absorption length $\kappa$  
\begin{equation}\label{Labs=}
 L_{abs}\equiv\frac{1}{\kappa}\equiv
\sqrt{\frac{\ell_{sc}\ell_{tr}} {3(1-a)}}.\end{equation} 

\subsection{Injection depth and the improved diffusion approximation}
\label{vbdb}

In the previous section we avoided the problem of how an incoming plane wave
becomes diffusive inside the medium. We introduced a ``trapping plane'' at the
injection depth $z_0\sim\ell$. The precise statement is that the solution of
the transfer equation, or Milne equation, of a semi-infinite medium starting at
$z=0$ behaves as $I(z)=z+z_0$ for $z\gg\ell$;  formally this expression
vanishes at $-z_0$ outside the medium.  This can also be expressed as \begin{equation}
I(0)=z_0 I'(0) \label{II'0}.\end{equation}

If there is an index mismatch between the system and its surroundings,  the
walls of the system will partially reflect the light. Therefore the light will
remain longer in the system. The transmission coefficient will become smaller
by a factor of order unity. This effect is of practical importance, as the
scattering medium usually has a different index of refraction from its
surroundings, usually air or liquid. Only in the special case of index matched
liquids the mismatch is minimized.

The first to point out the importance of internal reflections were 
\citeasnoun{lagendijk1}. They also noted that $z_0$ changes. For a
one-dimensional medium they give the expression \begin{equation}
z_0=\frac{1+{\overline R}}{1-{\overline R}}\ell_{sc} , \end{equation} in which
${\overline R}$ is the mean  reflection coefficient.

For the three-dimensional situation, an analogous result was worked out
by \citeasnoun{zhu}:
For a system in which only the $z$ dependence is relevant,
Eq.~(\ref{IIJ}), relating the specific intensity $\I$ to the radiation density
 $I$ and the current density ${\bf J}$, reads
\begin{equation} \I(z,\mu)=I(z)+\frac{3}{v}\mu J_z(z), \end{equation}
with velocity $v=\ell_{sc}/t_{sc}$, where $t_{sc}$ is the scattering time. 
It follows that
\begin{equation} J_z(z)=\frac{v}{4\pi}\int_{-1}^{1}\mu \,
\d\mu\int_0^{2\pi}\d\phi \, \I(z,\mu) . \end{equation}
From this one reads off that the radiation current per unit of
solid angle $\d\Omega=\d\mu\, \d\phi$ equals 
\begin{equation} \frac{\d J_z}{\d\Omega}=v\cdot\mu\cdot\frac{\I}{4\pi}
=\frac{1}{4\pi}\left\{ \mu vI(z)+3\mu^2J(z)\right\}. \end{equation}
The total radiation at depth $z=0$ in the positive $z$ direction is thus
\begin{eqnarray} \label{J+z}
J_{+z}(0)&=&\frac{v}{4\pi}\int_0^1\mu\,\d\mu \,2\pi\,
\I(0,\mu)\nonumber\\
&=&\frac{v}{4}I(0)+\frac{1}{2}J_z(0)\nonumber\\
&=&\frac{v}{4}I(0)-\frac{v\ell_{tr}}{6}I'(0), \end{eqnarray}
where the diffusive current (\ref{hulp}) has been inserted.

In the absence of internal reflections 
$\d J_z/\d\Omega$ must vanish for all $\mu>0$. If we impose that $J_{+z}$
vanishes and if we compare with Eq.~(\ref{II'0}), we obtain
\begin{equation} z_0=\frac{2}{3}\ell_{tr}. \end{equation}
For isotropic scattering this expression is not far from the exact
value $z_0=0.71044\ell$ that follows from the radiative transfer
equation, see Sec.~\ref{theoexact}.

Let us now assume that the refractive index of the scattering medium
$n_0$ differs from that of its surroundings $n_1$. The ratio 
\begin{equation} m=\frac{n_0}{n_1}. 
\end{equation} exceeds unity for a dry medium in air, 
but can be smaller than unity if the medium is in between glass plates with 
a high refractive index. In both cases internal reflections will appear
at the interface.
The reflection coefficient is given by 
\begin{equation} R(\mu)=\left| \frac{\mu-\sqrt{m^{-2}-1-\mu^2}}
                 {\mu+\sqrt{m^{-2}-1-\mu^2}}\right|^2. \end{equation}
$R(\mu)$ equals unity in the case of total reflection, 
namely, when the argument of the
square root becomes negative. We can now calculate the current that 
is internally reflected at the interface $z=0^+$:
\begin{eqnarray} J_{+z}^{\rm refl}
&=&-\int_{-1}^0 \d\mu\int_0^{2\pi}\d\phi\,R(\mu)
\frac{\d J_z}{\d\Omega}
=-\int_{-1}^0 \frac{\d\mu}{2} R(\mu)\left(v\mu I(0)+
3\mu^2 J_z(0)\right)\nonumber\\
&=&\frac{v}{2}C_1 I(0)-\frac{3}{2}C_2J_z(0)
=\frac{v}{2}C_1I(0)+\frac{v\ell_{tr}}{2}C_2I'(0), 
\end{eqnarray}
where
\begin{equation} \label{C1=C2=}
C_1=\int_0^1 \d\mu \mu R(\mu), \qquad C_2=\int_0^1 \d\mu \mu^2 R(\mu). \end{equation}
By equating this angular average to the angular average
(\ref{J+z}), Zhu, Pine and Weitz derive Eq.~(\ref{II'0})
with 
\begin{equation} \label{z0===}
 z_0=\frac{2}{3}\,\frac{1+3C_2}{1-2C_1}\,\ell_{tr} . \end{equation}
It was later pointed out by \citeasnoun{thmn5} that 
this result becomes exact in the limit of large index mismatch
($m\to 0$ or $m\to\infty$). The reason is that then the interfaces
are good mirrors, so that the outer world is not seen and hence 
also close to the mirrors the intensity is diffusive.
 
In the derivation of Eq.~(\ref{z0===}) the phase function has only been used
in the expression for the diffusion coefficient. This led to the occurrence
of the transport mean free path. One would therefore expect that
Eq.~(\ref{z0===}) remains valid for arbitrary anisotropic scattering in the
limit of large index mismatch.
This can be explained as follows: The interfaces act as good mirrors.
Therefore many scatterings also occur close to the wall before the radiation
can exit the medium. After many scatterings the radiation has become
isotropic anyhow. Evidence for this point will be given in
 Sec.~\ref{theoanisotropy}.

\section{Microscopics: wave equations, $t$ matrix, and cross sections}
\label{theomicroscopy}

\subsection{ Schr\"odinger and scalar wave equations}
Light scattering is described by the Maxwell equations. The vector 
character of the amplitudes leads to a tensor character of the
intensity. In regard to sound waves or Schr\"odinger waves, this 
introduces extra complications, that we do not address here.
For discussions of the vector case see, e.g.,
\citeasnoun{vanderhulst2}, \citeasnoun{mackintoshjohn1},
\citeasnoun{peters}, and \citeasnoun{amic2}. The vector character is 
especially important in case of multiple scattering of light
in a Faraday active medium in the presence of a magnetic field.
For fundamental descriptions, see
\citeasnoun{mackintoshjohn2} and \citeasnoun{vtigmaynnieuw}.
This field also includes the so-called photonic Hall effect,
a light current that arises, under these conditions, perpendicular
 to the magnetic field.
From studies of the diffuse intensity ~\citeasnoun{Nunp93} 
imagined such a  mechanism to exist.
This was indeed shown by ~\citeasnoun{vTiggHall}, and confirmed 
experimentally by ~\citeasnoun{RikkvT}.
Another interesting application is multiple scattering 
of radiation emitted by a  relativistic charged particle 
in a random environment, see ~\citeasnoun{Gevorkian}
for diffusion, and ~\citeasnoun{GevorkianN} for enhanced backscatter.

Acoustic waves and spinless electrons are described by scalar waves
and this turns out to be a good approximation for light as well.
The Schr\"odinger equation reads in units in which ${\hbar^2}/{2m}=1$ 
\begin{eqnarray}
-\nabla^2\Psi + V\Psi &=& E\Psi .
\end{eqnarray}
As potential we choose a set of point scatterers.
\begin{equation} V({\bf r}) = -\sum_i u\delta({\bf r}-{\bf R}_i) . 
\end{equation}
Here $-u$ is the bare scattering strength and ${\bf R_i}$ are the locations of
the scatterers.

Acoustic waves are described by the classical wave equation
\begin{eqnarray}
\label{klassiek1}
\nabla^2\Psi - \frac{\varepsilon({\bf r})}{c^2}
\frac{\partial^2}{\partial t^2}\Psi
&=&0, \end{eqnarray}
where $\varepsilon$ is the normalized local mass density and
$c$ is the speed of sound in a medium with $\varepsilon=1$.
We shall apply the same equation to light, thereby neglecting 
its vector character. $\varepsilon$ is then the dielectric constant, and
$c$ is the speed of light in vacuum where $\varepsilon=1$.
For monochromatic waves $ \Psi({\bf r},t) = \Psi({\bf r})\eexp{i \omega t}$
the time derivatives are replaced by frequency $i\omega$. 
In practice this applies to a stationary experiment with a monochromatic
beam. For classical waves we can also introduce point
scatterers by setting $ \varepsilon({\bf r}) = 1+\sum_i \alpha
\delta({\bf r}-{\bf R}_i)$, where $\alpha$ is the polarizability of the 
scattering region that we approximate by a point.
The Schr\"odinger equation and the scalar wave equation then
both take the form
\begin{eqnarray}
\label{beiden}
&-\nabla^2\Psi - u\sum_i \delta({\bf r}-{\bf R}_i) \Psi = k^2\Psi    ,
\\ \nonumber
&\mbox{where } \left\{
 \begin{array}{ll}
 u=\mbox{constant} & \mbox{in the Schr\"odinger
equation; $k=\sqrt{E}$} \\
 u=\alpha k^2 & \mbox{in the classical wave equation; $ k=\omega/c $.} 
 \end{array} \right.    
\end{eqnarray}
Many static results can be derived without specifying which kind  of waves are
discussed. However, the dynamics of acoustic waves will be different from
those of Schr\"odinger waves due to the frequency dependence of $u$.

\subsection{The $t$ matrix and resonant point scatterers}
To elucidate the notion of the $t$ matrix,
we start the microscopic description with a simple example: 
scattering of an incoming beam from one scatterer in one dimension.
The quantum-mechanical wave equation in $d=1$ with a scatterer 
in $x=x_0$ reads
\begin{eqnarray} \label{1dim}
-\Psi''(x)-u\delta(x-x_0)\Psi(x) &=& E\Psi(x) \qquad \mbox{ with }E = k^2 .
\end{eqnarray}
Assuming that a wave $\eexp{ikx}$ comes in from $x=-\infty$, we look
for a solution of the form 
\begin{eqnarray}
\Psi = \left\{
 \begin{array}{ll}
 \eexp{ikx}+A\eexp{-ik(x-x_0)} &\mbox{ $x \leq x_0$} \\
 B\eexp{ik(x-x_0)} &\mbox{ $x\geq x_0$} .
 \end{array}
\right.
\end{eqnarray}
The constants $A$ and $B$ are determined by the continuity of $\psi$ and
the cusp in $\psi'$ at $x=x_0$. The solution can then be represented as 
\begin{eqnarray}
\Psi &=& \eexp{ikx}-\frac{u}{u+2ik} \eexp{ik|x-x_0|}\eexp{ikx_0} 
\nonumber \\
&=& \eexp{ikx} + G(x,x_0)\,t\, \eexp{ikx_0}.
\label{t-1d}
\end{eqnarray}
The first term is just
the incoming wave. The second term represents the wave 
at the scatterer in $x_0$,
where it picks up a scattering factor $t$, 
\begin{eqnarray}\label{t1d} t &=& \frac{u}{1-iu/(2k)}, \end{eqnarray}
and is transported to $x$ by the medium without scatterers. 
The Green's function in the medium without scatterers, 
defined as $(-\partial^2/\partial x^2 -k^2)G(x,x')
=\delta(x-x')$, reads
\begin{eqnarray}
\label{1dgreenx}
G(x,x')=\frac{\eexp{ik|x-x'|}}{-2ik}.
\end{eqnarray}
Note that we have chosen the sign of $i\epsilon$ such that
the Fourier representation of the pure Green's function reads
$G(p)=1/(p^2-k^2-i \epsilon)$. 

\subsection{The $t$ matrix as a series of returns}

In the above example Eqs. (\ref{t1d}) and (\ref{1dgreenx})
show that the $t$ matrix can be expressed as
\begin{eqnarray}
\label{t-g}
t(x_0)=\frac{u}{1-uG(x_0,x_0)}   .
\end{eqnarray}
Expansion in powers of $u$ yields the ``Born series'',
a series 
that have a clear physical interpretation: the term of order $u^k$ describes
waves that arrive at the scatterer, return to it $k-1$ times, and then 
leave it for good:
\begin{eqnarray}
\label{t-expansie}
t &=& \; \; \; \; \; u \; \; \; + \; uGu\;+uGuGu+uGuGuGu+... \\
\bullet &=& \! \! \! \! \! 
\unitlength=1.0mm
\special{em:linewidth 0.4mm}
\linethickness{0.4mm}
\begin{picture}(66.00,5.00)
\thicklines
\put(22.00,1.00){\line(1,0){6.00}}
\put(39.00,1.00){\line(1,0){6.00}}
\put(55.00,1.00){\line(1,0){9.00}}
\put(66.00,1.00){\makebox(0,0)[lc]{+ ...}}
\put(49.00,1.00){\makebox(0,0)[lc]{+}}
\put(32.00,1.00){\makebox(0,0)[lc]{+}}
\put(15.00,1.00){\makebox(0,0)[lc]{+}}
 \put(6.80,0.12){$\times$}
\put(26.80,0.12){$\times$}
\put(20.80,0.12){$\times$}
\put(37.80,0.12){$\times$}
\put(43.80,0.12){$\times$}
\put(40.80,0.12){$\times$}
\put(53.80,0.12){$\times$}
\put(56.80,0.12){$\times$}
\put(59.80,0.12){$\times$}
\put(62.80,0.12){$\times$}
\thinlines
\put(25.00,3.50){\oval(6.00,3.00)[t]}
\put(40.50,3.50){\oval(3.00,3.00)[t]}
\put(43.50,3.50){\oval(3.00,3.00)[t]}
\put(56.50,3.50){\oval(3.00,3.00)[t]}
\put(59.50,3.50){\oval(3.00,3.00)[t]}
\put(62.50,3.50){\oval(3.00,3.00)[t]}
\end{picture}
\end{eqnarray}
The curved lines in the figure indicate that scattering occurs from
the same scatterer (so far there is only one).
The $t$ matrix is indicated by a circle
in the figure. For an extended scatterer
 this expression reads 
\BA\label{TVT}
t({\bf r},{\bf r}')&=&-V({\bf r})\delta({\bf r}-{\bf r}')+V({\bf r})G({\bf r},{\bf r}')V({\bf r}') 
\nonumber \\ &&-\int d^3{\bf r}'' V({\bf r})
G({\bf r},{\bf r}'') V({\bf r}'') G({\bf r}'',{\bf r}') V({\bf r}') + \ldots \EA
Except for the case of point scatterers, this iterative solution of 
the single scatterer problem is not very helpful. 
The
problem of vector wave scattering from a sphere was solved by
\citeasnoun{mie}, but the result is quite involved, too involved for
our multiple scattering purposes. 
The approach shows, however, that the $t$ matrix
depends on the Green's function $G$, which describes propagation in
the medium without that scatterer. However, the Green's
function will depend on further details
of the scattering medium, such as the presence of walls or other scatterers.
Generally, it cannot be taken from the literature but has to be calculated for the
problem under consideration:
{\it the $t$ matrix describes scattering in a local environment.}
The standard infinite-space $t$ matrix 
found in the literature only applies to that specific situation.

\subsection{Point scatterer in three dimensions}

In three dimensions the Green's function reads 
\begin{equation}\label{Garound0}
 G({\bf r},{\bf r'})= \frac{\eexp{ik|{\bf r}-{\bf r'}|}}{4 \pi |{\bf
r}-{\bf r'}|}
\approx \frac{1}{4\pi|{\bf r}-{\bf r'}|} + \frac{ik}{4\pi}
+ O(|{\bf r}-{\bf r'}|) , \end{equation}
which is closely related to the Yukawa potential for hadron-hadron
interactions. The divergence at ${\bf r}'={\bf r}$ will cause some problems 
in taking the point limit.

\subsubsection{Second-order Born approximation}

For weak scatterers one often truncates the Born series. A complication is that
the Green's function in three dimensions diverges for 
${\bf r} \to {\bf r'}$, which makes
 the Born series (\ref{t-expansie}) ill defined for point 
scatterers. In reality these divergences are cut off by the physical
size of the scatterer, so they play no role for weak scattering.
One thus keeps the first-order term and 
the imaginary part of the second-order term; the
regularized real part of the second-order Born term will be small
compared to the first-order term. This leads to the second-order Born
approximation
\begin{eqnarray} \label{2eordeborn}
t = u+i\, u^2 {\rm Im}\, G({\bf r},{\bf r})   .
\end{eqnarray}
For the system under consideration this becomes
\begin{equation} t=u+iu^2\frac{k}{4\pi}.\end{equation}
The fact that ${\rm Im}\,t >0$ signifies that $t$ matrix still describes
scattering [see Eq.~(\ref{sigmaex})].
More detailed aspects of the scatterer, such as resonances, are not taken 
into account.
For electrons the second order Born approximation is often applied.
Instead of working with point scatterers one usually considers a 
Gaussian random potential with average zero
$\langle V({\bf r}) \rangle=0$ and correlation $\langle
V({\bf r}) V({\bf r}') \rangle =u^2 \delta ({\bf r}-{\bf r}')$.
 This leads exactly 
to $t = iu^2\,{\rm Im}\,G({\bf r},{\bf r})$$=$$iu^2k/4\pi$. 
For light, however, 
the second-order Born approximation is less applicable if 
resonances occur.

\subsubsection{Regularization of the return Green's function}

For finite-size scatterers the divergence of the Green's function at coinciding
points is not a severe problem. The physical scatterer always has some finite
radius, $1/\Lambda$, that cuts off any divergency. When we wish to consider
point scatterers, the divergence does cause a mathematical problem. In
contrast to high energy physics, cutoffs in condensed matter physics represent
physical parameters. Indeed, we shall identify $\Lambda$ as an internal
parameter of the point scatterer, that fixes the resonance frequency.

The return Green's function $G({\bf r}\to{\bf r'})$ resulting from 
(\ref{Garound0})
diverges. It should be regularized as
\begin{eqnarray}
\label{greg1}
G^{\rm reg}({\bf r},{\bf r})&=& \Lambda + \lim_{{\bf r}'\rightarrow {\bf
r}}
\left[ G({\bf r},{\bf r'})-\frac{1}{4\pi |{\bf r}-{\bf r'}|} \right]
=\Lambda + \frac{ik}{4\pi} . \end{eqnarray}
Another method, often used in quantum field theory, is to introduce a
large momentum cutoff,
\begin{eqnarray}
\nonumber
G({\bf r},{\bf r})&=& \int \frac{\d^3{\bf p}}{(2\pi)^3} G({\bf p})
\\ \nonumber
&=& \int \frac{\d^3{\bf p}}{(2\pi)^3}\frac{1}{{ p}^2} +
 \int \frac{\d^3{\bf p}}{(2\pi)^3}
 \{ G({\bf p}) - \frac{1}{{ p}^2} \}
\\
G^{\rm reg}({\bf r},{\bf r})&=& \Lambda +
\int \frac{\d^3{\bf p}}{(2\pi)^3}
\{ G({\bf p}) - \frac{1}{p^2} \}
= \Lambda + \frac{ik}{4\pi} .
\end{eqnarray}
This subtraction is also possible if there are other scatterers or walls.
Different regularization schemes define different point scatterers.
The physical result is largely insensitive to such details. 

\subsubsection{Resonances} 

If we insert the regularized return Green's function  in Eq.~(\ref{t-g}), we
can write the $t$ matrix 
in the same form as in the d=1 case, by introducing the effective
scattering length $U_{\rm eff}$, 
\begin{eqnarray}
\nonumber
t &=& \frac{u}{1-uG^{\rm reg}({\bf r},{\bf r})}
\nonumber \\ &=& \frac{u}{1-u(\Lambda+\frac{ik}{4\pi})}
\nonumber \\&\equiv& \frac{U_{\rm eff}}{1-U_{\rm eff}\frac{ik}{4\pi}}
\label{ueff},  \EA
where
 \BA
U_{\rm eff} &\equiv& \frac{u}{1-u\Lambda}.
\end{eqnarray}
For Schr\"odinger waves, the regularization 
brings just a shift in $u$ as compared to the  second-order Born approximation.
This has no further consequences, since $u$ and $\Lambda$ are constants, and 
so is $U_{\rm eff}$. For scalar waves there is an important
difference, 
since then 
$u$ is frequency dependent [c.f. Eq.~(\ref{beiden})] : $u = \alfa k^2$.
Let us call $k_*$ the wave number at resonance. 
If we identify\footnote{That can be done provided $\alfa >0$, that is to say,
when $\varepsilon_{\rm scatterer} >
 \varepsilon_{\rm medium} $.}
$\Lambda \equiv 1/(\alfa k_*^2)$, we get
\BA
t&=&\frac{\alfa k^2k_*^2}{k_*^2-k^2-ivk_*^2k^3/4\pi} \\
U_{\rm eff} &=& \frac{\alfa k^2}{1-k^2/k_*^2} \label{ueff2} .
\end{eqnarray}
One sees that $U_{\rm eff} \to \infty$ for $k \to k_*$.
Then $t$ becomes equal to $t_* = 4\pi i/k=2i\lambda$. 
As $\lambda \gg 1/\Lambda$ this shows resonance 
with strong scattering: the effective scattering length
$|t|=2\lambda$ is much larger than 
the physical size of the scatterer $\sim 1/\Lambda$.

The resonance is an {\em internal} resonance of the scatterer, 
comparable with the s-resonance of a Mie sphere.
It is strongly influenced by the environment in which the sphere
is embedded. 
For small frequencies
 $t\approx \alfa k^2$, leading to the Rayleigh law
 $\sigma \sim \omega^4$ for $\omega \rightarrow 0$. 

\subsubsection{Comparison with Mie scattering for scalar waves}
We compare the above result of regularization with an exact result.
The $t$ matrix for scalar s-wave scattering reads
 (see, e.g. \citeasnoun{merzbacher}, page 238).
\begin{equation}
t=\frac{4\pi \eexp{-2ika}}
{mk \cot (mka) -ik} -\frac{4\pi \eexp{-ika} \sin ka}{k},
\end{equation}
where $a$ is the radius of the sphere and $m$ the ratio between 
refractive indices of the sphere and the outside.
The first resonance occurs at wave number $k_*=\pi/(2ma)$. It becomes 
sharp if one takes small
 $a$, large $m$, such that $k_*$ remains fixed. 
For $k$ close to $k_*$ one gets
\begin{equation}
t=4\pi \left( \frac{\pi^4}{32m^2 a^3}(\frac{1}{k^2}-\frac{1}{k_*^2})-ik
\right)^{-1} \label{tmie} .\end{equation}
For our point scatterer (\ref{ueff}) we have
\begin{equation}
t=\frac{U_{\rm eff}}{1-iU_{\rm eff}k/4\pi}=4\pi \left( \frac{4\pi}{\alfa}
(\frac{1}{k^2}-\frac{1}{k_*^2}) -ik \right)^{-1} \label{tmie2} , \end{equation}
where the scattering length is
\begin{equation} \alfa=\int d^3 r (\epsilon(r)-1)=\int_{r<a}d^3 r [m^2-1]
=\frac{4\pi a^3}{3} (m^2-1). \end{equation} 
In the limit of small $a$ this becomes $ \alfa=\frac{3\pi a^3 m^2}{3}$.
Comparing Eq.~(\ref{tmie}) with 
Eq.~(\ref{tmie2}) we find the prefactors $4\pi/\alfa $ and $\pi^4 /32 m^2 a^3$.
The difference is a factor $\pi^4/96\approx 1.0147$, thus the results 
coincide within 2\%.
In Eq.~(\ref{greg1}) we called $1/\Lambda$ a measure of the radius
of the scatterer. We find, if
$m \gg 1$, \begin{equation} \frac{1}{\Lambda}=\frac{k_*^2
(m^2-1)a^3}{3}\approx \frac{\pi^2}{12} a \approx 0.822 a . \end{equation}
Indeed $1/\Lambda$ is a good measure of the radius.  We have thus found a
simple expression for the $t$ matrix which incorporates the essential physics
near the s-wave resonance. Finally, we mention that  Mie scatterers absorb
radiation if the refractive index has a small imaginary part. For point
scatterers absorption is described by giving $u$ a small negative imaginary
part.

\subsection{Cross sections and the albedo}

In the literature several cross sections are encountered.
Here we discuss the three most important.
Let a plane wave be incident in direction $\n'$.
As mentioned above, the total wave scattered from one scatterer is
\begin{equation} \Psi({\bf r})=\eexp{ik\n'\cdot{\bf r}}
+\int d^3{\bf r}'d^3{\bf r}'' G({\bf r},{\bf r}')t({\bf r}',{\bf r}'')
\eexp{ik\n'\cdot{\bf r}''} \label{mcw20},
\end{equation}
with a real-space representation of the $t$ matrix [see Eq.~(\ref{TVT}].
Its Fourier transform is called the off-shell $t$ matrix, 
\begin{equation} t(\p,\p')=\int d^3{\bf r}
d^3{\bf r}'\eexp{-i\p\cdot{\bf r}+i\p'\cdot{\bf r}'}
t({\bf r},{\bf r}')\label{mcw21} .\end{equation}
Let us assume that the center of the scatterer is located at the origin.
Far away it holds that 
\begin{equation} G({\bf r},{\bf r}')\approx 
\frac{\eexp{ikr-ik\n\cdot{\bf r}'}}{4\pi r}
\qquad\quad (\n\equiv\frac{{\bf r}}{r}). \end{equation}
We insert this in Eq.~(\ref{mcw20}) and with (\ref{mcw21}) we find that
the scattered wave has the form 
\begin{equation} \Psi_{sc}({\bf r})\approx 
\frac{\eexp{ikr}}{4\pi r}t(k\n,k\n') .
\end{equation} 
The {\it scattering cross section} is defined as the scattered
 intensity integrated over a sphere, normalized by the incoming intensity
\begin{eqnarray}
\sigma_{sc}=\int_{4\pi} r^2d\n
\left| \frac{t(k\n,k\n')}{4\pi r} \right|^2
=\frac{1}{(4\pi)^2} \int_{4\pi}d\n \vert t(k\n,k\n')\vert^2 . 
\end{eqnarray} 
Notice that the $t$ matrix is only needed for momenta
$|\p|=k$ (``far field'', ``on the mass shell'', ``on shell $t$ matrix'`). 
One often denotes it by $t(\n,\n')$.
An {\it isotropic point scatterer} thus has scattering cross section
\begin{equation} \sigma_{sc}=\frac{\bar t t}{4\pi} . 
\label{sigmasc}\end{equation}

The second important quantity is the extinction cross section. It
tells how much intensity is lost from the incoming beam.
Let us assume that a plane wave
is incident along the $z$ axis from $z=-\infty$. We consider the intensity
in a small solid angle around the $z$ axis for large positive $z$.
Because then $r\approx z+(x^2+y^2)/2z$, the total wave is given by
\begin{equation} \Psi({\bf r})=\eexp{ikz}+t(\n,\n)\frac{\eexp{ikz}}{4\pi z}
\eexp{ik(x^2+y^2)/2z} . \end{equation}
The intensity for large $z$ is
\begin{equation} \Psi^*\Psi=1+{\rm Re}\frac{t(\n,\n)}{2\pi z}
\eexp{ik(x^2+y^2)/2z} .\end{equation}
We integrate this over $x$ and $y$ inside an area $A$
perpendicular to the $z$-as. The condition that $x$ and $y$ be much smaller
than $z$ makes the integrals Gaussian, after which the $z$ dependence 
disappears:
 \begin{equation} \int_A dxdy \Psi^*\Psi=A-\sigma_{ex}.\end{equation}
The {\it extinction cross
section} is the surface over which the incoming beam has to be 
integrated to collect an equal amount of intensity. It is equal to
\begin{equation}\label{sigmaex} 
\sigma_{ex}(\n)=\frac{{\rm Im}\,t(\n,\n)}{k}. \end{equation}

For a point scatterer one has $t(\n,\n)=t$.
The albedo of the scatterer is the ratio of scattered
and extinct intensity, \begin{equation} \label{albedo} 
a=\frac{\sigma_{sc}}{\sigma_{ex}}= 
\frac{k \bar t t}{4\pi {\rm Im} t } . \end{equation}
For pure scattering $a=1$; this is called the {\it optical theorem}. If
absorption is present the absorption cross section can be defined as $
\sigma_{abs}= \sigma_{ex}-\sigma_{sc}$ The albedo then equals \begin{equation}
a=\frac{\sigma_{sc}} {\sigma_{sc}+\sigma_{abs}} . \end{equation}

For an extended spherical scatterer the extinction cross section is 
angle dependent. One defines the 
earlier encountered phase function in terms of the $t$ matrix as
\begin{equation} p(\cos\Theta)=p(\n\cdot\n')
=\frac{k \mid t(\n,\n')\mid^2}{4\pi {\rm Im}t},\end{equation} 
with \begin{equation} \int \frac{d \n}{4\pi} p(\n, \n')=a. \end{equation}
Note that, because of the far-field construction, the optical theorem cannot be
applied immediately in a system with many scatterers. Instead one has to impose
the {\it Ward identity}, which is its generalization.

\section{Green's functions in disordered systems}
\label{greense} \label{theogreensfies}

After our microscopic treatment of a single scatterer, we now consider
scattering from many scatterers. The Green's function of a given sample depends
on the realization of disorder: the location and the orientation of scatterers.
Averaged over disorder it is called the {\it amplitude Green's function}. It
describes, on the average, unscattered propagation, such as an
incoming beam or the wave scattered from any given scatterer. It should be
contrasted with the {\it diffuse intensity}, which is the (multiple) scattered
intensity that will be discussed below. For a introduction to Green's
functions in disordered systems see \citeasnoun{economouboek}.

\subsection{Diagrammatic expansion of the self-energy}

The amplitude Green's function is related to two important concepts: 
the density
of states and the self-energy. Let us consider waves with frequency $\omega=c
k_0$. It will be seen that the presence of many scatterers will change the
``bare'' wavenumber $k_0$ into the ``effective wavenumber'' 
\begin{equation} K=k+\frac{i}{2\ell_{sc}}. \end{equation}
This describes a phase velocity $v_{ph}=\omega/k=ck_0/k$ .

For an electron in a random potential
the definition of the amplitude Green's function is 
\BA G_{{\bf r},{\bf r}'}&=&\prod_{{\bf r}''}
 \int \d V_{{\bf r}''} p(V_{{\bf r}''}) g_{{\bf r},{\bf r}'} , \EA
with \begin{equation} g_{{\bf r}, {\bf r}'}=
\left( \frac{1}{ p^2-E+V} \right)_{{\bf r},{\bf r}'}. \end{equation}
The effect of the random potential is to introduce in the bare 
Green's function $g(p)$ the self-energy $\Sigma$,
\begin{equation} \label{Gselfen} G(p)=\frac{1}{p^2-E-\Sigma(p)}.
\end{equation}
Exact averaging over the disorder 
is only possible in particular cases in one dimension 
and cannot be done in general.
\footnote{Exactly solvable models exist in one dimension.
 B.I. Halperin considers Gaussian white noise potential on a line
\cite{halp}; Th.M. Nieuwenhuizen considers exponential 
distributions on a 1D lattice \cite{thmn6,thmn7}.}
Therefore, we employ a diagrammatic approximation to calculate the self-energy.
Suppose that we have a system with randomly located strong scatterers
(water drops in fog, lipid particles in milk, TiO$_2$ particles in a liquid
or solid sample). In the limit of point scatterers the potential
becomes $V({\bf r})=-\sum_{i=1}^N u \delta({\bf r}-{\bf R}_i)$. The
Green's function of this problem is represented diagrammatically 
in Fig.~\ref{gvg}.

$G_0$ denotes a bare propagator, that is to say, the propagator in the medium
without scatterers. The quantity needed is $g$, the Green's function of the
random medium, also called dressed propagator. To calculate all diagrams would
amount to solving the problem exactly, which is usually not possible. We shall
therefore assume that the density of scatterers $n$ is small. This will allow
us to set up a perturbative expansion of $\Sigma$.
We average over all disorder configurations, that is to say, all
positions (and orientations) of the $N$ scatterers in the volume $V$.
We sort the diagrams (see Fig.~\ref{sigma}) and get
\BA
G=\langle g \rangle &=&
G_0+G_0 \Sigma G_0+G_0 \Sigma G_0 \Sigma G_0 +\ldots \nn
&=& G_0+G_0\Sigma G\label{Dysoneqn}\\ 
 &=&\frac{1}{G_0^{-1}-\Sigma}=\frac{1}{-\nabla^2-k_0^2-\Sigma}     .
\EA
The relation (\ref{Dysoneqn}) is the Dyson equation.
The lowest order approximation to $\Sigma$ was already calculated in 
Sec.~\ref{theomicroscopy}, as it is proportional to the $t$ matrix.
Due to the averaging over the scatterer positions, there is a also 
a density dependence, which can already be calculated from the first
order Born term,
 \begin{equation} -\Sigma^{(1)}=-u\int
\frac{d^3\R_1}{V}
\cdots \frac{d^3\R_N}{V} \sum_{i=1}^N \delta({\bf r}-\R_i)=-u\sum_{i=1}^N \int
\frac{d^3{\bf R}_i}{V} \delta({\bf r}-\R_i) = -\frac{uN}{V} =-nu.
\end{equation} 
The same argument holds for all orders in the Born series, so one obtains
to ${\cal O}(n)$
\begin{equation} \Sigma=nt .\end{equation} 
This lowest order approximation is also known as the {\it independent-
scatterer approximation}. 
In Fig.~\ref{sigma} it corresponds to the approximation
$\Sigma=\bullet $. The effective wave number is extracted by comparing the
dressed and bare propagators 
\begin{equation} K=\sqrt{k_0^2+\Sigma}\equiv k+\frac{i}{2\ell}, \end{equation}
which leads to the complex index of refraction 
\begin{equation} m=\frac{K}{k_0}\approx \sqrt{1+\frac{nt}{k_0^2}}. 
\end{equation}

\subsection{Self-consistency}

It is impossible to calculate all diagrams. Often it is feasible, however, to
take into account, without much extra effort, all higher order contributions of
a certain type. In the case of the self-energy we can take $t=u/[1-u G(r,r)]$,
instead of $t=u/[1-u G_0 (r,r)]$. This is called a self-consistent approach.
Now the last diagram in the series for $\Sigma$ of Fig.~\ref{sigma} is
included. Similar terms with any number of intermediate dots are also accounted
for. (One has to be careful to avoid overcounting of diagrams, however).
Nevertheless, some two-scatterer diagrams have still been neglected at this
self-consistent one-scatterer level.

Physically the self-consistent method is very natural: it describes that
propagation from one  scatterer to another does not happen in empty space, but
in a space filled with other scatterers. Therefore self-consistency is a
fundamental concept, and it will serve to satisfy conservation laws (Ward
identities). Indeed, without self-consistency there is no exact cancellation in
the Ward identities. For instance, in the second-order Born approximation one
will derive  Eq.~(\ref{diffbulkeqn}) with $1-a\sim u^3$, formally describing
absorption or even creation of intensity, in situations where intensity should
be strictly conserved. In order to work with such approaches one must neglect
$u^3$ terms. In the more physical self-consistent approach such {\it ad hoc}
manipulations are not needed and not allowed.

The self-consistent $t$ matrix in the independent scatterer
approximation,
\begin{equation}\label{tuL=}
t=\frac{u}{1-u\Lambda-\frac{i}{4\pi}u \sqrt{k_0^2+nt}} \label{tzelf},
\end{equation}
can lead to a real value of $t$. This amounts in this approximation to a
 gap in the density of states \cite{polishchuk}.
Probably there is no real gap, but still a small density of states,
which may lead to Anderson localization \cite{polishchuk}.

\section{Transport in infinite media: isotropic scattering}\label{transport}

Just as the amplitude Green's function follows from solving the Dyson 
equation, the intensity follows from solving the Bethe-Salpeter  equation. In
this section we restrict ourselves to an approximate form of this equation:
the so-called ladder approximation. It will be seen that this is corresponds
to the independent scatterer approximation at the intensity level.
The ladder approximation allows a
microscopic derivation of the radiative transfer equation discussed
in Sec.~\ref{theomesoscopy}.

In infinite media there are no boundaries which simplifies the analysis. 
In practice this situation applies to cases where the distance
to boundaries is many mean free paths, so in the bulk of the
multiple scattering medium. We first consider this situation and 
derive the diffusion equation from microscopics. In doing so, we find 
expressions for the diffusion coefficient and the speed of transport.

\subsection{Ladder approximation to the Bethe-Salpeter equation}

For the transport of energy we must consider the intensity. We have to 
multiply the Green's function by its complex conjugate. In other words,
we must multiply the retarded Green's function by the advanced one.
This must be done before averaging over disorder.
We indicate this in Fig.~\ref{ladderfig} 
by drawing the expansion for $g$ (first line) and drawing the one 
for $g^\ast$ below it.

When scattering from a certain scatterer occurs more than once, 
we indicate this again by dashed lines. Visits to scatterers are included
in $g$ or $g^\ast$, as before, but it may also happen that 
both $g$ and $g^\ast$ scatter from a common scatterer. This 
leads to the connection lines between the upper and lower propagators
in the figure. 

Of special importance are the ladder diagrams, depicted in
Fig.~\ref{ladderfig}(a), which describe the diffuse intensity 
in the independent
scatterer approximation. They lead to the classical picture of propagation of
intensity from one common scatterer to another. As the intermediate
propagation takes place in a medium with many other scatterers, each
intermediate $g$ or $g^\ast$ line visits many new scatterers: they are the
dressed propagators of the previous section.

Alternative names for the ladder diagrams are: ladder sum, diffuson,
particle-hole channel. They are constructed in three steps: (i) Sum all
immediate returns to the scatterers. As was explained in
Sec.~\ref{theomicroscopy}, this replaces the bare scattering potential $u$ by
the $t$ matrix. (ii) Keep only those diagrams in which  $g$ and $g^\ast$ visit
a certain series of common scatterers once and only once and in the same
sequence. The intensity $g^\ast g$ propagates from one scatterer to another. 
(iii) Use the dressed Green's function as intermediate propagator. As a result
in between the common scatterers both $g$ and $g^\ast$ visit any amount of
other scatterers.

Let $I({\bf r})= |\Psi ({\bf r})|^2$ denote the intensity arriving 
at point ${\bf r}$.
According to Fig.~\ref{ladderfig} it can be decomposed into terms without
scattering, terms with one common scattering, term with two, etc:
\BA
\label{laddereqnlo}
I({\bf r})&=& |\Psi_{in} ({\bf r})|^2 
+ n\bar{t}t \int d^3 {\bf r}' |G({\bf r}-{\bf r}')|^2
 |\Psi_{in} ({\bf r}')|^2 \nonumber\\
&+& (n\bar{t}t)2 \int d^3 {\bf r}' d^3{\bf r}''
 |G({\bf r}-{\bf r}')|^2 |G({\bf r}'-{\bf r}'')|^2 
|\Psi_{in} ({\bf r}'')|^2+\cdots
\label{mcw45} 
\EA
which can be written as an integral equation
\begin{equation} \label{laddereqn}
\L({\bf r})=n\bar{t} t |\Psi_{in} ({\bf r})|^2 
+n\bar{t}t \int d^3 {\bf r}' G({\bf r}-{\bf r}') 
G^*({\bf r}-{\bf r}')\L({\bf r}').
\label{mcw4} \end{equation}
where $ \L({\bf r})=n\bar t t I({\bf r})$. 
The ladder propagator $\L({\bf r})$ is the intensity that
leaves point ${\bf r}$ after being scattered.

In the second-order Born approximation the same scatterer is at most visited
twice consecutively, with the result that the $t$ matrices in the above ladder
equations are replaced by the scattering potentials $u$.

\subsection{Diffusion from the stationary ladder equation}

In the bulk, that is to say far from boundaries, the source term
in the ladder equation vanishes. Eq.~(\ref{mcw4}) therefore takes the form
\beq\label{SMbulkeq}
\L({\bf r})= n\bar{t}t\int\mid G({\bf r}')\mid^2 \L({\bf r}+{\bf r}')
d^3{\bf r}'. \label{blad}
\eeq
Now assume that $\L({\bf r})$ varies slowly on the scale of one mean free
path and expand
\begin{equation}
\L({\bf r}'+{\bf r})=\L({\bf r})+{\bf r}'\cdot\nabla \L({\bf r})+
\frac{1}{2}{\bf r}'{\bf r}':\nabla \nabla \L({\bf r}). \label{mvr}
\end{equation}
where $:$ denotes a tensor contraction.
Inserting this in Eq.~({\ref{blad}) yields three contributions. 
The first term is
\begin{equation}
n\bar{t}t\L({\bf r})\int d^3{\bf r}' \frac{\eexp{-r'/\ell}}{(4\pi r')^2}=
n\bar{t}t\frac{\ell}{4\pi}\L({\bf r}) = a\L({\bf r}).
\end{equation}
In the prefactor we have recognized the albedo $a$ 
(see Sec.~\ref{theomicroscopy}).
The second term vanishes due to symmetry. The third term yields 
\begin{equation}
\frac{1}{2}n\bar tt\int d^3{\bf r}' G^*({\bf r}') G({\bf r}')
 {\bf r}'{\bf r}':\nabla\nabla \L({\bf r})
= \frac{1}{2}nt\bar{t}\times\frac{1}{3}\times 2\ell^3 \nabla^2 \L({\bf r}).
\end{equation}
Inserting this in Eq.~(\ref{SMbulkeq}) yields the stationary
diffusion equation 
\begin{equation} \label{diffbulkeqn}
\nabla^2 \L({\bf r}) = \frac{3(1-a)} {\ell^2} \L({\bf r})
\label{difff} . 
\end{equation}
in agreement with eqs. (\ref{tdepdiffeq}) and (\ref{Labs=}),
since for isotropic scattering it holds that $\ell_{sc}=\ell_{tr}$.

\subsection{Diffusion coefficient and the speed of transport}

Here we give a simple derivation of the nonstationary diffusion equation. In
doing so, we automatically encounter the speed of transport. 
We again consider isotropic scatterers. In an infinite system it is useful
to consider the Fourier-Laplace transformed ladder equation
\begin{eqnarray} 
\L(\q,\Omega)&=&S(\q,\Omega)+n\bar t(\omega_-)t(\omega_+)
\int\frac{d^3\p}{(2\pi)^3}
G(p_+,\omega_+)G^*(p_-,\omega_-)\L(\q,\Omega)
\nonumber\\ &\equiv& S(\q,\Omega)+M(\q,\Omega)\L(\q,\Omega)\nonumber\\
&=&\frac{S(\q,\Omega)}{1-M(\q,\Omega)} \label{mcw15}, \end{eqnarray}
where $S$ is the transformed source term; its precise form is of no 
interest here. We introduced
\begin{equation} \p_\pm=\p\pm\frac{1}{2}\q,\qquad 
\omega_\pm=\omega\mp \frac{1}{2}\Omega.
 \end{equation}
The parameters $\p$ and $\omega$ are ``internal'' or ``fast'' variables, which
involve one period of the wave. $\q$ and $\Omega$ are macroscopic or slowly
varying parameters. They describe variations over distances 
much larger than the
wavelength and times much larger than the oscillation period.

At $\Omega=0$, the bulk kernel has the property
\begin{equation} \label{A1q=}
M(\q,\Omega=0)=n\bar tt\int \frac{\d^3p}{(2\pi)^3}G(\p+\q)G^\ast(\p)
=\frac{{\rm arctan}(q\ell)}{q\ell} . \end{equation}
We want to know the behavior for large distances and
times, so for small $\q$ and $\Omega$.
We expand to orders $\Omega$ and $q^2$. Denoting
 $G\equiv G(\p,\omega)$, it holds that
\begin{eqnarray} G(\p+\frac{1}{2}\q,\omega-\frac{1}{2}\Omega)&\approx&
\left(p^2+\p\cdot\q+\frac{1}{4}q^2-
\frac{\omega^2}{c^2}+\Omega\frac{\omega}{c^2}-nt+\Omega\frac{n}{2}\,
\frac{\d t}{\d\omega}\right)^{-1} 
 \\
&\approx& G +\left[ -(\p\cdot\q)-\frac{1}{4}q^2-
\Omega(\frac{\omega}{c^2}+\frac{n}{2}\,\frac{\d t}{\d\omega} ) \right] G^2
+(\p\cdot\q)^2G^3 .\nonumber \end{eqnarray}
Inserting this in Eq.~(\ref{mcw15}) implies for the kernel
\begin{eqnarray} M(\q,\Omega)&=&n{\bar t}t\left\{1-\frac{1}{2}
\Omega(\frac{\d\log t}
{\d\omega}-\frac{\d\log\bar t}{\d\omega}) \right\}\times \\
&& \left\{ I_{11}
-\Omega(\frac{\omega}{c^2}+\frac{nt' }{2})I_{21}
+\Omega(\frac{\omega}{c^2}+\frac{n\bar t'}{2})I_{12} \right. \nonumber \\
&&
\left.
-\frac{q^2}{4}(I_{12}+I_{21})+\frac{k_0^2q^2}{3}(I_{31}+I_{13}-I_{22})
\right\}. \nonumber\end{eqnarray}
Here we have defined the integrals 
\begin{equation} I_{kl}=\int\frac{d^3\p}{(2\pi)^3} G^k(\p)G^{*\,l}(\p).\end{equation}
They are calculated in the appendix. Inserting their values yields
\begin{equation} M(\q,\Omega)= a-i\Omega \tau_{sc}-\frac{1}{3}q^2\ell^2.\end{equation}
As discussed by \citeasnoun{albada2}, $\tau$ has two contributions,
\begin{equation} \tau=\frac{\ell}{c}+\tau_{dw}\equiv\tau_{sc}+\tau_{dw}.\end{equation}
The first term is the ``scattering time'' or 
``time of travel'' $\tau_{sc}=\ell/c$.
The explicit expression for the ``dwell time'' 
$\tau_{dw}$ follows as
\begin{equation} \tau_{dw}= {\rm Im}\frac{\d\log t}{\d\omega}
+\frac{2\pi}{k_0\bar tt}{\rm Re}\frac{\d t}{\d\omega}.\end{equation}
We insert the $t$-matrix of a point scatterer, Eq.~(\ref{ueff}), yielding
\begin{eqnarray} \tau_{dw}&=&{\bar tt} \left( \frac{1}{4\pi cU_{\rm eff}}
+\frac{k_0U_{\rm eff}'} {4\pi U_{\rm eff}^2}\right)
+\frac{2\pi({\bar tt})^2}{k_0{\bar tt}}
\left(\frac{ U_{\rm eff}'} {U_{\rm eff}^4}-\frac{ k_0^2U_{\rm eff}'}
{16\pi^2 U_{\rm eff}^2}
-\frac{k_0}{8\pi^2 cU_{\rm eff}}\right) \nonumber\\
&=&\frac{k_0\bar tt}{8\pi U_{\rm eff}^2} 
\left( 1+\left[\frac{4\pi}{k_0U_{\rm eff}}
\right]^2 \right) \frac{\d U_{\rm eff}}{\d\omega}.\end{eqnarray}
Notice that the terms without $U_{\rm eff}'$ have compensated each other.
This cancellation follows more generally from a Ward identity \cite{albada2}. 
For electrons in a random potential $\tau_{dw}$ therefore vanishes.

For acoustic waves and light waves the situation is more interesting.
Since $U_{\rm eff}$ depends explicitly on frequency, $U_{\rm eff}'$
does not vanish. Both terms are additive, leading to a finite
dwell time $\tau_{dw}$. Using Eq.~(\ref{ueff2}) this yields at resonance
\begin{equation} \tau_{dw}=\frac{2\pi}{\omega}\,\cdot
\frac{1}{\alfa k^3}=\frac{{\rm period}}{\rm coupling}.
\end{equation} 
This dwell time becomes longer when the coupling to the environment, that is,
the normalized scattering strength $k^3\alfa=4\pi k^3a^3 (m^2-1)/3$, becomes
weaker. Therefore this is an important experimental  effect. The speed of
transport, 
 \begin{equation} v=c\frac{\tau_{sc}}{\tau_{sc}+\tau_{dw}}, 
 \end{equation} can be
substantially smaller than the speed of light when realistic values of $n$ are
inserted in this formula. 

For resonant atoms at fixed positions, the reduction of speed may be as large
as $10^6$ ~\cite{VdovinGalitskii,thmn11}. As a final result the diffusion
coefficient $D=v\ell/3$ reduces, as is observed experimentally. For the general
situation and for more details, see \citeasnoun{albada3} and 
\citeasnoun{lagendijk2}. 

With this result for $\tau$ 
we find for the ladder propagator (\ref{mcw15}) at small $\q,\Omega$
\BA \L(\q,\Omega)&=& \frac{3S(\q\approx {\bf 0},t=0)}{\ell^2} 
\frac{1}{q^2+\kappa^2+i\tOmega}.
\EA
involving the reduced external frequency 
\begin{equation} \tOmega\equiv \frac{\Omega}{D}.\end{equation}
This is exactly the propagator of Sec.~\ref{theomacroscopy} with absorption 
length $L_{abs}\equiv 1/\kappa\equiv\ell/\sqrt{3(1-a)}$.

One often considers the Schwarzschild-Milne equation with 
stationary $\delta$ source, $S({\bf r},t)$$=$
$n\bar t t\delta({\bf r})$$=$$4\pi\ell^{-1}\delta({\bf r})$.
In the diffusion approximation this yields 
\begin{equation}\label{difladder} \L(\q,\Omega)=\frac{12
\pi}{\ell^3} \frac{1}{q^2+\kappa^2+i\tOmega} \label{mcw16} .\end{equation}
This form is commonly called the ``diffuson.''

\section{Transport in a semi-infinite medium} \label{theoseminf}

In this section we consider in detail the transport equation for a very thick
slab. We follow the approach of \citeasnoun {thmn5}. We shall also consider the
case of a mismatch  in refractive index between the medium and the outside.

\subsection{Plane wave incident on a semi-infinite medium}

In many situations the index of refraction of the scattering medium differs
from that of its surroundings. An example is TiO$_2$ particles suspended in
a liquid. Boundaries between media of different indexes act partly as mirrors.
They cause a direct reflection of the incoming light and re-inject part of the
multiple scattered light that tries to exit the system. As long as the system
is optically thick, 
they lead to effects of order unity, no matter how small the
ratio $\lambda/\ell$, thus they are important in a quantitative analysis.

We shall calculate the angular resolved intensity profile for the case of a
plane wave incident on a perfectly flat interface. This will create a specular
reflection at the interface. The situation of nonspecular reflections from
nonideal interfaces also has some practical relevance.

Consider a semi-infinite medium. For $z>0$ there is a scattering medium with
refractive index $n_0$. For $z<0$  there is a dielectric with index $n_1\equiv
n_0/m$.  Our notation is indicated in Fig. \ref{skin.eps}. The system is
governed by Eq.~(\ref{beiden}).  We first determine the incoming wave in the
scattering medium.  We consider a system with scatterers in the half-space
$z>0$. We replace the action of the scatterers by a self-energy term. The
average wave equation reads, after Fourier transformation of the transverse
vector to $\qt=(q_x,q_y)$, 
\beqa \label{8.1}
\frac{d^2}{dz^2}\Psi(z)+ P^2 \Psi(z) & =& 0, \mbox{\ \ } z>0 \\
\frac{d^2}{dz^2}\Psi(z)+ p^2 \Psi(z) & =& 0, \mbox{\ \ } z<0 \\
 P^2= k_0^2 - \qt^2 +nt, & & p^2=k_1^2 - \qt^2 \label{Pp} .
\eeqa
A plane wave of unit amplitude, incident from $z<0$, causes
a reflected wave for $z<0$ and a refracted wave for $z>0$:
\beqa\label{psiin=}
\Psi_{in}({\bf r}) & = &
\eexp{i
{\bf k}^a_\perp\cdot\rhoo+ip_az}-\frac{P_a-p_a}{P_a+p_a}\eexp{i
{\bf k}^a_\perp \cdot\rhoo -ip_az} \mbox{\ \ \ } (z<0) \nonumber\\
& = & \frac{2p_a}{P_a+p_a}\eexp{i {\bf k}^a_\perp\cdot\rhoo+iP_az}
\mbox{\ \ \ } (z>0) \label{eqpsiin} .\eeqa
The prefactors in Eq.~(\ref{psiin=})
follow from the requirement of continuity of $\Psi$ and its derivative.
To lowest order in the density we find for the real and imaginary parts of $P$ 
\begin{equation}
P = k_0\cos \theta' + i  \frac{1}{2\ell\cos \theta'}     ,
\end{equation}
where $\theta'$ is the angle of the refracted incoming wave with respect to 
the $z$ axis. 

The source term in the scattering medium, the unscattered
incoming intensity, can be written as
\beq
I_{in}=\left|
\Psi_{in}\right|^2=\mid\frac{2p_a}{P_a+p_a}\mid^2\eexp{-z/\ell\cos\theta_a}
= \frac{p_a}{P_a}T_a\eexp{-z/\ell\mu_a}.
\eeq
Inside the random medium the unscattered intensity is damped
exponentially (Lambert-Beer law).
To leading order in $1/k\ell$,
\begin{equation}
T_a=\frac{4p_aP_a}{(P_a+p_a)^2}=1-R(\mu_a), \mbox{\ \ \ }
R(\mu)=\left|\frac{\mu-\sqrt{\mu^2-1+1/m^2}}
  {\mu+\sqrt{\mu^2-1+1/m^2}}\right|^2,
 \label{mcw22}
\eeq
where $R(\mu)$ is the angular reflection coefficient for scalar waves,
and $R=1$ for total reflection, which occurs when $m>1/\sqrt{1-\mu^2}$.
For vector waves the same equation applies in the $s$-wave
channel \cite{amic2}.
In this expression for $R(\mu)$ the imaginary part of $P_a$,
of order $1/k\ell$, has been neglected. It has been shown
by \citeasnoun{thmn5} 
that we must neglect this to ensure flux conservation.
 
We use the optical depth 
$ \tau={z}/{\ell} $
(not to be confused with the average time per scattering 
$\tau=\tau_{\rm sc}+\tau_{\rm dw}$) and introduce $\Gamma (\tau)$ as
\beq\label{Gammadef}
I(z)=\frac{\ell}{4\pi}\L(z=\tau\ell)=\frac{p_a}{P_a}T_a\Gamma (\tau). \eeq
$\Gamma$ is the diffuse intensity per unit of intensity entering
inside the medium. 
In terms of $\Gamma$ the ladder equation becomes dimensionless
\beq
\Gamma(\tau)=\eexp{-\tau/\mu_a}+\int_0^\infty
 \d\tau'M(\tau,\tau')\Gamma(\tau'),
\label{Milne} \eeq
where $M(\tau,\tau')$ follows from the square of the amplitude Green's 
function. 
In contrast to Eq.~(\ref{SMslab}) $G$ now consists of two terms for $z,z'>0$:
a direct term and a term involving reflection for the boundary $z=0$:
\beq
G(z,z',\qt)=\frac{i}{2P}\{\eexp{iP\mid z-z'\mid}+
\frac{P-p}{P+p}\eexp{iP(z+z')}\}.
\eeq
We shall later need $G$ for $z<0$, $z'>0$ for which 
\beq\label{Goutgoing}
G(z,z',\qt)=\frac{i}{P+p}\eexp{-ipz+iPz'}.
\end{equation}
Inside the random medium, $\mid G\mid^2$ thus has four terms. 
The cross terms oscillate quickly and can
be omitted. We thus keep the previous bulk contribution to the kernel
$M_B$ and the new layer kernel $M_L$, $M=M_B+M_L$.
For the bulk kernel we find 
 \begin{equation}
M_B(z,z')= 4\pi \int \d^2\rhoo\mid G(z,\rhoo;z',\rhoo')\mid^2.
\end{equation}
It holds that
\begin{equation}
M_B(z,0)=\int \d x \d y \frac{1}{4\pi}
\frac{\eexp{-\sqrt{x^2+y^2+z^2}/\ell}}{x^2 +y^2+z^2}.
\end{equation}
Since
\begin{equation} 
r^2=z^2+\rho^2=z^2/\mu^2\Rightarrow 
\rho \d\rho=-z^2 \d\mu/\mu^3.
\end{equation}
one gets
\BA\label{Mbulk}
M_B(z,0) & = &4\pi \int \d x\d y \frac{1}{(4\pi r)^2} \eexp{-z/\ell\mu} =
\int_0^\infty \frac{\rho \d\rho}{2r^2}\eexp{-z/\ell\mu} \nonumber \\
&=& \int_0^1 \frac{\d\mu}{2\mu} \eexp{-z/\ell\mu},
\EA
yielding for the bulk kernel [cf, Eq.(\ref{SMslab})]
\begin{equation}\label{Mbulk=}
M_B(\tau,\tau')=\int_0^1\frac{\d\mu}{2\mu}\eexp{-|\tau-\tau'|/\mu}
=\frac{1}{2}E_1(|\tau-\tau'|), \end{equation}
where $E_1$ is an exponential integral.
In the same fashion the layer term is
\BA\label{Mlayer}
M_L(\tau,\tau')&=&4\pi \int \d x \d y\left| \frac{i}{2P} \frac{P-p}{P+p}e^
{iP(z+z')}\right|^2 \nonumber \\
&=&\int_0^1\frac{\d\mu}{2\mu}\eexp{-(\tau+\tau')/\mu}R(\mu).
\EA
It consists of three effects: exponential decay of intensity that goes from the
point ${\bf r}'$ with angle $\theta$
 towards the wall having an optical path length
$\tau'/\mu$; a reflection factor $R(\mu)$; a further decay over an optical
path length $\tau/\mu$ between the wall and the observation point ${\bf r}$.

We now calculate the angular resolved reflection.
For a plane wave incidence the solution of Eq.
(\ref{mcw4}) is $\L(z,\rhoo)=\L(z)$. We can write for the reflected 
intensity at point $(z',\rhoo')$ with $z'>0$
\beq\label{Irefl}
I_R(z',\rhoo')=\int \d z \d^2\rhoo \: \L(z,\rhoo)\mid
G(z,\rhoo;z',\rhoo')\mid^2 \label{Ir}.
\eeq
Here $\L(z)$ is related to $\Gamma(\tau)$ via $z=\ell\tau$
 and Eq.~(\ref{Gammadef}).

It is useful to consider first a finite area $A$ and
to continue this area periodically. We can then integrate out the 
$\rhoo$ dependence; later we shall take the limit $A\rightarrow\infty$.
Using the periodicity of the surface one can express the Green's function as
\beq
G(z,\rhoo;z',\rhoo')=\frac{1}{A}\sum_{\qt}
 G(z;z';\qt)\eexp{i\qt\cdot(\rhoo-\rhoo')}.
\label{mcw10} \eeq
Since the incoming plane wave is infinitely broad, the diffuse intensity
$I$ does not depend on $\rhoo$. The integration over $\rhoo$ 
is simple:
\BA
\int \d^2\rhoo \mid G(z,\rhoo;z',\rhoo')\mid^2
=\frac{1}{A}\sum_\qt GG^* 
\approx \int \frac{\d^2\qt}{(2\pi)^2} \mid
G(z,z',\qt)\mid^2 .    \label{eqextra4} \EA
The last step holds in the limit $A\rightarrow\infty$. If we denote the
outgoing $\qt$ by ${\bf q}_\perp^b$, we can express $\d^2\qt$
as
\begin{equation}
\d^2{\bf q}_\perp^b=q_\perp^b \d q_\perp^b \d\phi_b
=k_1^2\sin\theta_b\cos\theta_b\d\theta_b\d\phi_b=
k_1^2\cos\theta_b\d\Omega_b. \label{eqextra3}
\end{equation}
Inserting Eqs.~(\ref{eqextra3}) and (\ref{eqextra4}) in Eq.~(\ref{Irefl}) 
yields a result that is independent of the observation point $(z',\rhoo')$,
\beq
I_R(z',\rhoo')=\int \d z\int \frac{k_1^2\cos\theta_b}{4\pi^2}\d\Omega_b
\frac{1}{\mid P_b+p_b\mid^2}\L(z)\eexp{-z/\ell\cos\theta_b'},
\eeq
where $\theta_b'$ is the direction of the radiation in the medium, that is
refracted into the outgoing direction $\theta_b$ [see Fig.~\ref{skin.eps}]; 
we also recall that $\mu_b=\cos\theta_{b}'$.
We finally find for the angular resolved reflected diffuse
intensity
\begin{eqnarray} A_R\equiv \frac{\d R(a\rightarrow b)}{\d \Omega_b}=
\frac{\d I_R}{\d\Omega_b}&=&\frac{k_1^2\cos\theta_b}{4\pi^2}
\frac{T_b}{p_bP_b}\int \d z\L(z)\eexp{-z/\ell\cos\theta_b'}.
\end{eqnarray}
Using Eqs.~(\ref{Gammadef}) and (\ref{Pp}) at $nt\to 0$, one can express 
the numerical prefactor of the integral
\begin{eqnarray}
\frac{T_b}{4P_b p_b}&\,&
\frac{k_1^2}{(2\pi)^2}\cos\theta_b4\pi\frac{p_aT_a}{P_a}
= \frac{k^2_1p_a}{4\pi p_b}\cos\theta_b \frac{T_aT_b}{P_aP_b}\nonumber
\\ &=& \frac{k^2_1}{4\pi}
\frac{\cos\theta_a}{\cos\theta_b}\cos\theta_b\frac{T_aT_b}{k_0\mu_ak_0\mu_b}
=\frac{\cos{\theta_a}}{4\pi m^2}\frac{T_aT_b}{\mu_a\mu_b}.
\eeqa
For the angular resolved diffuse reflection of a semi-infinite medium
we thus find 
\begin{equation}
A_R(\theta_a,\theta_b)=\frac{\cos{\theta_a}}{4\pi m^2}\frac{T_aT_b}{\mu_a\mu_b}
\gamma(\mu_a,\mu_b),
\label{mcw5}
\eeq
with the generalized {\it bistatic coefficient}
\beq
\gamma(\mu_a,\mu_b)=\int_0^\infty d\tau\Gamma_S(\mu_a,\tau)\eexp{-\tau/\mu_b}=
\int_0^\infty d\tau d\tau' G_S(\tau,\tau')\eexp{-\tau/\mu_a}\eexp{-\tau'/\mu_b}
\label{mcw6}.
\end{equation}
In the absence of index mismatch ($m=1$) one has $T_{a,b}=1$,
$\mu_{a,b}=\cos\theta_{a,b}$, so the prefactor in Eq.~(\ref{mcw5})
becomes $1/(4\pi\cos\theta_b)$. 
Fig.~\ref{jmlar} shows numerical results for $A_R$ 
for perpendicular incidence $(\theta_a=0)$, see \cite{thmn5}.

\subsection{Air-glass-medium interface}

We now consider a semi-infinite scattering medium, separated from the air by
another dielectric, such as glass, of thickness $d$. For $z<-d$ there is air,
for $-d<z<0$  glass, and for $z>0$ the scattering medium.  A plane wave of unit
amplitude comes in from $z=-\infty$ with wave vector $(\qt,k_z)$. The
perpendicular component $\qt$ is conserved at the interfaces.  We denote by
$p_{i}$ the $z$-component of the wave vectors in the  three sectors $i=0,1,2$,
where $i=0$ corresponds to the scattering medium, $i=1$ to air, and $i=2$ to
glass. The wave numbers in the three media are $k_0=\omega/c_0$,
$k_1=\omega/c_1$ ,and $k_2=\omega/c_2$, 
respectively, where $c_i$ is the speed of
propagation in the medium $i$.
 The incoming, refracted, and specularly reflected
waves are given by
\beqa
\Psi({\bf r}) = \left \{ \begin{array}{ll}
{\rm e}^{i\qt\cdot\rhoo+ip_{1}z}+r{\rm e}^{i\qt\cdot\rhoo-ip_{1}z} &
(z<-{\rm d}) \nonumber \\
t_{1}{\rm e}^{i\qt\cdot\rhoo+ip_{2}z} + r_1{\rm e}^{i\qt\cdot\rhoo
-ip_{2}z} &(-{\rm d}<z<0) \nonumber \\
t{\rm e}^{i\qt\cdot\rhoo+ip_{0}z} & (z>0) .
\end{array}
\right.
\end{eqnarray}
Here $r$ is the reflection amplitude of the system, $t$ the
transmission amplitude, and $p_i=\sqrt{k_i^2-\qt^2}$.
Continuity requirements at
$z=-{\rm d}$ and $z=0$ yield $r,\,r_1$ and $t,\,t_1$:
\beqa
r &=& \frac{(p_{0}+p_{2})(p_{1}-p_{2})-(p_{0}-p_{2})(p_{1}+p_{2}){\rm
e}^{2ip_{2}{\rm d}}}
{(p_{0}+p_{2})(p_{1}+p_{2})-(p_{0}-p_{2})(p_{1}-p_{2}){\rm e}^{2ip_{2}{\rm
d}}} \nonumber \\
t &=& \frac{4p_{1}p_{2}{\rm e}^{i(p_{2}-p_{1}){\rm d}}}
{(p_{0}+p_{2})(p_{1}+p_{2})-(p_{0}-p_{2})(p_{1}-p_{2})
{\rm e}^{2ip_{2}{\rm d}}}.
\eeqa
The reflection and transmission coefficients of the double interface are
\beq |r|^{2}\;\;\; ; \;\;\; 
\frac{p_{0}}{p_1}|t|^{2},\eeq
respectively.
As we assume the thickness of the glass plate is not smooth within
one wavelength, these expressions must be averaged
over the spread in thickness. This amounts to averaging over
the phase $2p_{2}{\rm d} \equiv \varphi$ and leads to the
average transmission and reflection coefficients
\beqa
T_{3} & =& 1-R_{3}=
\frac{p_{0}}{p_{1}}\int^{\pi}_{-\pi}\frac{d\varphi}{2\pi} |t|^{2}.
\end{eqnarray}
The integrand is of the form $A/(B-C\cos\phi)$.
The final result is
\begin{equation} T_{3}=1-R_3=
\frac{4p_{0}p_{1}p_{2}}{(p_{0}+p_{1})(p_{0}p_{1}+p_{2}^{2})}.
\end{equation}
We can check this in special cases.
Inserting $p_{2}=p_{1}=p$ and $p_{0}=P$ indeed reduces to previous the
result for two media.

We can describe this system by replacing $T(\mu)$ in previous equations
by $T_3(\mu)$ and replacing the reflection coefficient $R(\mu)$ in the 
Milne-kernel by $R_3(\mu)$. 
As before, $\mu\equiv p_0/k_0$ is the cosine of the angle $\theta'$ 
between the radiation and the $z$ axis.

Specular reflections in the glass have now been taken into account.
For a not too narrow beam this is useful for thin glass plates. 
For a not so broad beam impinging
on a medium with thick plates, multiple reflections 
from the glass interfaces can results in components that fall outside the
incoming beam and even outside the medium \cite{ospeck}.

\subsection{Solutions of the Schwarzschild-Milne equation}
\label{oplossing}

We consider properties of the transport equation in a semi-infinite space
\cite{thmn5}.  The Milne equation (\ref{Milne}) has a special solution 
$\Gamma_S(\mu_a;\tau)$, while the associated homogeneous equation without
source term has a solution $\Gamma_H(\tau)$.  We are interested in the
asymptotic behavior of  $\Gamma_S(\mu_a;\tau)$ and $\Gamma_H(\tau)$. 
Deep in the bulk $(\tau\gg 1)$ one expects a slow variation
of $\Gamma(\tau')$ at the scale of one mean free path 
($|\tau-\tau'|={\cal O}(1)$) and one can expand 
\beq
\Gamma(\tau') = \Gamma(\tau) + (\tau'-\tau)\Gamma'(\tau) +
\frac{1}{2}(\tau'-\tau)^2\Gamma''(\tau) + \cdots
\eeq
When this is inserted in Eq.~(\ref{Milne}), one finds for 
$\tau\gg 1$ 
\beq
\Gamma(\tau) \approx \Gamma(\tau) + \frac{1}{3} \Gamma''(\tau) +
\cdots
\label{gamtau}
\eeq
Here we used that for large $\tau$
\beq
\frac{1}{2} \int_0^1\frac{\d\mu}{2\mu}\int_0^\infty\d\tau'
\eexp{-\vert\tau-\tau'\vert / \mu}(\tau-\tau')^{2} \approx \frac{1}{3}.
 \label{eendrieint}
\eeq
The term $O(\Gamma')$ vanishes due to
symmetry of the $(\tau'-\tau)$-integral.
Eq.~(\ref{gamtau}) thus yields again the diffusion behavior
 $\Gamma''(\tau)\approx 0$.
We therefore consider the homogeneous solution $\Gamma_H$ 
and the special solution $\Gamma_S$ with the asymptotic behaviors
\beqa
\left\{ \begin{array}{l} \Gamma_H(\tau)\approx\tau+\tau_0 \\
\Gamma_S(\tau;\mu_a)\approx \tau_1(\mu_a) \end{array} 
\qquad(\tau\to\infty). \right. \label{gammaasymp}
\end{eqnarray}
Corrections occur due to the interface and decay exponentially 
in $\tau$. For a numerical solution one may introduce
$\delta\Gamma_S(\tau)=\Gamma_S(\tau)-\tau_1(\mu_a)$ and $\delta\Gamma_H(\tau)=
\Gamma_H(\tau)-\tau-\tau_0$ and integrate them
from $\tau=\infty$. The requirement that they vanish there, determines
$\tau_0$ and $\tau_1$.
The constant $\tau_0$ depends only on the index ratio $m$,
while $\tau_1(\mu_a)$ also depends on the direction of the incoming beam.

We define the Green's function $G_S(\tau,\tau')$
as the solution of the inhomogeneous equation
\beq\label{specialG}
G_S(\tau,\tau')=\delta(\tau-\tau')+\int_0^\infty \d\tau''
\{M_B(\tau,\tau'')+M_L(\tau,\tau'')\}G_S(\tau'',\tau'), \label{greenmil}
\eeq
subject to the boundary condition 
$G_S(\infty,\tau') < \infty $. This function is symmetric, 
$G_S(\tau,\tau')=G_S(\tau',\tau)$, and has the limit
\beq
\lim_{\tau'\to\infty}G_S(\tau,\tau')={1\over D}\Gamma_H(\tau).
\label{greenlim}
\eeq
The latter equality can be proven by taking the limit $\tau'\to\infty$
in Eq.~(\ref{greenmil}).
The delta function
vanishes and the remaining equation is the same as the one for
$\Gamma_H(\tau)$. Therefore $G_S(\tau,\tau'\to\infty)$ is proportional
to $\Gamma_H(\tau)$. The multiplicative prefactor can be fixed
by expanding $G_S(\tau,\tau')$ in $\tau'-\tau$ and inserting this 
in the right-hand side of Eq.~(\ref{greenmil}). 
Using Eq.~(\ref{eendrieint}) again, one finds for $\tau,\tau'\gg 1$
\beq
0 = \delta(\tau-\tau') + \frac{1}{3} \frac{\d^2}{\d\tau^2} G_S(\tau,\tau').
\eeq
The solution is $G_S(\tau,\tau')=3 \, {\rm min}(\tau,\tau')$ in the regime
$(\tau$, $\tau'$, $\vert\tau-\tau'\vert\gg 1)$. The diffusion
coefficient $D$ in Eq.~(\ref{greenlim}) reads $1/3$ in reduced units,
that is to say, $D=v\ell/3$ in physical units.

From Eq.~(\ref{greenmil}) it follows that
\beq
\Gamma_S(\tau;\mu)=\int_0^\infty \d\tau' G_S(\tau,\tau')\eexp{-\tau'/\mu},
\label{gammas}\eeq
and, in particular, using eqs. (\ref{gammaasymp}) and (\ref{greenlim}),
\beq\label{GGG}
\tau_1(\mu)=\lim_{\tau\to\infty} \Gamma_S(\tau;\mu) = 
{1\over D}\int_0^\infty\Gamma_H(\tau)\eexp{-\tau/\mu}\d\tau. \label{taueen}
\eeq
The physical interpretation of $\tau_1(\mu)$ is the
limit intensity ($z=\tau\ell\to\infty$) of a semi-infinite medium.

Numerical values of the injection depth $\tau_0$, the normalized limit
intensity $\tau_1(1)$ and the normalized bistatic coefficient $\gamma(1,1)$ can
be found in Table~\ref{table1} for various values of the index ratio $m$.

In \citeasnoun {thmn5} it was verified explicitly that in this approach flux
is conserved. This derivation will not be reproduced here.
We refer the interested reader to the original paper.

\section{Transport through a slab}

We now discuss the transmission properties of an optically thick slab with
isotropic scatterers. We derive the ``ohmic'' or diffusive scaling behavior
$T\sim \ell/L$ mentioned in Sec.~\ref{theomacroscopy}, 
and give the full angular
dependence of the transmission and reflection. This result is then used to
calculate the resistance of an idealized conductor.

\subsection{Diffuse transmission}

We consider a medium with finite thickness $L\gg\ell$. The medium has optical
thickness $ b=L/\ell\gg 1$. For the moment we wish to neglect boundary effects.
Therefore we are restricted to positions 
not too close to the boundary (10 mean free paths is a
good measure, as the corrections decay  exponentially). The solution for 
$\Gamma(\tau)$ is a linear combination of the special and the homogeneous
solutions,
\begin{eqnarray}\label{beginSM}
\Gamma(\tau) &=& \Gamma_S(\tau) - \alpha \Gamma_H(\tau) 
\mbox{ for $0\leq \tau \leq 10$, }\nn 
&=& \tau_1(\mu) - \alpha(\tau +\tau_0) \mbox{
for $\tau>10$}.
\end{eqnarray}
Near the other boundary it holds similarly that
\begin{eqnarray}\label{eindSM}
\Gamma(\tau) &=& 
\alpha' \Gamma_H(b-\tau) \mbox{ for $0\leq b-\tau \leq 10$, }
\nn
&=& \alpha'(b-\tau+\tau_0) \mbox{ for $b-\tau > 10.$}
\end{eqnarray}
As they have the same functional form, both shapes can be matched in 
the bulk of the sample. This yields
\beq \label{al=al=}
\alpha=\alpha'= \frac{\tau_1(\mu)}{b+2 \tau_0}.
\eeq
Inserting this value in Eqs. (\ref{beginSM}) and (\ref{eindSM}) gives 
the intensity
anywhere in the slab, expressed in terms of $\Gamma_S$
and $\Gamma_H$ of the semi-infinite problem. Notice
the important role played by the diffusive behavior in the bulk.
The Schwarzschild-Milne equation has brought the precise behavior
at scales of one mean free path from the boundaries.

To calculate the angular transmission profile,
we follow the derivation for the diffuse reflection [see Eq.~(\ref{mcw5}].
The expression for the differential transmission 
coefficient per unit solid angle $d\Omega_b$ of a beam incident under angle 
$\theta_a$ is given analogously as
\beq \label{dTabdO=} \frac{\d T(a\rightarrow b)}{\d\Omega_b}
 = \frac{\cos\theta_a T_a T_b}{4\pi m^2 \mu_a \mu_b}
\int_0^b \d \tau \Gamma(\tau) \eexp{-(b-\tau)/\mu_b}.
\eeq
Using the solution for $\Gamma(\tau)$ we rewrite the integral 
\beqa
\int_0^b \d \tau \Gamma(\tau) \eexp{-(b-\tau)/\mu_b} &\approx& \alpha
\int_{-\infty}^b
\d \tau \Gamma_H(b-\tau) \eexp{-(b-\tau)/\mu_b} \nonumber \\ &=& \alpha
\int_0^\infty \d \tau
\Gamma_H(\tau) \eexp{-\tau/\mu_b} \nonumber 
\\ & = &\frac{\tau_1(\mu_a)}{b+2\tau_0} \tau_1(\mu_b) D,
\eeqa
where we used Eq.~(\ref{taueen}). We have derived the angle-dependent
differential transmission coefficient for a slab of optical 
thickness $b=L/\ell$,
\beq\label{Tabiso}
\frac{\d T(a\rightarrow b)}{\d\Omega_b}\equiv \frac{A^T(\theta_a,
\theta_b)}{b+2\tau_0} \equiv \frac{\cos \theta_a T_a T_b}{12\pi m^2 \mu_a
\mu_b (b+2 \tau_0)} \tau_1(\mu_a) \tau_1(\mu_b).
\eeq
As the intensity at the side of incidence ($z=0$), equals $\Gamma(\tau)
=\Gamma_S(\tau) - \alpha \Gamma_H(\tau)$,
 it is clear that the transmission term
$\alpha \Gamma_H(b-\tau)$ arises at the cost of the reflection. Therefore also
flux conservation is also  satisfied for an optically thick slab.

In Fig.~\ref{jmlat} we show $A^T(\theta_a,\theta_b)$, the normalized angular
resolved transmission for perpendicular incidence $(\theta_a=0)$, for several
values  of the index ratio $m$.

\subsection{Electrical conductance and contact resistance}

For metallic conductors in the mesoscopic regime the conductance
is given by the Landauer formula \cite{landauer}
\beq
G = \frac{2e^2}{h} \sum_{a,b} T_{ab}^{flux},
\eeq
where $h/e^2 \approx 25k\Omega$ is the quantum unit of resistance. 
This equation simply counts the weight of the channels that contribute
to transmission. In the above description the wave number $k_0$ 
can be replaced by the Fermi wave number vector $k_F$. 
The analog of the contrast in refractive index is now
the potential difference between the conductor
and the contact regions ($V_1 \neq V_0)$.
In our formalism the conductance is given by 
\beq
G = \frac{2e^2}{h} \sum_{a,b} T_{ab}^{flux} =
 \frac{2e^2}{h} \sum_{a,b} T_{ab}
\frac{\cos \theta_b}{\cos\theta_a} = \frac{2e^2}{h}\,\frac{k_F^2 A}
{3 \pi (b+2\tau_0)}, \label{eqGextra} \eeq
with $T_{ab} =A^T(\theta_a,\theta_b)/(b+2\tau_0)$. Herewith one gets 
\beq
G= \frac{A \sigma_B}{L+2z_0} \mbox{ , } \sigma_B = \frac{2 e^2
k_F^2 \ell}{3 \pi h},
\eeq
where $z_0=\tau_0 \ell$ and $\sigma_B$ is the ``Boltzmann'' value for the 
bulk conductivity. We distinguish a bulk resistance and a
contact resistance $R_c$,
\beq
R=\frac{1}{G} = \frac{L}{A \sigma_B} +2 R_c \mbox{ , }
R_c=\frac{3 \pi \hbar}{2 e^2 A k_F^2} \tau_0.
\eeq
The number of modes can be estimated as
$N \approx A k_F^2$. The above expression shows that $R_c$
is proportional to the dimensionless thickness $\tau_0$ and
inversely proportional to the number of channels.
The nontrivial part is coded in $\tau_0$. 
It depends on the potential drop $V-V_1$, but not on the density 
of scatterers. Because
\beq
k_F^2 - V_1 = k_1^2 \mbox{ , } k_F^2 - V_0 = k_0^2,
\eeq
we can make an analogy with light scattering
by putting $k_0^2=m^2 k_1^2 $: 
\beq
m^2 \Longrightarrow \frac{k_F^2-V_0}{k_F^2-V_1}.
\end{equation}

\section{The enhanced backscatter cone}
\label{terugstrooikegel}

So far we have seen mainly effects that are largely diffusive and that could to
some extent also be derived from particle diffusion. 
The enhanced backscatter
cone is the clearest manifestation of interference due to the wave 
nature of the
light. The wave character manifests itself most clearly in loop
processes. The
advanced and retarded waves can go around in two ways, in the 
same direction and
in the opposite. This leads to an enhanced return  to the origin,
 which is the
basic mechanism for Anderson localization (Sec.~\ref{chintro}).

The optical enhanced backscatter in the exact backscatter direction has the
same characteristics as a closed loop. It brings two possibilities, thus a
factor of 2, for all scattering series involved. For Faraday 
active media it will be suppressed in a
magnetic field. Away from the backscatter direction there is 
partial extinction.

\subsection{Milne kernel at nonzero transverse momentum}
We follow the discussion of \citeasnoun{thmn5}.
The backscatter diagrams are closely related to the standard
 ladder diagrams.
The only difference is the crossed attachment of the incoming and
outgoing lines to the first and last scatterers.
These diagrams are called maximally crossed diagrams (Fig.~\ref{figcoop}).

Let the incoming wave be denoted by $a$ and the outgoing wave by $b$.
Their wave vectors have components
$ ({\bf k}_{\perp}^a,p_a),\; ({\bf k}_{\perp}^{b},p_b )$.
The product of incoming and outgoing waves at the first scatterer is
\begin{equation} \Psi^{\ast}\Psi = \frac{p_aT_a}{P_a}
{\rm e}^{i({\bf k}^{a}_{\perp}-
{\bf k}^{b}_{\perp})\cdot\roo + i(p_a-p_a^*)z }\;,
\eeq
with $ \roo = (x,y) $.
We define the transverse wave vector \begin{equation}{\bf Q}
\equiv ({\bf k}^{a}_{\perp}-{\bf k}^{ b}_{\perp}). \end{equation}
For perpendicular incidence one has
($\theta_a=0$; ${\bf k}^{a}_{\perp}={\bf 0}$).
We consider the regime of angles close to the backscatter direction
 (\(\theta_b\simeq0\)).
Then it holds that $ |{\bf Q}| \simeq k_{1}\theta_{b} $
so that \beq \Psi^{\ast}\Psi =\frac{T(1)}{m}
\eexp{i{\bf Q}\cdot\roo} \eexp{-z/\ell}\;. \eeq

Consider the diffuse intensity $I$ in the backscatter cone.
It is given by 
\beq I({\bf r}) =\frac{T(1)}{m}
\eexp{i{\bf Q}\cdot\roo} \eexp{-z/\ell} +
\frac{4\pi}{\ell}\int 
\d{\bf r}\, '|G({\bf r}-{\bf r}')|^{2} I({\bf r}') . \eeq
Notice that -dependence occurs only in the source term of this integral
equation. Inserting $I({\bf r})={\rm e}^{i{\bf Q}\cdot\roo}I(z,Q)$ 
yields the Milne equation for $I(z,Q)$,
\beq I(z,Q) =\frac{T(1)}{m}{\rm e}^{-z/\ell}
 + \int \frac{\d z'}{\ell}M_C(z,z',Q) I(z',Q),\eeq
with the Q-dependent Milne cone kernel
\beq M_C(z,z',Q) = 4\pi\int \d^{2}\rho'{\rm e}^{i{\bf Q}{\bf\cdot}(\roo\,
'-\roo)}|G({\bf r},{\bf r}')|^{2}
\;.\eeq
The normalized intensity of the maximally crossed diagrams
satisfies
\BA \label{MilneQcone}
\Gamma_C(\tau,Q)&=& \frac{m I(z,Q)}{T(1)} \nonumber \\
\Gamma_C(\tau,Q)&=& \eexp{-\tau} +\int \d\tau' M_C(\tau,\tau',Q)
\Gamma_C(\tau',Q) .\EA
The Milne kernel again has a bulk and a layer term,
 \begin{equation} M_C(\tau,\tau',Q)=M_B(\tau,\tau',Q)+M_L(\tau,\tau',Q).
\end{equation}
Inserting $|G({\bf r})|^{2}= {\rm e}^{-r/\ell} /(4\pi
r)^{2}$ 
yields
\beq M_{\rm B}(z,z',Q) = 4\pi\int\rho \,\d\rho\, \d\phi 
\frac{1}{(4\pi)^{2}((z-z')^{2}+\rho^{2})}{\rm e}^{i Q \rho\cos\phi}
{\rm e}^{-\sqrt{(z-z')^{2}+\rho^{2}}/\ell}
\label{mbulk}\;.\eeq
In terms of the variables \( \tau = z/\ell \) and \( \mu = \cos\theta'
\) , where $\theta'$ is the angle between ${\bf \rho}$ and the $z$ axis,
we find that
\[ \frac{\sqrt{(z-z')^{2}+\rho^{2}}}{\ell} = \frac{|\tau-\tau'|}{\mu}\;
\mbox{ , }
\frac{\rho \d\rho}{\ell^{2}}= -\frac{|\tau-\tau'|^{2}}{\mu^{3}}\d\mu\;. \]
Inserting this in Eq.~(\ref{mbulk}) yields
\beqa
M_{\rm B}(|\tau - \tau'|,Q)=\int^{1}_{0}
\frac{\d\mu}{2\mu}\int^{\pi}_{-\pi}\frac{\d\phi}{2\pi}\; 
{\rm e}^{iQ \ell \cos\phi|\tau-\tau'|\sqrt{\mu^{-2}-1}}{\rm
e}^{-|\tau-\tau'|/\mu} \nonumber \\
=
\int^{1}_{0}\frac{\d\mu}{2\mu}
{\rm J}_0\left(Q \ell |\tau-\tau'|\sqrt{\mu^{-2}-1}\right)
\; {\rm e}^{-|\tau-\tau'|/\mu} \;,
\label{mb}
\eeqa
where J$_0$ is the zeroth-order Bessel function.
A similar analysis yields for the layer term
\beq
M_{\rm L}(\tau + \tau',Q) = 
\int_{0}^{1}\frac{d\mu}{2\mu}R(\mu){\rm J}_{0}\left(Q\ell (\tau+\tau')
\sqrt{\mu^{-2}-1}\right){\rm e}^{-(\tau+\tau')/\mu} \;.
\label{ml} \eeq
This form is useful at small and large $Q$.
  
\subsection{Shape of the backscatter cone}
The intensity in the backscatter direction consists of two parts: a diffuse
background $A^{\rm R}$, discussed in Sec.~\ref{theoseminf} and a
contribution $A^{\rm C}$ from the maximally crossed diagrams. In the case of
perpendicular incidence we have found for the background contribution
Eq.~(\ref{mcw5})
\beq
A^{\rm R}(0,0) = \frac{T(1)^{2}\gamma(1,1)}{4\pi m^{2}}\;,
\eeq
where the normalized bistatic coefficient $\gamma$ is given by 
\beqa
\gamma(\mu_{a},\mu_b) & = & \int^{\infty}_{0}\d\tau\,{\rm
e}^{-\tau/\mu_{b}} \Gamma_{\rm S}(\tau,\mu_{a}) \label{GrGamma} \\
 &=& \int^{\infty}_{0}\d\tau \int_0^\infty \d\tau'\,{\rm e}^{-\tau/\mu_{b}\:
 - \:\tau'/\mu_{a}}G_{\rm S}(\tau,\tau')\; .
\label{gamma} 
\eeqa
The contribution of the maximally crossed diagrams, normalized by the
diffuse background, is given by
\beq \label{209}
A^{\rm C}(Q) = \frac{\gamma_{\rm C}(Q)- \gamma_{\rm TR}}{\gamma(1,1)}
\;, \end{equation}
where $\gamma_{\rm TR}$ is the amplitude of the paths that are 
identical to their time reversed analogs. Such paths do not yield
a time-reversed contribution, and this should thus be subtracted 
from the first term. For low scatterer density, $\gamma_{\rm TR}$ 
consists of the single scattering event (in, scatter, out). 
This process yields [cf. Eq.~(\ref{mcw6})]
\beq
\gamma_{\rm TR} = \int^{\infty}_{0}\d\tau \int_0^\infty \d\tau'\,{\rm 
e}^{-\tau-\tau'}\delta(\tau-\tau') = \frac{1}{2}\;.
\eeq
The nontrivial term is defined as
\beq
\gamma_{\rm C}(Q) =\int_0^\infty \d\tau\eexp{-\tau}\Gamma_C(\tau,Q)=
 \int^{\infty}_{0}\d\tau \int_0^\infty \d\tau'\,{\rm 
e}^{-\tau-\tau'}G_{\rm C}(\tau,\tau',Q)\;, \label{gammac}
\eeq
where $G_C$ is the solution of Eq.(\ref{MilneQcone}) with source
$\delta(\tau-\tau')$ instead of ${\rm e}^{-\tau}$ 
and vanishing for $\tau\to\infty$.
In Fig. \ref{jmlac} we present the numerical results for $A^C(Q)$
for several values of the index ratio $m$. 

\subsection{Decay at large angles}

From Eqs.~(\ref{mb}) and (\ref{ml}) it follows that
 $M_{\rm B}$ and $M_{\rm L}$ decay quickly as functions of $\tau$
when $Q$ is large. Physically this happens because the enhanced
backscatter is suppressed at large angles due to dephasing.
 For large angles we can therefore restrict Eq.~(\ref{gammac})
to low order scattering. To second-order 
we have $G_{\rm C}(\tau,\tau';Q) = 
 \delta(\tau-\tau') + M_{\rm C}(\tau,\tau';Q)$. This yields
\beqa
\gamma_{\rm C}(Q)= \frac{1}{2} + \int^{\infty}_{0}\d\tau\,\d\tau'\,{\rm 
e}^{-\tau-\tau'}(M_{\rm B}(|\tau-\tau'|;Q)+M_{\rm L}(\tau+\tau';Q))\;.
\label{gambml} \eeqa
For the bulk term 
\beqa
\int^{\infty}_{0}\d\tau\,\d\tau'\,{\rm 
e}^{-\tau-\tau'}M_{\rm B}(|\tau-\tau'|;Q) 
\!&\stackrel{(\ref{mb})}{=}& \!
\int_0^1 \frac{\d\mu}{2\mu} \int_{-\pi}^\pi \frac{\d\phi}{1+1/\mu-iQ \ell
\cos \phi\sqrt{1/\mu^2-1}} \nonumber \\
\!&=& \!
\int^{1}_{0}\frac{\d\mu}{2}\frac{1}{\sqrt{(\mu+1)^{2}+Q^{2}\ell^2(1-\mu^{2})}}
\nonumber \\
 &\stackrel{Q large}{\approx}& \frac{\pi}{4Q\ell}.
\eeqa
Inserting Eq.~(\ref{ml}) yields for the layer term in Eq.~(\ref{gambml})
\beqa
\int^{\infty}_{0}&\d\tau&\,\d\tau'\,{\rm e}^{-\tau-\tau'}
M_{\rm L}(\tau+\tau';Q)
\nonumber \\ &=& \int^{1}_{0}\frac{\d\mu}{2}\int^{\pi}_{-\pi}
\frac{\d\phi}{2\pi}R(\mu)\frac{\mu}
{(\mu+1-iQ\ell \cos\phi\sqrt{1-\mu^{2}})^{2}}
\nn
&=&\frac{1}{2}\int^{1}_{0}\d\mu\,\mu 
R(\mu)\frac{\mu+1}{[(\mu+1)^{2}+Q^{2}\ell^2(1-\mu^{2})]^{3/2}}.
\end{eqnarray}
The integrand decays quickly for large values of
 $Q^{2}\ell^2(1-\mu^{2})$. Setting $\mu=1-x/( Q^{2}\ell^2) $
we obtain for the layer term
\beqa
\frac{R(1)}{Q^{2}\ell^2}\int^{\infty}_{0}\frac{\d x}{(4+2x)^{3/2}} = 
\frac{R(1)}{2Q^{2}\ell^2} \;.
\eeqa
This yields finally
\beq
A^{\rm C}(Q) \simeq \frac{T(1)^{2}}{4\pi m^{2}}\left(\frac{\pi}{4Q\ell} + 
\frac{R(1)}{2Q^{2}\ell^2}\right), 
\eeq
for large $ Q $. The effect of the layer term is essentially
 the square of the bulk term.
This is because the paths involved are essentially twice 
as long (Fig.~\ref{back2})

\subsection{Behavior at small angles}

Assuming a linear $Q$ dependence, 
we expand $\Gamma_{\rm C}$ around $Q = 0$
\begin{equation}
\Gamma_{\rm C}(\tau;Q) \simeq 
\Gamma_{\rm C}(\tau;Q=0)-Q\tilde{\Gamma}_{\rm C}(\tau;Q) .
\label{rondQ=0}
\eeq
Inserting this in Eq.~(\ref{MilneQcone}) and subtracting the $Q=0$
terms yields, to leading order in $Q$,
\beq
\tilde{\Gamma}_{\rm C}(\tau;Q) = \int^{\infty}_{0}\d\tau'\,M_{\rm C}(\tau,
\tau';Q)\tilde{\Gamma}_{\rm C}(\tau';Q) \;.
\eeq
As this is a homogeneous equation, it holds that
$\tilde{\Gamma}_{\rm C}(\tau;Q)=\alpha \Gamma_{\rm H}(\tau;Q)
$,so that Eq.~(\ref{rondQ=0}) becomes
\beqa\label{alpha}
\Gamma_{\rm C}(\tau;Q) & =& \Gamma_{\rm S}(\tau) - \alpha Q
\Gamma_{\rm H}(\tau) 
\stackrel{\tau\gg 1}{\approx} \tau_{1}(1)- \alpha Q\ell \tau \label{alfa2} .
\end{eqnarray}
Deep inside the scattering medium ($\tau \gg 1$) 
the diffusion approximation holds, implying
\beqa
 \Gamma_{\rm C}''(\tau;Q) = Q^{2}\ell^2\Gamma_{\rm C}(\tau;Q) 
 \Rightarrow \Gamma_{\rm C}(\tau;Q) = \tilde{\alpha}{\rm e}^{-Q\ell\tau}\;.
\eeqa
Matching this for small $Q\ell\tau$ with Eq.~(\ref{alfa2}) one finds
$ \tilde{\alpha} = \alpha = \tau_{1}(1) $.
Using Eq.~(\ref{GrGamma}) this yields finally 
\beqa\label{peakcone}
\gamma_{\rm C}(Q) & =& \gamma_{\rm C}(0) - 3Q\ell\tau_{1}^{2}(1) \nonumber \\
& =& \gamma_{\rm C}(0)\left(1-\frac{Q}{\Delta Q}\right)\;,
\eeqa
where we used Eq.~(\ref{taueen}) with $D= \frac{1}{3}$. 
The normalized opening angle $\Delta Q$ is 
 \beq\label{3.40}
\Delta Q = 3\frac{\gamma_{\rm C}(0)}{\ell \tau_{1}^{2}(1)}.
\end{equation}
Note that $\gamma_C(0) \equiv \gamma(Q=0; \mu_a=\mu_b=1)$. 
The linearity of the peak of the cone is a result of diffusion,
that is to say, of long light paths. The complicated expression
for the opening angle shows that the skin layer plays
an important quantitative role. The reason hereto is obvious:
the light has to traverse it.

\section{Exact solution of the Schwarzschild-Milne equation}
\label{Exact Solution}\label{theoexact}

Some useful formulae were given in previous sections for calculation of the
diffuse intensity. However, the Schwarzschild-Milne equation can be solved
exactly in the absence of internal reflections \cite{chandrasekhar}. We
summarize the approach of \citeasnoun{thmn5} here and  generalize the
Schwarzschild-Milne equation to include internal reflections.

\subsection{The homogeneous Milne equation}
\label{S3.1}

The starting point of this analysis is the integral form
of the radiative transfer equation, i.e., the Schwarzschild-Milne
equation, Eq.~(\ref{Milne}).
In the absence of internal reflections, $M_L=0$,
the remaining kernel $M_B$ depends only on the difference $(\tau-\tau')$,
giving the Milne equation the structure of a convolution equation.
Because of its half-space geometry the problem is still nontrivial. 

We consider first the homogeneous Milne equation. Its solution 
$\Gamma_H(\tau)$ has the asymptotic behavior $\Gamma_H(\tau)=\tau+\tau_0$
[Eq.~(\ref{gammaasymp})].
We define the Laplace transforms of $M_B(\tau,0)$ and of 
$\Gamma_H(\tau)$ as
\begin{eqnarray}
m(s)&=&\int_{-\infty}^\infty M_B(\tau,0)\ e^{s\tau} {\rm d}\tau
\qquad(-1<{\rm Re}\,s<1)
\label{3.1} \\
g_H(s)&=&\int_0^\infty \Gamma_H(\tau)\ e^{s\tau} {\rm d}\tau=D\,
\tau_1(\mu=-1/s)\qquad({\rm Re}\, s<0)
.\label{3.2}
\end{eqnarray}
In the case of isotropic scattering the Milne kernel Eq.~(\ref{Mbulk}) leads to
\begin{equation}
m(s)={1\over 2s}\ln{1+s\over 1-s}. \label{3.3}
\end{equation}
The small-$s$ behavior of $m(s)$
\begin{equation}
m(s)=1+Ds^2+{\cal O}(s^4)\qquad(s\to 0) , \label{3.4}
\end{equation}
determines the dimensionless diffusion constant $D=1/3$.

It turns out that the problem can be solved for any symmetric Milne kernel.
The homogeneous Milne integral equation is equivalent to
\begin{equation}
\phi(s) g_H(s)=\int_{-i\infty}^{i\infty}
{{\rm d}t\over 2\pi i}\ {m(t)g_H(t)\over t-s}
\qquad(-1<{\rm Re}\, t<{\rm Re}\, s<0)
,\label{3.5}
\end{equation}
with $\phi(s)=1-m(s)$,
and the asymptotic behavior Eq.~(\ref{gammaasymp}) is equivalent to
\begin{equation}
s^2 g_H(s)=1-\tau_0 s+{\cal O}(s^2)\qquad(s\to 0).
\label{3.7}
\end{equation}

Equation~(\ref{3.5}) can be solved in closed form, 
for an arbitrary bulk kernel.
Let us consider first the following ``rational case,''
where $M_B(\tau,0)$ is a finite superposition of $N$ decaying exponentials,
namely,
\begin{equation}
M_B(\tau,0)=\sum_{a=1}^N {w_ap_a\over 2} e^{-p_a\vert\tau\vert}
,\label{3.8}
\end{equation}
with weights $w_a >0$ and decay rates (inverse decay lengths)
$p_a >0$.
We then have 
\begin {equation}
m(s)=\sum_{a=1}^N {w_a p_a^2 \over p_a^2-s^2},\qquad
\phi(s)=-s^2\sum_{a=1}^N {w_a \over p_a^2-s^2}
.\label{3.9}
\end{equation}
By comparing with Eq.~(\ref{3.4}), we obtain the normalization conditions
$ \sum_a w_a=1$, $\sum_a {w_a /p_a^2}=D $. 
Equation~(\ref{3.5}) can be evaluated by means of residue calculus.
It loses its integral nature and becomes
\begin{equation}
\phi(s) g_H(s)=-\sum_{a=1}^N {w_ap_a g_H(-p_a) \over 2(p_a+s)}
\label{3.11}.\end{equation}
In order to solve Eq.~(\ref{3.11}), we first write the rational function
$\phi(s)$ in factorized form,
\begin{equation}
\phi(s)=\frac{\prod_{\alpha=1}^{N-1} (1-z_\alpha^2/s^2) }
{ \prod_{a=1}^N (1-p_a^2/s^2)} . \label{3.12}
\end{equation}
The $(N-1)$ zeros of $\left[ \phi(s)/s^2 \right]$ in the variable $s^2$
have been denoted by $z_\alpha^2$, with ${\rm Re} z_\alpha >0$,
for $1\le\alpha\le N-1$.
The normalization of Eq.~(\ref{3.12}) has been fixed by using $\phi(\infty)=1$.

An alternative expression for the diffusion constant $D$
is derived by setting $s$ zero in Eq.~(\ref{3.12})
and comparing with Eqs.~(\ref{3.4}) and (\ref{3.9}).
We get
\begin{equation}
D=\frac{ \prod_{\alpha=1}^{N-1} z_\alpha^2 }{\prod_{a=1}^N p_a^2 }.
\label{3.13}
\end{equation}

The solution of Eq.~(\ref{3.11}) is of the form
\begin{equation}
\phi(s) g_H(s)=\frac{{\cal N}_0(s)}{\prod_{a=1}^N(p_a+s)}.
\label{3.14}
\end{equation}
with ${\cal N}_0(s)$ a polynomial of degree $(N-1)$,
which can be determined as follows.
Consider the value $s=-z_\alpha$:
we have $\phi(-z_\alpha)=0$, whereas $g_H(-z_\alpha)$ is finite.
Equation~(\ref{3.14}) shows therefore that $s=-z_\alpha$ is a zero
of the polynomial ${\cal N}_0(s)$.
Since there are exactly $(N-1)$ such values,
the solution Eq.~(\ref{3.14}) is determined up to a normalization,
which can be fixed by using Eq.~(\ref{3.7}).
We thus obtain
\begin{equation}
s^2 g_H(s)=\frac{\prod_{a=1}^N (1-s/p_a) }
{ \prod_{\alpha=1}^{N-1} (1-s/z_\alpha)} . \label{3.15}
\end{equation}
This equation can be recast for ${\rm Re} s<0$ as
\begin{equation}
s^2 g_H(s)=\exp \left\{ \sum_{a=1}^N \ln(1-s/p_a)-\sum_{\alpha=1}^{N-1}
\ln(1-s/z_\alpha) \right\} .
\label{3.16}
\end{equation}
The sum over the poles and zeros of the function $\left[ \phi(s)/s^2 \right]$,
with positive real parts, can be rewritten as the following complex integrals:
\begin{equation} \label{3.17}\matrix{
s^2 g_H(s)&=\exp\left\{\int_{-i\infty}^{+i\infty}{{\rm d}z\over 2\pi i}
\ \left[ {\phi'(z)\over\phi(z)}-{2\over z} \right]
\ \ln(1-s/z) \right\}\nonumber\hfill \cr
&=\exp\left\{ -s \int_{-i\infty}^{+i\infty}{{\rm d}z\over 2\pi iz(z-s)}
\ \ln\left[ -{\phi(z)\over Dz^2} \right] \right\}.\hfill\cr}
\end{equation}
This result gives the solution of the problem
for an arbitrary bulk Milne kernel.

The thickness $\tau_0$ of the skin layer can be evaluated
by comparing the results Eqs.~(\ref{3.15}) and (\ref{3.17}) with the expansion 
(\ref{3.7}). We get
\begin{equation}
\tau_0
=\sum_{a=1}^N {1\over p_a} -\sum_{\alpha=1}^{N-1} {1\over z_\alpha}
=\int_{-i\infty}^{+i\infty}{{\rm d}z\over 2\pi i z^2}
\ \ln\left[ -{\phi(z)\over Dz^2} \right] .
\label{3.18}
\end{equation}
The value for the case of point scattering is obtained by inserting
into Eq.~(\ref{3.18}) the expression for $\phi(z)$ 
coming from Eq.~(\ref{3.3}).
By making the change of variable $z=i\tan\beta$, we get
\begin{equation}
\tau_0={1\over\pi} \int_0^{\pi/2} {{\rm d}\beta\over\sin^2\beta}
\ \ln{ \tan^2\beta \over 3(1-\beta{\rm cot}\beta) }=0.710446090\ldots
\label{3.19}
\end{equation}
We recover a well-known numerical value
\cite{chandrasekhar,vanderhulst2,vanderhulststark}.

\subsection{The inhomogeneous Milne equation}
\label{S3.2}
The above method can be generalized to the inhomogeneous 
situation \cite{thmn5}. 
For point scattering one obtains the result \begin{equation}
\tau_1(\mu)=\mu\sqrt{3}
\ \exp \left\{-{\mu\over\pi} \int_0^{\pi/2}{\rm d}\beta
\ {\ln (1-\beta{\rm cot}\beta) \over \cos^2\beta+\mu^2\sin^2\beta} \right\}
.\label{3.30}
\end{equation}
This quantity is maximal at normal incidence $(\mu=1)$,
where it assumes the somewhat simpler form
\begin{equation}
\tau_1(1)=\sqrt{3}
\ \exp \left\{-{1\over\pi} \int_0^{\pi/2}{\rm d}\beta
\ \ln (1-\beta{\rm cot}\beta) \right\} =5.03647557\ldots \label{3.31}
\end{equation}

Finally, it is worth noticing that Eq.~(\ref{3.16}) also implies the following
remarkably simple result:
\begin{equation}
\gamma(\mu_a,\mu_b)
={\tau_1(\mu_a)\tau_1(\mu_b)\over 3(\mu_a+\mu_b)} .
\label{3.32}
\end{equation}
which is particular to the situation
in which there are no internal reflections.
In the presence of mismatch the functional
form changes. See Eq.~(\ref{gmambmism}) for the limit of strong mismatch.

\subsection{Enhanced backscatter cone}
\label{S3.3}

Exact results can also be derived for the backscatter cone in the absence of
internal reflections. Consider, for the sake of simplicity, the
enhanced backscatter of a normally incident beam $(\mu_a=1)$. The
$Q$ dependent Milne equation (\ref{MilneQcone}) can be solved by means of the
Laplace transformation.

In the case of point scattering, one gets for the backscatter amplitude
\cite{gorod}
\begin{equation}
\gamma_C(Q)={1\over 2} \exp\left\{ -{2\over\pi}\int_0^{\pi/2} {\rm d}\beta
\ \ln\left( 1-{\arctan\sqrt{Q^2\ell^2+\tan^2\beta}\over \sqrt{Q^2\ell^2
+\tan^2\beta}} \right) \right\} \label{3.38}.
\end{equation}

For small values of the reduced wave vector $Q$,
the result Eq.~(\ref{3.38}) assumes the general form Eq.~(\ref{peakcone}),
corresponding to the triangular shape of the backscatter cone.
The value for $Q=0$ reads
\begin{equation}
\gamma(1,1)={\left[ \tau_1(1) \right]^2 \over 6}=4.22768104 \label{3.39},
\end{equation}
in agreement with Eq.~(\ref{3.32}).
In agreement with Table~\ref{table1}, one has $\Delta Q =1/2 \ell $ 
[see Eq.~(\ref{peakcone})].

We end this section by mentioning the expression for the backscatter
cone amplitude for isotropic scattering by point scatterers with
 an arbitrary albedo,
\begin{equation}
\gamma_C(a;Q)={a\over 2} \exp\left\{ -{2\over\pi}\int_0^{\pi/2}
 {\rm d}\beta
\ \ln\left( 1-a{\arctan\sqrt{Q^2\ell^2+\tan^2\beta}\over
\sqrt{Q^2\ell^2+\tan^2\beta}} \right) \right\}
.\label{3.50}
\end{equation}
This result shows that, as soon as the scattering albedo
$a$ is smaller than unity the backscatter amplitude
is an analytic function of $Q^2$, i.e., the cusp at $Q=0$ disappears.
This confirms the physical intuition that the triangular shape of the cone
is due to the existence of arbitrarily long diffusive paths.
It is therefore a characteristic of the problem with unit albedo,
i.e. with no absorption, in a half-space geometry.
When the reduced wave vector $Q$ and the strength of absorption
$(1-a)$ are both small, we observe the following scaling behavior:
\begin{equation}
\gamma_C(a;Q) \approx \frac {\gamma_C(1;0)}{1+2\sqrt{Q^2\ell^2+3(1-a)}} .
\label{3.51}\end{equation}
The denominator of this expression is not reproduced quantitatively
by the diffusion approximation [see for example Eq.~(71) of 
\citeasnoun{vandermark}],
although it pertains to the long distance physics of the problem.
We also notice that the prefactor of the square root is nothing but
the reciprocal of the value (\ref{3.40}) of $\ell\Delta Q$. Indeed,
Eq.~(\ref{3.32}) yields 
$\gamma_C(0)\equiv \gamma(1,1)=\tau(1)^2/6$, by which Eq.~(\ref{3.40})
reduces to $\Delta Q=1/(2\ell)$.

\subsection{Exact solution for internal reflections in diffusive media}
\label{theoanisotropy}

The solution for internal reflections, $m\neq 1$, can be simply expressed in 
the solution without internal reflections, $m=1$.
Let us look at
$\Gamma^{(m)}_H$, the homogeneous solution for the situation
with index ratio $m$. If we define
\begin{equation} \label{gH=}
g_H(\mu)=\frac{R(\mu)}{2\mu}\int_0^\infty \d\tau' e^{-\tau'/\mu}
\Gamma_H^{(m)}(\tau'),
\end{equation}
then we get from the homogeneous version of Eq.~(\ref{Milne})
\begin{equation} 
\Gamma^{(m)}_H(\tau)= M_B\ast\Gamma_H^{(m)}(\tau)+
\int_0^1\d\mu g_H(\mu)e^{-\tau/\mu} .
\end{equation}
We observe that comparison with Eq.~(\ref{Milne}) 
immediately gives the solution
\begin{equation} 
\Gamma_H^{(m)}(\tau)=\Gamma_H^{(1)}(\tau)+
\int_0^1 \d\mu g_H(\mu)\Gamma_S^{(1)}(\tau;\mu) . \end{equation}
Using the definition Eq.~(\ref{gH=}) we have to solve the inhomogeneous
equation
\begin{equation} \label{gHeq}
g_H(\mu)=\frac{R(\mu)\tau_1^{(1)}(\mu)}{6\mu}
+\frac{R(\mu)}{2\mu}
\int_0^1 \d\mu'\gamma^{(1)}(\mu,\mu') g_H(\mu'), \end{equation}
where we used the identities \cite{thmn5}
\begin{equation} 
\int_0^\infty d\tau\Gamma_H^{(1)}(\tau)e^{-\tau/\mu}=\frac{1}{3}
\tau_1(\mu)\end{equation}
\begin{equation} \gamma^{(m)}(\mu_b,\mu_a)=
\int_0^\infty d\tau\Gamma_S^{(m)}(\tau,\mu_a)e^{-\tau/\mu_b}.
\end{equation}
 
In a similar fashion we find for the special solution
\begin{equation} 
\Gamma_S^{(m)}(\tau;\mu_a)=\Gamma_S^{(1)}(\tau;\mu_a)+
\int_0^1 \d\mu g_S(\mu,\phi;\mu_a)\Gamma_S^{(1)}(\tau;\mu)\end{equation}
with 
\begin{equation} \label{gSeq}
g_S(\mu;\mu_a)=\frac{R(\mu)}{2\mu}\gamma^{(1)}(\mu;\mu_a)
+\frac{R(\mu)}{2\mu}\int_0^1 \d\mu'\gamma^{(1)}(\mu,\mu') g_S(\mu';\mu_a).
\end{equation}
Following the same steps we obtain the
enhanced backscatter intensity. 
We must solve
\begin{eqnarray} \label{gQeq}
g(\mu,\phi;\mu_a,Q)&=&\frac{R(\mu)}{2\mu}
\gamma^{(1)}\left(\frac{\mu}{1+iQ\ell\sqrt{1-\mu^2}\cos\phi},\mu_a;Q\right)
+\frac{R(\mu)}{2\mu}\int_0^1 \d\mu'\int_{-\pi}^\pi\frac{d\phi'}{2\pi}
\\
&\times&\gamma^{(1)}\left(\frac{\mu}{1+iQ\ell\sqrt{1-\mu^2}\cos\phi},
\frac{\mu'}{1+iQ\ell\sqrt{1-\mu^{'2}}\cos\phi'};Q\right)
 g(\mu',\phi';\mu_a,Q), \nonumber\end{eqnarray}
where~\cite{luckup}
\begin{equation} \gamma^{(1)}(\mu_a,\mu_b,Q)
=\frac{\tau_1^{(1)}(\mu_a,Q)\tau_1^{(1)}(\mu_b,Q)} {3(\mu_a+\mu_b)},
\end{equation}
involves
\begin{eqnarray} 
\tau_1^{(1)}(\mu;Q)&=&\mu\sqrt{3}
\exp\left\{-\frac{\mu}{\pi}\int_0^{\pi/2}
\frac{\d\beta} {\cos^2\beta+\mu^2\sin^2\beta}
\ln\left(1-\frac{ \arctan{\sqrt{Q^2\ell^2+\tan^2\beta}}}
{\sqrt{Q^2\ell^2+\tan^2\beta^2}}\right)\right\},\end{eqnarray}

We find the following expressions for the dimensionless injection depth
$\tau_0$, the limit intensity $\tau_1$, the differential reflection
$dR/d\Omega$,
and $B(Q)$, the intensity of the backscatter cone normalized to the diffuse
background, at outgoing angle $\theta_b$ coded in
$Q=k\ell(\theta_a-\theta_b)$:
\begin{eqnarray} \tau_0^{(m)}&=&
\tau_0^{(1)}+\int_0^1\d\mu g_H(\mu)\tau_1^{(1)}(\mu)\\
\tau_1^{(m)}(\mu_a)&=&\tau_1^{(1)}(\mu_a)+\int_0^1\d\mu g_S(\mu;\mu_a)
\tau_1^{(1)}(\mu)\\
\frac{\d R}{\d\Omega_b}(\theta_b;\theta_a)&=&\frac{\cos\theta_a}{4\pi m^2}
\,\frac{T(\mu_a)T(\mu_b)}{\mu_a\mu_b}\gamma^{(m)}(\mu_a,\mu_b)
\nonumber\\ 
\gamma^{(m)}(\mu_a,\mu_b)&=&
\gamma^{(1)}(\mu_a,\mu_b)+
\int _0^1\d\mu g_S(\mu;\mu_a)\gamma^{(1)}(\mu,\mu_b)\\
B(\mu_a,Q)&=&\frac{1}{\gamma^{(m)}(\mu_a,\mu_a)}
\left\{\gamma^{(1)}(\mu_a,\mu_a;Q)-\frac{\mu_a}{2}\right.\\
&+&
\left.\int_0^1 \d\mu\int_{-\pi}^\pi\frac{d\phi}{2\pi}
\gamma^{(1)}\left(\frac{\mu}{1+iQ\ell\sqrt{1-\mu^2}\cos\phi},\mu_a;Q\right)
 g(\mu,\phi;\mu_a,Q)\right\}\nonumber ,
\end{eqnarray}
respectively. 
The subtracted term in $B(Q)$ comes from single scattering, which does
not contribute to enhanced backscatter.

Equations (\ref{gHeq}), (\ref{gSeq}), and (\ref{gQeq}) can be solved
numerically. In many cases only a few iterations are needed. 
For $m$ close to unity the leading corrections for $\tau_0$ and $\tau_1(\mu)$
are given by the first term in the equation for the $g$'s.

Explicit results can be obtained in the limit of large  index mismatch
\cite{thmn5}. The point is that for $R(\mu)=1$ the integral kernel becomes
$(1/2\mu)\gamma^{(1)}(\mu,\mu')$, which has a unit eigenvalue with right
eigenfunction $g(\mu')=1$ and left eigenfunction $g(\mu)=2\mu$.  For a large
index mismatch this eigenvalue will bring the leading effects. The results will
be discussed in the next section.

\subsection{Exact solution for very anisotropic scattering}

As discussed in Sec.~\ref{theomesoscopy} the main effect of anisotropic
scattering is that the scattering mean free path $\ell=1/n\sigma$ is replaced
by the transport mean free path $\ell_{tr}= \ell /( 1-\langle\cos
\Theta\rangle)= \ell \tau_{tr}$. As a result, the injection depth decomposes as
$z_0=\tau_0\ell_{tr}$; likewise the opening angle of the backscatter cone is
proportional to $1/\ell_{tr}$.

For anisotropic scattering the transport equation cannot be solved exactly in
general. Several special cases were discussed by \citeasnoun{vanderhulst2}. He
found that in absence of index mismatch the effects of anisotropy are
typically not very large. \citeasnoun{amic} showed that the limit of strongly
forward scattering can be solved exactly. This extremely anisotropic case is
expected to set bounds for more realistic anisotropic scattering kernels. The
approach of \citeasnoun{amic} goes along the lines of previous sections.
However, certain constants occurring there are replaced by functions of an
angle, which are only partly known. For details of the lengthy derivation we
refer the reader to the original paper. 
The main results are presented in Table 
\ref{table2}, compared with the case of isotropic scattering. 

We can compare the universal results in the very anisotropic scattering regime,
for some of the quantities listed in Table \ref{table2}, with the outcomes of
numerical approaches. \citeasnoun{vanderhulst2}
 has investigated  the dependence
of various quantities on anisotropy for several commonly used phenomenological
phase functions, including the Henyey-Greenstein phase function
\begin{equation}\label{HGphasefion}
p(\Theta)={1-g^2\over(1-2g\cos\Theta+g^2)^{3/2}} . \end{equation} The data on
the skin layer thickness show that, as a function of anisotropy,
 $\tau_0$ varies
from 0.7104 (isotropic scattering) to 0.7150 (moderate anisotropy), passing a
minimum of 0.7092 (weak anisotropy); see  \citeasnoun{vanderhulst2}. The trend
shown by these data suggests that the universal value 0.718211 is actually an
absolute upper bound for $\tau_0$. Numerical data concerning
 $\tau_1(1)$ is also
available. \citeasnoun{vdHulstNotes} has extrapolated two series of data
concerning the Henyey-Greenstein phase function which admits a common limit for
very anisotropic scattering $(g\to 1)$. This limit reads in our notation
$\tau_1(1)/4=1.284645$, whereas \citeasnoun{vdHulstNotes} gives the
 two slightly
different estimates $1.273\pm 0.002$ and $1.274\pm 0.007$. The agreement is
satisfactory, although one cannot entirely exclude that the observed 0.8\%
relative difference is a small but genuine nonuniversality effect. Indeed, the
Henyey-Greenstein phase function might belong to another universality
 class than
appropriate for the approach of \citeasnoun{amic}.  
The same remark applies to a
less complete set of data \cite{vdHulstNotes} concerning the intensity
$\gamma(1,1)$ of reflected light at normal incidence.

In the presence of index mismatch the very anisotropic scattering limit can be
solved exactly along the same lines as for isotropic scattering. The formalism
becomes very heavy, however. In the limit of a large index mismatch the leading
behaviors are expected to be the same as for isotropic scattering.
 The reason is
that a large mismatch implies that the radiation that has entered
 the scattering
medium will  often be reflected internally before it can exit the medium.
Therefore it undergoes many scatterings, so it loses the nature (anisotropic or
not) of the individual events. This was verified explicitly by
 \citeasnoun{amic}

\section{Large index mismatch} \label{groot}

We have seen that the occurrence of internal reflections complicates
the solution of the Schwarzschild-Milne equation. Yet in the limit of 
a large index mismatch the boundaries will act as good mirrors. 
They re-inject the radiation so often that it is also diffusive close
to the boundaries. It turns out that, to leading order, the radiative transfer 
problem is greatly simplified. 
 
Following \citeasnoun{thmn5} we consider the  Schwarzschild-Milne equation in
the regime of large index mismatch ($m \to 0$ , or $ m \to \infty$). In both
cases the reflection coefficient $R(\mu)$ is close to unity and
the effect of $T(\mu)=1-R(\mu)$ is small. We can thus expand in powers of $T$.
We first consider the Green's function of the  Schwarzschild-Milne equation
Eq.~(\ref{specialG})  and write the kernel as \begin{eqnarray}
M(\tau,\tau') &=& M_{\rm B}(\tau-\tau')+M_{\rm L}(\tau+\tau')\\ 
&=& M_{\rm B}(\tau-\tau')+M_{\rm B}(\tau+\tau')-N(\tau+\tau')\\ 
\label{definN}
N(\tau+\tau')&=&\int_0^1 \frac{\d \mu}{2\mu} 
\eexp{-\frac{\tau+\tau''}{\mu}}T(\mu),
\EA
where $N$ is of order $\int \d\mu T(\mu)\ll 1$. This allows an expansion of
Eq.~(\ref{specialG}) in powers of $N$.
If $N=0$ (which happens for $m \to 0$ or $m \to \infty$),
the kernel $M$ has unit eigenvalue, and Eq.~(\ref{specialG})
has eigenfunction $G_S(\tau)=C_S={\rm const.}$, showing that for
a medium with perfect mirrors the intensity inside is independent
of the conditions outside.
For small $N$ the value of $C_S$ will depend sensitively on $N$.
Let us write, for $N \ll 1$, the Green's function as
\begin{eqnarray}
G_S(\tau,\tau')&=&C_S+G_0(\tau,\tau')+G_1(\tau,\tau')
\\ \nonumber
&&C_S \gg 1 ,\;G_0 \approx 1 ,\; G_1 \ll 1 .
\end{eqnarray} 
We insert this in Eq.~(\ref{specialG}):
\begin{eqnarray}
C_S+G_0+G_1&=& \delta(\tau-\tau')+C_S\nonumber \\ &&+\int_0^\infty\d \tau''
\{M_{\rm B}(\tau -\tau'') + M_{\rm B}(\tau+\tau'')\}(G_0+G_1) \nonumber \\
&& -\int_0^\infty \d \tau'' N(\tau+\tau'')(C_S+G_0+G_1) .
\end{eqnarray}
Comparing terms of equal order yields
\begin{eqnarray}
C_S&=& C_S
\\ \nonumber
G_0(\tau,\tau')&=&\delta(\tau-\tau')+\int_0^\infty \d \tau''
\{M_{\rm B}(\tau -\tau'') + M_{\rm B}(\tau+\tau'')\}G_0(\tau'',\tau')
\nonumber \\ &&-C_S\int_0^\infty \d \tau''N(\tau+\tau'')
\\ \nonumber
G_1(\tau,\tau')&=&\int_0^\infty \d \tau''
\{M_{\rm B}(\tau -\tau'') + M_{\rm B}(\tau+\tau'')\}G_1(\tau'',\tau')
\nonumber \\
&&-\int_0^\infty \d \tau''N(\tau+\tau'')G_0(\tau'',\tau'),
\end{eqnarray}
where we have neglected terms of order
 $N G_1\sim {\cal T}^2 $.
 Integrating these results over $\tau$ yields
an expression for $C_S$ and a compatibility condition to ensure that
 $G_0$ has a unique solution:
\begin{eqnarray}
C_S=\left\{\int_0^\infty \int_0^\infty \d \tau \d \tau'' 
N(\tau +\tau'') \right\}
^{-1} \\ 
\int_0^\infty \d \tau \d \tau'' N(\tau +\tau'') G_0(\tau'',\tau')=0
\; \mbox{ for all } \tau' . \end{eqnarray}
We can simplify the expression for $C_S$ by introducing
the angular averaged flux-transmission coefficient
$\T$ :
 \begin{eqnarray}
C_S&=& \left\{ \int_0^\infty \int_0^\infty \d \tau \d \tau''
\int_0^1 \frac{\d \mu}{2\mu} \eexp{-\frac{\tau+\tau''}{\mu}}T(\mu)
\right\} ^{-1} \\ \nonumber
&=& 2 \left\{ \int_0^1 \mu \d \mu T(\mu)
\right\} ^{-1} \equiv \frac{4}{\T} , \nonumber
\EA
where
\BA
\T&\equiv &2\int_0^1 \mu \d \mu T(\mu)
= 2\int_0^{\pi/2} \cos \theta \sin \theta\, \d \theta\, T(\cos
\theta) . \end{eqnarray}
Note that $\mu=\cos \theta$ stands for the projection factor of flux 
in the $z-$direction. The expression for $\T$ can be worked out
explicitly using
\begin{eqnarray}
T(\mu)&=&\frac{4\mu \sqrt{\mu^2-1+m^{-2}}}{(\mu+\sqrt{\mu^2-1+m^{-2}})^2} .
\end{eqnarray}
For $m<1$ we introduce $\epsilon^2 = m^{-2} -1$ 
\begin{eqnarray}
\T&=&2\int_0^1 \mu \d \mu
\frac{4\mu \sqrt{\mu^2+\epsilon^2}}{(\mu+\sqrt{\mu^2+\epsilon^2})^2}
\label{mvr2} . \end{eqnarray}
When we substitute $\phi$$=$$\sinh^{-1}({\mu}/{\epsilon})$ this becomes
\begin{eqnarray}
\T&=&8\epsilon^2\int_0^{\phi_+}\d \phi
\frac{\sinh^2 \phi \cosh^2\phi}{\eexp{2\phi}}
\\ 
&=& 2\epsilon^2 \int_0^{\phi_+} \d \phi \sinh^2 2\phi \eexp{-2\phi}
=\frac{\epsilon^2}{2} \int_0^{\phi_+} \d \phi
\left[ \eexp{2\phi}-2\eexp{-2\phi}+\eexp{-6\phi} \right] \\
&=&\frac{\epsilon^2}{2}\left\{ \frac{\eexp{2\phi_+}}{2}
+\eexp{-2\phi_+}-\frac{\eexp{-6\phi_+}}{6}-\frac{4}{3} \right\} .
\end{eqnarray}
Using
\begin{equation}
\sinh\phi_+=\frac{1}{\epsilon}=\frac{m}{\sqrt{1-m^2}}, \qquad
\cosh\phi_+=\frac{1}{\sqrt{1-m^2}}.
\end{equation}
we find $\exp{(2\phi_+)}=(1+m)/(1-m)$. This yields
\begin{eqnarray}
\T&=&
\frac{4m(m+2)}{3(m+1)^2}\qquad (m\le 1) .
\end{eqnarray}
The leading behavior for 
$m \to 0$ could have been found more quickly by taking the leading behavior
of the numerator and denominator of Eq.~(\ref{mvr2}).
\begin{eqnarray}
\T \approx 2\int_0^1 \mu d\mu \frac{4\mu\epsilon}{\epsilon^2}
= \frac{8}{3}m\qquad (m\rightarrow 0). \end{eqnarray}

Now we do the same for $m>1$. We must not forget the Brewster angle:
for $\mu < \sqrt{1 - m^{-2}}$ there occurs total reflection.
We now introduce $\epsilon^2 = 1-m^{-2}$ and carry out the
integration from $\epsilon$ to $1$:
\begin{eqnarray}
\T&=&8\int_\epsilon^1 \mu^2 \d \mu \frac{\sqrt{\mu^2-\epsilon^2}}
{(\mu+\sqrt{\mu^2-\epsilon^2})^2}
\\ &=&\frac{4(2m+1)}{3m^2(1+m)^2}\qquad (m\ge 1) .
\end{eqnarray}
The leading behavior for large $m$ could again
have been found by keeping leading terms:
\begin{eqnarray}
\T \approx \frac{8}{3m^3}; \qquad (m\rightarrow \infty).
\end{eqnarray}
Further we notice that $\T=1$ for $m=1$, as it should be.

\subsection{Diffuse reflected intensity}
We can now derive simply the leading behavior of the generalized
bistatic coefficient:
\begin{eqnarray}\label{gmambmism}
\gamma(\mu_a,\mu_b)&=&\int_0^\infty \d \tau \d \tau'
G_S(\tau,\tau') \eexp{-\tau/\mu_b} \eexp{-\tau'/\mu_a}
\\
&=& \int_0^\infty \d \tau \d \tau'
C_S \eexp{-\tau/\mu_b} \eexp{-\tau'/\mu_a} + {\cal O}(\T^0)
\\
&=& \frac{4 \mu_a\mu_b}{\T} + {\cal O}(\T^0) .
\end{eqnarray}
We can use this to find an approximation for the reflected 
intensity
$A^{\rm R} (\theta_a,\theta_b)$ in the case $m\gg1$.
\begin{eqnarray}
A^{\rm R}&=&\frac{\cos \theta_a}{4\pi m^2} \frac{T_aT_b}{\mu_a\mu_b}
\gamma(\mu_a,\mu_b).
\end{eqnarray}
When $m\gg1$, $T(\mu) \approx 4 \mu/m$ (since $\theta$ 
cannot exceed the Brewster angle, which is small at large $m$),
implying
\begin{eqnarray}
A^{\rm R}&=&\frac{\cos \theta_a}{4\pi m^2} \frac{4 \cos \theta_a}{m}
\frac{4 \cos \theta_b}{m} \frac{4}{\T}
\approx \frac{6}{\pi m} \cos^2 \theta_a \cos \theta_b .
\end{eqnarray}

\subsection{Limit intensity and injection depth}

The limit intensity $\tau_1(\mu)$ of a semi-infinite space
can also be calculated. From Eq.~(\ref{GGG}) it follows that
\begin{eqnarray}
\tau_1(\mu)&=&\int_0^\infty \d \tau' G_S(\infty,\tau')\eexp{-\tau'/\mu}
\\ 
&=&\frac{1}{D} \int_0^\infty \d \tau' \Gamma_H(\tau')\eexp{-\tau'/\mu}
\\
&=&\lim_{\mu_b \to \infty} \frac{\gamma(\mu,\mu_b)}{\mu_b}
\\
&=& \frac{4\mu}{\T} .
\end{eqnarray}
The injection depth $z_0=\tau_0 \ell$ can also be calculated.
If we take the (diffusion) approximation $\Gamma_H=\tau_0+\tau\approx\tau_0$
and insert it in the above equality we find
\begin{eqnarray}
\tau_1(\mu)=\frac{\tau_0}{D}\mu&=&\frac{4\mu}{\T}
\\ \label{mcw7}
\tau_0&\approx&\frac{4}{3\T} .
\end{eqnarray}

Just as in the calculation of $G_S$ we can write the homogeneous
solution $\Gamma_H$ as the sum of three parts:
\begin{eqnarray}
\Gamma_H(\tau)&=&\frac{4}{3\T} + \Gamma_0(\tau)+\Gamma_1(\tau) \\ 
&& \Gamma_0(\tau)={\cal O}(1),\qquad \Gamma_1(\tau) ={\cal O}(\T) .
\end{eqnarray}
We further require that $\Gamma_0$ behaves as $\tau
+\tau_{00}$ for 
$\tau \to \infty$. The constant $\tau_{00}$ is a ${\cal O}(1)$ correction
to $\tau_0$. 
For $m \to \infty$ we can use definition Eq.~(\ref{definN})
and that $\mu \approx 1$ since the Brewster angle is small:
\begin{eqnarray}
N(\tau + \tau')&=&\int_0^1 \frac{\d \mu}{2\mu} 
\eexp{-\frac{\tau+\tau''}{\mu}}T(\mu)
\approx \frac{\T}{4} \eexp{-(\tau+\tau')} .
\end{eqnarray}
Inserting this in the equation for $\Gamma_H$ and comparing
 terms up to order $\T^2$, we obtain
\begin{eqnarray}
\Gamma_0&=-& 
\eexp{-\tau}+\int_0^\infty \d \tau' 
[M_{\rm B}(\tau-\tau')+M_{\rm B}(\tau+\tau')]
\Gamma_0 , \end{eqnarray}
while the equation for $\Gamma_1$ yields the compatibility relation
\begin{equation} 
\int_0^\infty d\tau \eexp{-\tau} \Gamma_0(\tau)=0 \label{mcw9}.
\end{equation}
These equations cannot be solved analytically. Numerical analysis
yields $\tau_{00}=-1.0357$ ~\cite{thmn5}, implying
\begin{eqnarray} \label{mcwtn8}
\tau_0^{\rm exact}=\frac{4}{3\T} - 1.0357 +{\cal O}(\T)
,\qquad (m\to\infty). \end{eqnarray}

In the limit $m\to 0$ the transmission coefficient is uniformly small,
$T(\mu) \approx 3{\cal T}\mu/2$. Therefore the term of Eq.~(\ref{specialG}) 
arising from the kernel $N$, defined in Eq.~(\ref{definN}), 
becomes equal to
\BA -\frac{3}{2}{\cal T}\int_0^\infty d\tau''\int_0^1\frac{d\mu}{2\mu}
\eexp{-(\tau+\tau'')/ \mu} \mu\frac{4}{{\cal T}} &=&-3\int_0^1 \mu\, d\mu
\eexp{-\tau/\mu}\times \label{mcw12} \nonumber \\
\int_0^1 \d\mu' \mu' \int_0^\infty d\tau \eexp{-\tau/\mu'} {\rm J}_0
(\tau,\mu')&=& 0 \nonumber \\
\Rightarrow \int_0^1 \d\mu \mu \int_0^\infty d\tau \eexp{-\tau/\mu}
\Gamma_0(\tau). \EA
The diffusive shape $\tau+\tau_{00}$ is the exact solution of this
equation, with $\tau_{00}=-3/4$.
This leads to
\begin{eqnarray}
\tau_0^{\rm exact}&=&\frac{4}{3\T}-\frac{3}{4}+{\cal O}(\T) 
\,\;\; \,(m\to 0).
\label{mcwtn13}
\end{eqnarray}
In Fig.~\ref{jmltau} numerical values of $\tau_1(1)$, $\gamma(1,1)$,
and $ 3\tau_0$ have been plotted as a function of $m$.
The solid line is $4/\T$,
while the dashed line is Eqs.~(\ref{mcwtn8})+(\ref{mcwtn13}).
Notice that these asymptotic expressions work quite well up to $m=1$.

\subsection{Comparison with the improved diffusion approximation}

We can compare the exact results for $\tau_0$ with the improved
diffusion approximation of Sec.~\ref{vbdb}
For $m\to\infty$ we can approximate $C_1$ and $C_2$ [see Eq.~(\ref{C1=C2=})]
as
\begin{equation}
C_1=\drd (1-\T) \; ,\; 
C_2\approx\drd-\int_0^1\mu\d\mu T(\mu) = \drd-\hlf \T    .
\end{equation}
From Eq.~(\ref{z0===}) it follows that 
\begin{equation}
\tau_0^{\rm diff}\approx \frac{4}{3\T}-1,
\end{equation}
which is very close to the exact value in Eq.~(\ref{mcwtn8}). Notice
that the same result follows when one inserts the diffusion approximation
$\Gamma_0(\tau)=\tau+\tau_{00}$ (for all $\tau$) in Eq.~(\ref{mcw9}).

For $m \to 0$  deviations from the diffusion approximation show up
only to second order in $\T$,
\begin{eqnarray}
\tau_0^{\rm diff}=\tau_0^{\rm exact}=\frac{4}{3\T}-\frac{3}{4}+
{\cal O}({\cal T})\qquad (m\rightarrow 0). \end{eqnarray}
To find corrections terms in powers of $\T$ is essentially as
 complicated as solving the Milne equation at finite $m$.
For $m=1$ (no internal reflections), the exact value is known:
$\tau_0^{\rm exact}=0.71044$. This can be
compared with the result for the diffusion
approximation found in Sec.~\ref{vbdb},
 \begin{eqnarray}
\tau_0^{\rm exact}&=&0.71044 \\
\tau_0^{\rm diff} & = &\frac{2}{3}=0.666666.
\end{eqnarray}

\subsection{Backscatter cone}

We now determine the shape of the backscatter cone in the presence of
strong internal reflections, in the regime $m \gg 1$. We have to solve
\begin{eqnarray}
\label{back1}
\Gamma_C(Q,\tau)&=&\eexp{-\tau}+
\int_0^\infty \d \tau' [M_{\rm B}(Q,\tau-\tau')+ 
M_{\rm B} (Q,\tau+\tau')]\Gamma_C(Q,\tau')
\nonumber \\
&&-\int_0^\infty \d \tau' N(Q,\tau+\tau')\Gamma_C(Q,\tau')    .
\end{eqnarray}
For small $Q$ the solution will have the form
\begin{eqnarray}
\Gamma_C(Q,\tau)\approx \gamma_C(Q)\eexp{-Q\tau}.
\end{eqnarray}
Integrating Eq.~(\ref{back1}) over $\tau$, using that the
 bulk-kernel $M_{\rm B}$ has an eigenvalue $1-\drd Q^2$, and neglecting
the
$Q$ dependence of $N$, we find 
\begin{eqnarray}
0&=&1+\int_0^\infty \d \tau \d \tau'
[M_{\rm B}(Q,\tau-\tau')-M_{\rm B}(0,\tau-\tau')+ \nonumber \\
&&M_{\rm B}(Q,\tau+\tau')
-M_{\rm B}(0,\tau+\tau')]\gamma_C(Q)\eexp{-Q\tau'}
\\
&&-\int_0^\infty \d \tau\d\tau' N(0,\tau+\tau')\gamma_C(Q)\eexp{-Q\tau}
\\
&=&1+\int_0^\infty \d \tau [-1+1-\drd Q^2]\gamma_C(Q)\eexp{-Q\tau}
-\frac{\T}{4}\gamma_C(Q),
\end{eqnarray}
yielding
\begin{equation}\label{gamaCQ}
\gamma_C(Q)=\frac{4}{\T+\frac{4}{3}Q} =\frac{4}{{\cal T}}\frac{1}{1+Q\tau_0}
=\frac{3}{Q+1/\tau_0} \label{mvr3} .\end{equation}
In passing, we have recovered the triangular shape of the backscatter cone.

Recently the width of the backscatter cone has been determined 
experimentally \cite{denouter2}. The index  ratio $m$ was varied using several
types of glass containers. The measurements have been compared with the 
theoris of \citeasnoun{lagendijk1}, of \citeasnoun{zhu} 
and of \citeasnoun{thmn5}. The first two results
have already been discussed in Sec.~\ref{vbdb}. The experiments confirmed the
theories of Zhu, Pine, and Weitz and of Nieuwenhuizen and Luck.

At first sight it may look strange that an experiment with light 
 (vector waves)
can be described by a scalar theory. \citeasnoun{thmn5} pointed out, however,
that for large $m$  only diffusive aspects survive. 
More subtle properties, such
as anisotropic scattering and the vector character lead to subdominant
effects. The latter was verified explicitly by \citeasnoun{amic2}, indicating
why the experiments of \citeasnoun{denouter2} are described so well by scalar
theories.

\section{Semiballistic transport}

Wave propagation through waveguides, quantum wires, films, and double-barrier
structures with a modest amount of disorder can be considered within the same
approach. The so-called {\it semiballistic regime} occurs when in the
transverse direction(s) there is almost no scattering, while in the long
direction(s) there is so much scattering that the transport is diffusive.
Various applications were noted by \cite{thmn12} and discussed
in detail by \citeasnoun{mosk}
The scattering has been considered both in the second order Born
approximation and beyond that approximation. In the latter case it was found
that attractive point scatterers in a cavity always have geometric resonances,
even for Schr\"odinger wave scattering. 

In the limit of long samples, the transport equation has been solved
analytically including geometries such as: waveguides, films, and tunneling
geometries such as Fabry-P\'erot interferometers and double barrier quantum
wells. The agreement with numerical data and with experiments  
is quite satisfactory. 
In particular, the analysis proved that the large,
gate-voltage independent resonance width of GaAs double-barrier systems,
observed by  \citeasnoun{gueret} can be traced back to scattering from the
intrinsically rough GaAs-AlGaAs interfaces.

In the analysis of these systems, the Schwarzschild-Milne equation is 
decomposed on the cavity modes, thus becoming an evolution equation 
in the $z$-direction for a matrix. The very same ideas as explained
above are used in the solution.
Due to the complication of the formalism we shall not present
details here.   
We refer the reader to the original works and the references therein.

\section{Imaging of objects immersed in opaque media}

A field of much current interest is the imaging of objects immersed in
diffuse media for the purpose of medical imaging. The advantage of imaging with
diffuse light is that it is  non-invasive and as opposed to conventional X-ray
tomography it does not cause radiation damage. One application is the detection
of  breast cancer, another application  is the location of brain regions
damaged by  stroke (a lot of blood, hence more absorption) or affected by
Parkinson's disease  (less than average blood level, hence more transparency). 
In these cases one can shine light at many locations, and measure the scattered
light at many exit points. This information can in principle be used to
determine the structure of the scattering medium and to detect
possibly abnormal behavior. It is therefore important to know  the scattering properties
of idealized objects hidden in opaque media. We shall discuss here some
instructive cases.

For light waves human tissue is a multiple scattering medium like the ones
discussed so far. (Yet the absorption length is short, typically a few tenths
of a millimeter). The diffuse transmission through tissue must already
have been noticed by our early ancestors 
when they held their hands in front of a candle. 
On distances much larger than the mean free path, light transport is
diffusive and is therefore governed by the Laplace equation. This observation
brings a deep analogy with electrostatics: an object deep inside an opaque
medium is analog to a charge configuration at large distance. It is well known
that the latter can be expressed in terms of its charge and dipole moment,
while higher multipole moments are often negligible. A similar situation occurs
in an opaque medium: small objects far from the surface can be described
adequately by their charge (which vanishes if there is no absorption) and
dipole moment (which describes to leading order scattering from the object). 
At large distances the effect of, for instance, a charge will depend on the
geometry of the system. As in electrostatics, mirror charges and mirror dipoles
can be used in simple geometries. However, to solve the boundary problem of
complicated geometries like a head is a hard problem. Close to the object, say,
within a few mean free paths, the full radiative transfer is needed,
making the problem more complicated than diffusion theory. 

~\citeasnoun{denouter} studied the diffuse image of an object embedded in a
homogeneous turbid medium and found that it is well described by the diffusion
approximation. This framework leads to a prediction of the  diffuse image of
the object, i.e., the profile of diffuse intensity transmitted through a thick
slab near the embedded object. The shapes of the predicted profiles are
universal since they depend only on large distance properties of the problem,
and were convincingly observed experimentally for the case of a pencil
(absorbing cylinder) and a glass fiber  (transparent cylinder). Further results
on imaging of objects within the framework of the diffusion approximation can
be found in \citeasnoun{SchotlandHaselgrove}, \citeasnoun{FengZeng},
\citeasnoun{ZhuWei}, and \citeasnoun{PaasschenstHooft}.

\subsection{Spheres}

In the diffusion approximation the problem is straightforward.
Close to the object (though still many mean free paths away)
the intensity can be written as
\begin{equation}\label{pqexp}
I({\bf r})=I_0({\bf r})+\delta I({\bf  r}),
\end{equation}
where $I_0({\bf r})$ is the intensity profile in the absence of the object.
\citeasnoun{luck98} have introduced the following notation for 
the disturbance  of the intensity
\begin{equation}\label{deltaI=}
\delta I({\bf r})=(q+{\bf p}\cdot \nabla)\frac{1}{|{\bf r}-{\bf r}_0|}
= \frac{q}{|{\bf r}-{\bf r}_0|}
-\p\cdot \frac{{\bf r}-{\bf r}_0}{|{\bf r}-{\bf r}_0|^3}  .
\end{equation}
$q$ is called the ``charge'' and ${\bf p}$ 
the ``dipole moment'' of the object. 
The linearity of the problem and the isotropy of the spherical defect
imply that the charge $q$ is proportional to the local intensity
$I_0({\bf r}_0)$ in the absence of the object,
while the dipole moment $\p$ is proportional to its gradient.
We therefore set
\begin{equation}
q=-QI_0({\bf r}_0),\quad\p=-P\nabla I_0({\bf r}_0),
\end{equation}
$Q$ is called the {\it capacitance} of the spherical object, and $P$ its {\it
polarizability}. These two parameters are intrinsic characteristics of the
object. The capacitance $Q$ is non-zero (and positive) only if the
sphere is absorbing.

For plane wave incidence on a slab geometry one has
the familiar expression $I_0({\bf r})=(L-z)I_0/L$.
A double set of mirror charges is needed
to make $I-(L-z)I_0/L$ vanish at $z=0$ and $z=L$ for all $\rho$:
\begin{eqnarray}
{I}&=&\frac{L-z}{L}I_0+q\sum_{n=-\infty}^\infty\left\{ 
\frac{1}{[(z-z_0+2nL)^2+(\rho-\rho_0)^2]^{1/2}}
-\frac{1}{[(z+z_0+2nL)^2+(\rho-\rho_0)^2]^{1/2}}\right\} \\
&-&p\sum_{n=-\infty}^\infty\left\{ 
\frac{z-z_0+2nL}{[(z-z_0+2nL)^2+(\rho-\rho_0)^2]^{3/2}}
+\frac{z+z_0+2nL}{[(z+z_0+2nL)^2+(\rho-\rho_0)^2]^{3/2}}\right\}.
\end{eqnarray}
The intensity $T(x,y)$ transmitted through the sample,
and emitted on the right side $(z=L)$ at the point $\rho=(x,y)$,
is proportional to the normal derivative of the diffuse intensity
at that point:
\begin{equation}\label{transmder}
T(x,y)=-K\ell\left.\frac{\partial I(x,y,z)}{\partial z}\right\vert_{z=L}.
\end{equation}
When there is no embedded object, the transmission is uniform 
$ T_0 = K I_0\ell/ L $.
In the presence of an object one finds
\begin{eqnarray}
T(\rho)=T_0\Bigg(
1&-&2Q(L-z_0)\sum_{n=-\infty}^{+\infty}
\frac{L-z_0+2nL}{\big[(L-z_0+2nL)^2+(\rho-\rho_0)^2\big]
^{3/2}}\Bigg.\nonumber\\
&-&2P\sum_{n=-\infty}^{+\infty}\Bigg.
\frac{2(L-z_0+2nL)^2-(\rho-\rho_0)^2}
{\big[(L-z_0+2nL)^2+(\rho-\rho_0)^2\big]^{5/2}}\Bigg).
\end{eqnarray}
The information on the type of scatterer is coded in
its capacitance $Q$  and  polarizability $P$. In experiments on
pencils and fibers, $2-d$ analogs of these profiles 
have been clearly observed experimentally~\cite{denouter}.
In the same fashion the reflected intensity can be estimated
by taking the derivative in Eq.~(\ref{transmder}) at $z=0$.
The rest of this section is devoted to determining $Q$ and $P$ in idealized
situations. This amounts to solving the scattering problem with the radiative
transfer equation in the presence of an object placed at ${\bf r}_0$.

For large objects the diffusion theory can be used which simplifies the problem.
Following \citeasnoun{denouter} we consider a macroscopic
object of radius $R$,
having diffusion coefficient $D_2$ and inverse absorption
length $\kappa$, while the medium is nonabsorbing and has
diffusion coefficient $D_1$. At the boundary of the object
one has 
\begin{eqnarray}
I_{out}(R^+)&=&I_{in}(R^-)\\
D_1\frac{\partial I_{out}}{\partial n}\Bigl|_{R^+}
&=&D_2\frac{\partial I_{in}}{\partial n}\Bigr|_{R^-}
\end{eqnarray}
Near the object one has
\begin{equation} 
I_{in}=A\, {\rm sinhc}(\kappa r)-\frac{3B}{\kappa^2r} [ {\rm sinhc} (\kappa r)
-\cosh(\kappa r )]\frac{z-z_0}{r},
\end{equation}
where sinhc$(x)$$\equiv$sinh$(x)/x$.
Matching the solutions gives 
\begin{equation}
Q=R\,\,\frac{D_2(\cosh\kappa R -{\rm sinhc}\, \kappa R)}
{D_1{\rm sinhc}\,\kappa R+D_2(\cosh\kappa R-{\rm sinhc}\,\kappa R)}
\end{equation}
For a totally absorbing sphere ($\kappa\to\infty$) the result $Q=R$
follows immediately by requiring in Eq.~(\ref{pqexp}) 
that $I(R)=0$.  The polarizability reads
\begin{equation}
P=R^3\,\frac
{D_1({\rm sinhc}\,\kappa R-\cosh\kappa R )
-D_2(2\cosh\kappa R-2{\rm sinhc}\,\kappa R -\kappa R \sinh\kappa R)}
{2D_1 ({\rm sinhc}\,\kappa R-\cosh\kappa R )+D_2(2\cosh\kappa R
 -2 {\rm sinhc}\,\kappa R -\kappa R \sinh\kappa R)}.
\end{equation}
In the absence of absorption this reduces to
\begin{equation} Q=0, \qquad P=R^3\,\frac{D_1-D_2}{2D_1+D_2}.
\end{equation}
 
\citeasnoun{LancasterNieuwenhuizen} consider the limit of small objects: a
point scatterer
with scattering cross section $\sigma_{sc}$, extinction cross section 
$\sigma_{ex}$, and albedo $a_e=\sigma_{sc}/\sigma_{ex}$. 
If diffuse intensity hits the scatterer once, but does not return to
it, $Q$ and $P$ read 
\begin{eqnarray}\label{Q1P1}
Q_1 = {3\over 4\pi}{\sigma_{abs}\over \ell}, \qquad
P_1 ={1\over 4\pi}{\sigma_{ex}\ell},
\end{eqnarray}
where $\sigma_{abs}=\sigma_{ex}-\sigma_{sc}$.
The intensity can also return an arbitrary number of times
to the scattererer. For point scatterers this boils down
to a geometric series expansion for $Q$ and $P$. \begin{eqnarray}
Q &=&\frac{Q_1}{ 1 + Q_1 B_1},\qquad
P = {P_1\over 1 + P_1 B_2},  \label{eq330}
\end{eqnarray}
where $B_1$ and $B_2$ are integrals describing the propagation of the diffuse
intensity through the medium  and the effect of hitting the scatterer another
time. These integrals are actually divergent in the limit $k\ell\to\infty$,
which leads to introduce extra internal parameters of the point scatterer.

As opposed to extended objects, for point scatterers the maximally crossed
diagrams also contribute. The reason is that exit and return point coincide, so
there is no de-phasing. The analysis finally yields 
\begin{eqnarray}
Q &=&\frac{Q_1}{ 1 + Q_1 (2B_1-B_1^{(1)})},\qquad
P = {P_1\over 1 + P_1 (2B_2-B_2^{(1)})}, 
\end{eqnarray}
which retains the structure of Eq. (\ref{eq330}). Note that there is an
additional factor 2, except for the diagrams with one intermediate common
scatterer, an effect that is accounted for by subtracting $B_{1,2}^{(1)}$.
This is very similar to the fact that the single-scatterer diagram does not
contribute to the enhanced back scatter cone, see eq. (\ref{209}).
\citeasnoun{LancasterNieuwenhuizen} also considered more realistic situations 
in an approach involving a partial wave expansion.

For extended objects it is a highly non-trivial task to go beyond the 
diffusion approximation. \citeasnoun{luck98} considered the application of the
radiative transfer equation to such cases. One thus has to solve the analog of
the scattering problem of quantum mechanics, for this transfer  equation. This
is generally a hard and still unsolved problem. However, there are three cases
where the inside of the scatterer plays a trivial role: a totally  absorbing
(black) object (having $\kappa=\infty$),  a transparent, non-absorbing object
(having $\kappa=0$, $D_2=\infty$), and a totally reflecting, non-absorbing
object (having $\kappa=D_2=0$). For spheres and cylinders of these types, the
scattering problem can be reduced to an integral equation in one variable,
which is solved exactly in the limits of small and large objects.  In the
latter case the results of diffusion theory are recovered, and the first order
corrections to diffusion theory were found, see Table \ref{table3} and
Figs.~\ref{jmlfig1} and ~\ref{jmlfig23}. The same approach can be extended, in
principle, to more general cases.

\subsection{Cylinders}

The use of the radiative transfer approach can be extended to other objects
\cite{luck98}. Consider a cylinder with radius $R$   with its axis parallel to
the $x$-axis. Assume the sample is infinitely long in the x-direction, so that
both $I_0({\bf r})$ and the disturbance $\delta I({\bf r})$ depend only on the
two-dimensional perpendicular component  ${\bf r}_\perp=(y,z)$. 
Eq.~(\ref{deltaI=}) is replaced by
\begin{equation}
\delta I({\bf r}_\perp)=(q+\p\cdot\nabla)\ln\frac{\tilde R}
{|{\bf r}-{\bf r}_0|}
=q\ln\frac{\tilde R}{|{\bf r}-{\bf r_0}|}
-\frac{\p\cdot ({\bf r}_\perp-{\bf r}_{0,\perp})}{|{\bf r}-{\bf r}_0|^2},
\end{equation}
The length scale $\tilde R$, requested by dimensional analysis,
is determined by conditions at the boundaries of the sample,
so that it is in general proportional to the sample size.
This sensitivity of $\delta I({\bf r})$ to global properties
of the sample arises because the logarithmic potential of a point charge
in two-dimensional electrostatics is divergent at long distances.
As a consequence, the capacitance $Q$ is not intrinsic to the cylinder.
An intrinsic quantity is its {\it effective radius} $R_{\rm eff}$,
defined by the condition that the cylindrically symmetric part
of the total intensity, including the charge term, 
vanishes at a distance $r=R_{\rm eff}$ 
from the axis of the cylinder.
We thus have $I_0({\bf r}_0)+q\ln(\tilde R/R_{\rm eff})=0$, hence,
with $q=-I({\bf r}_0)Q$,
\begin{equation}
Q=\frac{1}{\ln\frac{\tilde R}{R_{\rm eff} }}.
\end{equation}
We observe that $R_{\rm eff}$ and $\tilde R$ have the dimension of a length,
while the polarizability $P$ has the dimension of an area.
As a consequence, the orders of magnitude $R_{\rm eff}\sim R$ and $P\sim R^2$
can be expected for a cylinder of radius $R$.
For the same physical situations as above, 
more accurate predictions can be found in Table \ref{table3}.
For a cylinder in a slab geometry the transmitted intensity $T(y)$ 
finally reads in terms of $Q$ and $P$
\begin{equation}
T(y)=T_0\left(1-Q\pi\left(1-\frac{z_0}{L}\right)
\frac{\sin\frac{\pi z_0}{L}}{\cos\frac{\pi z_0}{L}+\cosh\frac{\pi y}{L}}
-P\frac{\pi^2}{L^2}\frac{1+\cos\frac{\pi z_0}{L}\cosh\frac{\pi y}{L}}
{\left(\cos\frac{\pi z_0}{L}+\cosh\frac{\pi y}{L}\right)^2}\right).
\end{equation}
where 
\begin{equation}
\tilde R=\frac{2L}{\pi}\sin\frac{\pi z_0}{L}
\end{equation} 
is indeed of order $L$, but also depends non-trivially on the aspect ratio
$z_0/L$. These shapes have been observed in the experiments of
~\citeasnoun{denouter}.

\section{Interference of diffusons: Hikami vertices}\label{chhik} 

So far our discussion has focused on the intensity aspects of wave transport.
We now consider interference effects, which are due to the inherent wave nature
of the problem. First we have to introduce the concept of interference
vertices. These vertices describe the interaction between diffusons. The
vertices are named after Hikami, who used them in 1981 \cite{hikami} for the
calculation of weak-localization corrections to the conductance. Their original
introduction dates back two years earlier to the work of \citeasnoun{gorkov}. 
Remember that each diffuson consists of two amplitudes. The vertices describe
the exchange of amplitudes between diffusons, which leads to correlations
between the diffusons. It is also possible to represent these processes using
the standard impurity technique diagrams, as described for instance by
\citeasnoun{abrikosov}, but the interference vertices prove to be more
convenient.

\subsection{Calculation of the Hikami four-point vertex} 

We start with the simplest vertex: the four point vertex, in which  two
diffusons interchange an amplitude.  The technique for calculating the vertex
by means of diagrams is well known \cite{abrikosov}: First, draw the diagrams;
second, write down a momentum for each line and use momentum conservation at
each vertex; and finally, integrate over the free momenta. The first step,
writing down all leading diagrams, requires some care. In Fig.~\ref{figh4sob}
we have drawn the diagrams in the Hikami box to second-order Born
approximation.  It is important to include {\em all} leading diagrams, as there
will be a cancellation of leading terms. These cancellations are imposed by
energy conservation. In the second-order Born approximation one neglects
scattering more than twice from the same scatterer; the (dashed) interaction
line indicates that the propagators scatter once on the same scatterer. 
Scattering processes involving just one propagator were already included, as
the propagators are dressed. We have not drawn all diagrams, one can imagine.
First, the box is attached to a diffuson ending with a scatterer (which is
consistent with the way we have defined the diffusons). As a result, common
scatterings between propagators on the same leg are not allowed in the vertex
as they are already included.  Secondly, diagrams with two or more scatterers
and parallel dashed lines are subleading. They contain loops with two
propagators of the {\em same} type, i.e., integrals like $\int \d^3 \p G(\p+\q)
G(\p)$, which turn out to be of higher order in $1/(k_0\ell)$.  Third diagrams
with two crossed dashed lines are also subleading. Finally, we again work in
the independent-scatterer approximation. 

As an
example we calculate the second right hand side diagram of Fig.~\ref{figh4sob}.
As usual in the diffusion approximation, the diffuse intensity 
is assumed to be slowly varying, which allows us
 to expand the Green's functions as
\BA G(\p+\q, \omega+\frac{1}{2}\Delta \omega) &\approx& \left[ p^2+2\p\cdot
\q+ q^2-\frac{w^2}{c^2} -\Delta \omega \frac{\omega}{c^2}-n t\right]^{-1}
\nonumber \\ &\approx & G- (2 \p\cdot \q+ q^2-\Delta \omega \frac{\omega}{c^2}
) G^2+ 4(\p\cdot\q)^2 G^3 , \label{eqgexp2} \EA 
with the shorthand notation $G=G(\p,\omega)$. The
momenta, numbered according to \fig{figh4sob}, point towards the vertex. We
number the Green's functions also according to the figure. Note that this
particular
diagram is a product of two loops. To lowest order in
$q^2$ we have 
\BA H_4^{(b)}& =& nu^2 \int \frac{\d^3 \p}{(2\pi)^3}
G_1(\p+\q_1,\omega+ \omega_1) G_3 (\p-\q_4, \omega+ \omega_3) G_4^*(\p,\omega+
\omega_4) \nonumber \\&& \times \int \frac{\d^3 \p'}{(2\pi)^3} G_1(\p'-\q_2,
\omega+ \omega_1) G_3 (\p'+\q_3,\omega+ \omega_3) G_2^*(\p',\omega+ \omega_2) 
\nonumber \\ 
&=& nu^2 \left[I_{2,1} + \frac{4k^2}{3} (q_2^2+q_3^2-\q_2\cdot
\q_3 )I_{4,1} +\frac{k}{c} (\omega_1+\omega_3-2\omega_4) I_{3,1} \right] 
\nonumber \\ && \times \left[ I_{2,1}+\ldots \right] , \EA where the 
${\bf q}_i$ are the momenta of the diffusons, the $\omega_i$ are the
frequency differences of the diffusons, and $u$ is the scatter potential.
The $I-$ integrals are given in Appendix \ref{seciint}. 
We take absorption terms into account to lowest order,
i.e., only in the leading contributions, the $I_{2,1}$ terms.
 This yields \begin{equation}
H_4^{(b)} = \frac{-\ell^3}{16\pi k^2} +\frac{\ell^5}{48\pi k^2}(-\q_2\cdot
\q_3-\q_1\cdot\q_4 +\sum q_i^2 +\kappa^2) +\frac{-i\ell^4}{16\pi k^2
c}(\omega_1+\omega_3-\omega_2-\omega_4). \end{equation} 
Here we made use of the fact 
that in second order Born approximation $nu^2=4\pi/\ell$.

A similar calculation gives $ 
H_4^{(a)} $ and $H_4^{(c)}$.
Defining the reduced frequencies of the legs as
 $\Omegamark_1=-(\omega_1-\omega_4)/D$,
$\Omegamark_2=-(\omega_1-\omega_2)/D $, 
$\Omegamark_3=-(\omega_3-\omega_2)/D $ 
and
$\Omegamark_4=-(\omega_3-\omega_4)/D $, 
we find for the sum of the three diagrams
\BA H_4 &=&
H_4^{(a)}+H_4^{(b)}+H_4^{(c)} \nonumber \\ &=&\frac{\ell^5}{96 \pi k^2} \left[
-2 {\bf q}_1 \!\cdot \!{\bf q}_3 -2 {\bf q}_2 \!\cdot \!{\bf q}_4 +
\sum_{i=1}^4 ({\bf q}_i^2 +\kappa_i^2 +i\Omegamark_i) \right]
 \delta(\sum \q_i).
\label{haha}\label{haha2} \EA 
This is the main result of this section. Note that the leading, constant
terms proportional to $\ell^3/k^2$ have canceled, 
which is closely related to
the current conservation laws~\cite{kane}. 

In Eq.~(\ref{haha2}) it is important
to keep track of the $q^2$ terms. When the vertex is attached to a diffuson,
the $q_j^2$ terms together with $\kappa$ and $\Omegamark$
 yield according to
the diffusion equation, $(q^2_j+\kappa^2+i\Omegamark_j)\L_j =12\pi /\ell ^3$. 
The constant factor in the right-hand side corresponds to
a delta function in real space. 
If we attach external diffusons to the 
Hikami box, this delta function is nonzero at roughly one 
mean free path from the surface and its contribution
can be neglected, as its effect 
is of order $\ell/L$. As a result, the absorption part and the
frequency-dependent part vanish:
\begin{equation} H_4(\q_1,\q_2,\q_3,\q_4)= -h_4 (\q_1 \!\cdot \q_3
+\q_2 \!\cdot \q_4 ) \delta(\sum \q_i)
 \; ,\label{hik1} \end{equation} 
where we have denoted $h_4 =\ell^5 /(48 \pi k^2)$.
In most calculations, however, we use a Fourier 
transform in the $z$ direction, because in our slab geometry the 
$(q_x,q_y,z)=(\qt,z)$ representation is the most 
convenient. Then the $q_z$ terms become differentiations:
\begin{equation} H_4=\frac{h_4}{2} \left[
2\partial_{z_1} \partial_{z_3}\!+\! 2 \partial_{z_2} \partial_{z_4}-2 {\qt}_1
\cdot {\qt}_3 \!-\!2 {\qt}_2 \cdot \qt_4 + \sum_{i=1}^4 (-\partial_{z_i}^2+
{\qtl}_i^2 +\kappa_i^2 +\Omegamark_i) \right]. \end{equation}
The differentiations work on the corresponding
diffusons and afterwards $z_i$ should be set $z$. Formally one has \BA && H_4
(z;z_1,z_2,z_3,z_4) \L_1(z_1)\L_2(z_2)\L_3(z_3)\L_4(z_4) \nonumber \\ &=&
\frac{h_4}{2} \left. \left[ 2\partial_{z_1} \partial_{z_3} + \ldots
\right] \L_1(z_1)\L_2(z_2)\L_3(z_3)\L_4(z_4) \right|_{z_i=z}, \label{mcwh4a}
\EA but we just write this as $ H_4(z) \L_1(z) \L_2(z)\L_3(z)\L_4(z)$.
Note that the vertices 
yield the spatial derivatives of the diffusons, that is,
their fluxes. 
This observation is helpful in estimating the influence
of internal reflections, which reduce the spatial 
derivatives of the diffusons.

\subsection{Six-point vertex: $H_6$}

We shall also need a second order diagram the six-point vertex $H_6$. Six
diffusons are connected to this diagram (Fig.~\ref{figh6}).
This diagram was calculated by \citeasnoun{hikami}. Again,
the dressings of the diagrams have to be added to the bare diagrams. Taking
rotations of the depicted diagrams into account, there are 16 diagrams in the
second-order Born approximation. (It is not allowed to dress the bare six-point
vertex with a scatterer that connects two opposite propagators. This dressing
gives a leading contribution also, but the resulting diagram is the same as 
a composed
diagram with two four-point vertices. Therefore, it should not be included
here, but it will enter in diagrams with two $H_4$ vertices.)
One finds 
\begin{eqnarray} H_6& =&\frac{-\ell^7}{96 \pi k^4} \left[{\bf q}_1 \!\cdot 
\!{\bf q}_2+{\bf q}_2\!\cdot \!{\bf q}_3+ {\bf q}_3\!\cdot \! {\bf q}_4+{\bf
q}_4\!\cdot \! {\bf q}_5 +{\bf q}_5 \!\cdot \!{\bf q}_6 +{\bf q}_6 \!\cdot
\!{\bf q}_1 \right. \nonumber \\ && \left. +\sum_{i=1}^6 ({\bf q}_i^2
+\frac{1}{2}\kappa_i^2+\frac{i}{2}\Omegamark_i)\right] \delta(\sum \q_i).
\label{eqhik6} \end{eqnarray} Apart from the frequency and
absorption terms included here, Hikami's original expression
can be derived from Eq.~(\ref{eqhik6}) using momentum conservation.

\subsection{Beyond the second-order Born approximation}\label{secbsob}

Going beyond the second-order Born approximation in 
the calculation of the diffuson diagrams, 
meant that in the diagrams the $t$ matrix replaced the potential $u$.
This resulted in a replacement of the mean free path: 
it became $\ell \equiv nt\tb /4\pi$ instead of $\ell \equiv nu^2 /4\pi$. 
However, for
the vertices the situation is more subtle. In the previous calculations we
worked to second-order of the scattering potential, thus neglecting diagrams
with more than two scatterings on the same scatterer. Including all orders
there are eight diagrams, Fig. \ref{hikbox}.

The calculation is very similar to that 
of the second-order Born approximation above. 
The only difference is that instead of $u$, 
the $t$ matrices $t$ and $\tb$ occur 
($t$ on
$G$'s, $\tb$ on $G^*$'s). After summing the expressions for the eight diagrams,
one finds the same result for the Hikami box as was found previously in second
order Born approximation, Eq. ~(\ref{haha}).
However, the definition of the mean free path is different: $\ell=nt\tb/4\pi$
instead of $\ell=nu^2/4\pi$. We conclude that, although extra diagrams were
present, only the mean free path was renormalized in 
extending from second-order
Born to the full Born series. This can also be shown on a much more generally
in the nonlinear sigma model \cite{lernerprive}. 
For the six-point vertex it was not checked explicitly whether the second and
full Born approximations give the same result 
(there are at least 64 diagrams), 
but we have no doubt they will.

\subsection{Corrections to the conductivity}\label{secweakloc}

In the original works of Gor'kov {\it et al.} and Hikami, the vertices were
introduced to calculate weak-localization corrections to the conductivity.
We have already mentioned that the maximally crossed diagrams
 give the largest correction
to the conductivity, increasing the return probability of the intensity. In the
Hikami vertex formalism these diagrams are drawn in Fig.~\ref{figghik}. The
correction to the diffusion constant diverges with
the sample size in one and two dimensions.
 This divergence indicates the absence of extended states in
one and two dimensions no matter how small the disorder. In three dimensions,
however, there is a transition to a localized state at a finite level of
disorder.  In this work we do not consider such loop effects. We suppose
that we are far from localization, so that we can restrict ourselves to the
leading processes, and the vertices show up only in the interaction between 
different diffusons.

\section{Short range, long range, and conductance
correlations: $C_1$, $C_2$, and $C_3$}\label{chc12} 

A nice feature of optical mesoscopic
systems, as compared to electronic systems, 
is that they permit to measure certain parts of the transmission signal
separately. Depending on whether integration over incoming or outgoing
directions is performed, one measures fundamentally different quantities. In
electronic systems one usually measures the conductance of the sample. This is
done by connecting incoming and outgoing sides to a clean electron ``bath.''
All electrons are collected and all angular dependence is lost. In contrast,
in optical systems one usually measures the angular resolved transmission, 
but angular
integration is possible. The correlation functions of the different
transmission quantities, their magnitudes, their decay rates, and the
underlying diagrams are all very different. 

In this section we discuss correlation functions for all three transmission
quantities. See also the review by \citeasnoun{berkovits3} for the  physical
meaning of the correlations.
We try to provide all quantitative 
factors relevant for an experiment. In later sections
the same approach will be extended to higher order correlations.
 
\subsection{Angular resolved transmission: the speckle pattern}

If a monochromatic plane wave shines through a disordered sample, one sees
large intensity fluctuations in the transmitted beam, so-called ``speckles.''
The speckle pattern is wildly fluctuating 
(as a function of the frequency of the
light or the outgoing angle). The typical diameter of a speckle spot on the
outgoing surface is $\lambda$. We are interested in the correlation between
speckle patterns of two different beams. The beams may have different incoming
angles, different frequencies, or different positions.

We denote the transmission from the incoming channel $a$ (wave coming in under
angles $\theta_a,\,\phi_a$) to the outgoing channel $b$ (waves transmitted into
angles $\theta_b,\,\phi_b$) as $T_{ab}$. 
There are short, long, and ``infinite''
range contributions to the correlation function of the speckles.
The frequency correlation functions can be classified as
\begin{equation} \frac{\langle T_{ab}(\omega)
T_{cd}(\omega+\Delta \omega) \rangle} { \langle T_{ab}(\omega)\rangle \langle
T_{cd}(\omega+\Delta\omega)\rangle} =1+C_1^{abcd}(\Delta
\omega)+C_2^{abcd}(\Delta \omega)+C_3^{abcd}(\Delta \omega)\label{eqdefcb}.
\end{equation} 
\citeasnoun{feng} originally put this expression forward as
an expansion in $1/g$, where $C_1$ was of order one, $C_2$ of order
$1/g$, and $C_3$ of order $1/g^2$. We return to this expansion below.
 We present a sketchy 
picture of the diagrams of the different correlations
 in Fig.~\ref{figc123}. The largest contributions
to transport are from the independent diffusons, yet correlations are
present, mixing diffusons with different frequencies or angles.

The unit contribution in Eq.~(\ref{eqdefcb}) just comes from the uncorrelated
product of average intensities. The $C_1$-term in the correlation function is
the leading correlation if the angular resolved transmission 
$T_{ab}$ is measured.
It is unity if the two incoming directions $a$ and $c$ are the same, the
outgoing directions $b$ and $d$ are the same, and $\Delta \omega$ is zero. If
one changes the frequency of the incoming beam, the speckles of the outgoing
beam deform and eventually the correlation vanishes exponentially. 
This effect is the short ranged $C_1$ contribution. Also of
interest is the case of the angular $C_1$ correlation with the frequency fixed:
If the angle of the incoming beam is changed, the speckles change due to two
effects. First, the speckle follows the incoming beam. This is also known as
the memory effect, which one can even see by the naked eye. Secondly, the
speckle pattern also deforms, i.e. it de-correlates, 
this is again the $C_1$ correlation.
The $C_1$ correlation is a sharply peaked function nonzero
 only if the angle of the
incoming and outgoing channel are changed by the same amount. Diagrammatically,
the $C_1$ factorizes in two disconnected diagrams, see \fig{figc123}. 

Theoretical studies of $C_1$ were first done by \citeasnoun{shapiro}.
Experimental work on $C_1$ as function of angle, explicitly showing the
memory effect, was done by \citeasnoun{freund}. \citeasnoun{albada1} studied
the frequency dependence of $C_1$. Later, the effects of
absorption and, especially, internal reflection were studied
 \cite{freund2,skin}.

\subsection{Total transmission correlation: $C_2$}

The $C_2$ correlation stands for the long range correlations 
in the speckle pattern. 
Consider again a single-direction-in single-direction-out experiment. Again
the two incoming angles are the same ($a=c$), yet this time we look at the
cross correlation of two speckles far apart. As there is an angle difference in
the outgoing channel, but not in the incoming beam, 
the $C_1$ term is now absent.
Instead there is a much weaker correlation. As the frequency shifts, this
correlation decays algebraically. This is the $C_2$ contribution. It describes
correlations between speckles far apart. 
In a single-channel-in single-channel-out
 experiment it is only possible to see the weak effects of these higher
order correlations in very strongly scattering media, i.e., if the system is
rather close to Anderson localization. This type of experiment was done by
 \citeasnoun{genack2}. The $C_2$ term can be measured more easily in a
setup using one incoming channel and integrating over all outgoing channels.
In an experiment the outgoing light is collected with an integrating sphere.
Therefore, one measures the total transmission $T_a=\sum_b T_{ab}$. In this
setup the $C_2$ correlation function, which has a much smaller peak value but
is long ranged, contributes for all outgoing angles. The $C_2$ term is now the
leading correlation as the sharply peaked $C_1$ term is overwhelmed, a precise
estimate of the $C_1$ contribution in $C_2$ was given in \cite{t3pre}. 
Only outgoing diffusons with no transverse momentum and no frequency 
difference in their amplitudes are 
leading in the total transmission, as then the phases of the
outgoing amplitudes cancel. From \fig{figc123}
one sees indeed that it holds for the $C_2$ diagram (outgoing lines of similar
style pair), but not for the $C_1$ diagram. The long range character of the
$C_2$ arises due to interference of the diffuse light paths. It is of order
$g^{-1}$ and corresponds to a diagram where the two incoming diffusons
interact through a Hikami vertex, where the diffusons exchange
amplitudes.

The long range $C_2$ correlation function was first studied by 
\citeasnoun{stephen1}. Zyuzin and Spivak introduced a Langevin approach to
simplify the calculation of correlation functions \cite{zyuzin1}.
\citeasnoun{pnini2} applied this method to calculate the correlation 
functions of light
transmitted through and reflected from disordered samples. The
$C_2$ correlation functions were measured in several experiments. For optical
systems Van Albada {\it et al.} performed measurements \cite{albada1,deboer}.
Microwave experiments were performed by \citeasnoun{genack1} and 
\citeasnoun{garcia}. In several
papers effects of absorption and internal reflections were studied, see
\citeasnoun{pnini1}, \citeasnoun{lisyansky}, \citeasnoun{zhu}, and
\citeasnoun{skin}. It was shown that absorption and internal reflection,
neglected in the earliest calculations, significantly reduce the correlations.

\subsection{Conductance correlation: $C_3$}

Finally, the $C_3$ term in \eq{eqdefcb} is dominant when the incoming beam is
diffuse, and one collects all outgoing light. Experimentally one could do this
using two integrating spheres. Then one measures, just as in electronic
systems, the conductance $g=\expect{T}=\sum_{a,b} T_{ab}$. In that measurement
contributions to the correlation are dominant, where the angles $a$ and $c$
are arbitrary far apart. Though $C_3$ is of order $g^{-2}$, it dominates over
the $C_1$ and $C_2$ terms as it has contributions from
 all incoming and outgoing angles. 
Therefore it is sometimes called the ``infinite'' range correlation. In
contrast to the $C_2$ term, now also the amplitudes of the incoming
diffusons must have opposite phases. 
This occurs in a diagram where the two incoming diffusons interact
twice. Note that a loop occurs, see \fig{figc123}. Apart from the normalization
to the average, $C_3$ equals the Universal Conductance
Fluctuations, or UCF. 

The UCF were an important discovery in the study of
mesoscopic electron systems. The electronic conductance of mesoscopic samples
shows reproducible sample-to-sample fluctuations. This was observed
 in experiments
on mesoscopic electronic samples by \citeasnoun{umbach}. The
theoretical explanation was given by \citeasnoun{altshuler2} and by
\citeasnoun{lee}. For reviews on the subject see
\citeasnoun{lee2} and the book edited by Altshuler {\it et al.} 
\cite{altshulerboekgeheel}. The discovery of these fluctuations
showed that interference effects are important in electronic systems, even far
from the localization transition.
The variance of the fluctuations is independent of the sample parameters such
as the mean free path, the sample thickness and, most remarkably, the average
conductance. Therefore, the fluctuations 
are called universal conductance fluctuations. 
The conductance
fluctuates when the phases of the waves in the dominant paths change. This
happens, of course, if one changes the position of the scatterers, e.g. by
taking another sample. One also may keep the scatterers fixed but apply a
magnetic field or vary the Fermi energy. In all these cases one modifies the
phases of scattered waves, so that different propagation paths become dominant.
The fluctuations are much larger than one would obtain classically by modeling
the system by a random resistor array, in which interference effects are
neglected. A classical approach is valid only on a length scale
 exceeding the phase
coherence length, where the fluctuations reduce to their classical value.

Unfortunately, despite some advantages of optical systems, the analog
of the UCF has not yet been observed in optical systems. Such experiments turn
out to be difficult. Although the magnitude of the fluctuations is universal,
they occur on a background of order $g$, where $g$ is the dimensionless
conductance (in optical experiments one typically has $g\sim 10^3$). The
relative value of the fluctuations to the background is thus $1/g$, so that the
$C_3$ correlation function is of order $1/g^2$, typically of order $10^{-6}$.
For electrons this problem is absent as smaller values of $g$ are achievable.
In the electronic case there have been many observations of the universal
conductance fluctuations.

A promising candidate for measuring $C_3$ is the microwave experiment of
\citeasnoun{stoytchev}. In such a set up one always has good control over the
channels. Moreover, one is able to study materials with large disorder, i.e.
down to $g=3$.

In principle the $C_2$ and $C_3$ correlations are present as a sub-leading term
in the angular correlation function, but there are also other contributions
of order $1/g$ and $1/g^2$, respectively, to the correlation
Eq.(\ref{eqdefcb}). The weak localization correction, see
Sec.~\ref{secweakloc},
is an example of a $1/g$ contribution to $C_1$. We have drawn a
 $1/g^2$ contribution to
$C_2$ in Fig.~\ref{figc2g2}. Such corrections are studied by 
\citeasnoun{garcia}.
Therefore, the interpretation of the $C$'s as an expansion in $1/g$ is
misleading. Rather we define the $C_1$, $C_2$, and $C_3$ as the
leading term in the correlation of $T_{ab}$, $T_a$, and $T$, respectively.
Equivalently, $C_1$ contains disconnected diagrams; incoming and outgoing
amplitudes automatically have the same pairing. 
The $C_2$ term contains connected
diagrams swapping the initial pairing, and $C_3$ contains all connected
diagrams in which incoming and outgoing pairings are identical.

\section{Calculation of correlation functions}

\subsection{Summary of diffuse intensities}
\label{rmpintro}

For the calculation of interference effects it is often too complicated to
take the precise behavior near the boundaries into account. We therefore 
approximate the diffuse intensity with tis bulk behavior, yet we will use the
prefactors and angular dependence, i.e. especially $\tau_0$ and
$\tau_1$, found with the radiative transfer approach.

The diffuse intensity created from a beam impinging in direction $a$ is
\begin{equation} {\cal L}^a_{\rm in}( z)= \frac{4\pi
\tau_1(\mu_a)T(\mu_a)} {k\ell A\mu_a}\,\frac{L-z}{L}. \label{eqlin}
\end{equation}
For uniform illumination of the sample the source term in the transport
equation becomes 
\begin{equation} S(z)= \sum_a S_a = A \int_0^k \frac{d\qt_a}{2\pi} \qtl_a S_a= 
\frac{2k}{\ell} \int_0^1 d\mu T(\mu) \eexp{-z/\ell \mu}. \end{equation}
Using this source term in the Schwarzschild-Milne equation gives the diffuse
intensity for a uniform illuminated sample
\begin{equation} \L_{\rm in} (z) = \frac{4k}{\ell} \frac{L+z_0-z}{L+2z_0}.
\label{eqdifin} \end{equation}
Defining 
\begin{equation}\epsilon_a=\frac{\pi T(\mu_a)\tau_1(\mu_a)}{k A\mu_a},
\label{eqdefepsilona}
\end{equation}
and comparing the pre-factors, one has \begin{equation}
\label{tfaa} {\cal L}_{\rm in}^a(z)=\epsilon_a {\cal L}_{\rm in}(z).
\end{equation}
Finally, for diffuse propagation between two points in the slab one has
\begin{equation}
-\frac{d^2} {d z^2} {\cal L}_{\rm int}(z,z';M)+M^2 {\cal L}_{\rm
int}(z,z';M)=\frac{12 \pi}{\ell^3} \delta(z-z') \label{difeq}
\label{eqdifeq}, \end{equation} with the solution
\begin{equation}\label{Lint2} {\cal L}_{\rm int}(z,z';M)=\frac{12
\pi}{\ell^3} \frac{\sinh M z_< \; \, \sinh M(L-z_>)}{M\sinh M L} .
\end{equation}
For angular resolved measurements the outgoing intensity in a certain 
direction $b$ is, in analogy with Eq.~(\ref{tfaa}), \begin{equation}
 \label{tfbb}
{\cal L}_{\rm out}^b(z)=\epsilon_b {\cal L}_{\rm out}(z) =\epsilon_b 
\frac{k}{\ell}\frac{z}{L}. \label{eqdifout} \end{equation}

The {\em angular transmission coefficient} reads
\BA \langle T_{ab} \rangle &=& = \frac{\pi
\tau_1(\mu_a) \tau_1(\mu_b) T(\mu_a) T(\mu_b) \ell}{3\mu_a\mu_b A (L+2z_0) }
= \epsilon_a\epsilon_b \expect{T} 
,\label{eqtabav} \\ 
\EA
The {\em total transmission coefficient} is
\BA
\langle T_a \rangle &=& \frac{\tau_1(\mu_a) 
T(\mu_a)\ell}{3 \mu_a(L+2z_0)} \label{eqtaav}.\EA
The conductance reads, cf. Eq.~(\ref{eqGextra})
\BA 
g\equiv \langle T\rangle &=& \int \d^3 {\bf r} \L_{\rm out}(z) S (z) 
\nonumber \\
&= &\frac{2 k A}{2\pi \ell} \int_0^\infty L_{\rm out}(z) \d z \int_0^1 \d\mu 
T(\mu) \eexp{-z/\mu \ell} \nonumber \\
&=&\frac{k^2
A\ell}{3 \pi(L+2z_0)}= \frac{4N\ell}{3(L+2z_0)}. \label{gtt} \EA

\subsection{Calculation of the $C_1$ correlation}

The calculation of $C_1$ is rather simple as it is just the product of 
two independent diffusons with exchanged partners. Both
diffusons consist of one amplitude from one beam and one complex
conjugated amplitude from the other beam. Due to momentum conservation
in each diffuson, the perpendicular momentum difference of the two
incoming beams, $\qt$, is equal to the difference of the outgoing
momenta. Therefore, the $C_1$ is only nonzero if incoming and outgoing
angles are shifted equally during the experiment.

We calculate the normalized product of a diffuson with ``squared mass''
$M^2=\qtl^2+\kappa^{2}+i\Omegamark$ and the complex conjugated diffuson.
This yields: \BA C_1(M)&=& \left|
\frac{T_{ab}(M)}{T_{ab} (\kappa) }\right|^2 = \left|
\frac{\L^{ab}(M)}{\L^{ab}( \kappa)} \right|^2 \nonumber
\\ &=&\frac{|M |^2}{\kappa^2} \left| \frac{(1+\kappa^2 z_0^2)\sinh \kappa
L+2\kappa z_0\cosh \kappa L} {(1+M^2 z_0^2)\sinh M
L+2M z_0 \cosh ML} \right|^2 \; . \EA
In Fig.~\ref{figc1} this function is drawn for various situations. In thick
samples where $M z_0 \ll 1$ and no absorption is present,
 the result reduces to ~\cite{feng} 
\begin{equation}
C_1(M) = \left| \frac{ML}{ \sinh M L} \right|^2 \; . \end{equation}
 For large angle or
frequency difference ($|M|L \gg 1$), $C_1$ decays exponentially,
 as $C_1(M)\sim
\exp(-2{\rm Re}(M)L)$. The spatial correlations of beams can be extracted from
the Fourier transform of the angular correlation function.

\subsection{Calculation of the $C_2$ correlation}\label{secc2}

The diagram of the long range correlation $C_2$ is depicted in 
Fig.~\ref{figc2diagram}. Remember that $C_1$ is zero if incoming and 
outgoing angles are changed unequally, but $C_2$ is connected, allowing
momentum flow from one diffuson to the other, therefore it is nonzero if
one changes incoming and outgoing angles by a different amount.
$C_2$ stems from the interaction between two diffusons, which exchange
partners somewhere inside the slab. The 
Hikami box describes this exchange. (For an estimate of 
$C_1$ correlations a $C_2$-type experiment see \citeasnoun{t3pre}.) We have
\BA C_2 & =&\frac{1}{\langle T_a\rangle \langle T_c\rangle }
 \int_0^L \d z
 \L^a_{\rm in}(z,M_1) \L^c_{\rm in}(z,M_3) H_4(z) \L_{\rm out}(z,M_2)
\L_{\rm out}(z,M_4) \; , \label{hikc2}
\EA
We have labeled the incoming diffusons
with odd numbers and the outgoing diffusons with even numbers, a
convention used throughout this work.
Attaching diffusons to the Hikami box allows a simplification of
the expression for the box. 
We calculate $C_2$ by inserting $ M_{2,4}^2=\qtl^2+\kappa^2 \pm i 
\Omegamark$,
 $M_1=M_3=\kappa$.
In the $z-$coordinate representation
the box reads, see Sec.~\ref{chhik},
\begin{equation} H_4(z_1,z_2,z_3,z_4)= h_4 (\partial_{z_1}
\partial_{z_3} +\partial_{z_2} \partial_{z_4}+\qtl^2 )
 \; . \label{hik2} \end{equation}
The above calculation for $C_2$
is similar to the one obtained in the Langevin approach, as one can see by
comparing Eq.~(39) of \cite{pnini2} to Eqs.~(\ref{hik2}) and
(\ref{hikc2}) in this section. In the Langevin approach one assumes
that there is a macroscopic intensity, which describes the average diffusion,
and an uncorrelated random noise current superposed. In our approach we have
a slowly varying diffuse intensity, while the point-like, uncorrelated
random interactions originate from the Hikami box.

We discuss some simplified cases. 
Consider first the case of angular correlations and no absorption
but taking the boundary effects into account:
\begin{equation}
C_2(\qtl)= \frac{3\pi L}{k^2 \ell A} F_2(\qtl L) \label{f2c2} \; ,
\end{equation}
in which we define the dimensionless function $F_2$ 
\BA
F_2(\qtl L)&=& [ \sinh
2\qtl L-2\qtl L+\qtl z_0 \; (6\sinh^2\qtl L-2\qtl^2L^2) \nonumber \\
&&+4 \qtl^2z_0^2 \; (\sinh 2\qtl L-\qtl L) +6\qtl^3 z_0^3 \; \sinh^2 \qtl L
+\qtl^4z_0^4 \; \sinh 2\qtl L ] \times \nonumber \\
&& [2\qtl L \left\{(1+\qtl^2 z_0^2)\sinh \qtl L+2\qtl z_0 \cosh \qtl 
L\right\}^{2}]^{-1}
 \; .\EA
which decays like $1/\qtl $ for large $\qtl$ ({\it i.e.} for large angles). 
However, it still should hold that $\qtl \ell \ll 1$,
 else we probe essentially 
interference processes within one mean free path from the surface, which is
beyond this approach.
Neglecting boundary effects, $z_0 \ll L$, one recovers 
 \cite{pnini1}
\begin{equation}
F_2(\qtl L)=\frac{\sinh(2\qtl L)-2\qtl L}{2\qtl L \sinh^2
(\qtl L)} \label{c2dq} \; , \end{equation}
with $F_2(0)=\frac{2}{3}$.
In Fig.~\ref{figc2} we have plotted the 
$F_2$ function for various situations.

We have studied the correlation as function of the two-dimensional momentum
$\qtl$. Similarly, we can study the real space correlations. 
The functional form in real
space is of roughly similar shape as \fig{figc2}, with typical decay length
$L$ in the $x,y$ direction \cite{stephen1}.
One can also obtain frequency correlations. Without absorption,
neglecting boundary effects, one finds \cite{deboer} \BA F_2 (\Omegamark) &=&
\frac{2}{\sqrt{2\Omegamark} \, L} \left(
\frac{\sinh(\sqrt{2\Omegamark}\,L)-\sin(\sqrt{2\Omegamark}\,L)}
{\cosh(\sqrt{2\Omegamark}\,L)-\cos(\sqrt{2\Omegamark}\,L)} \right) 
 \label{c2dw} \; . \EA

Of special interest is the top of the
correlation function. By definition the top of the correlation function
corresponds to the second cumulant, or second central moment,
of the total transmission distribution
function. With a plane wave as incoming beam, all transverse momenta
are absent, we also neglect absorption. 
These assumptions correspond to all $M$
being equal zero in Eq.~(\ref{hikc2}).
The diffusons are simple linear functions, given
by Eq.~(\ref{eqlin}) and (\ref{eqdifout}). Neglecting internal
reflections, one obtains for the second cumulant \cite{stephen1}
\begin{eqnarray} \langle \langle T_a^2 \rangle
\rangle&=&\frac{\expect{T^2_a} -\expect{T_a}^2}{\expect{T_a}^2}
\nonumber \\ &=& \frac{1}{\langle T_a
\rangle^2} \int \int {\rm d}x {\rm d}y \int_0^L {\rm d}z H_4 {\cal L}^a_{\rm
in}(z)eqlin
{\cal L}_{\rm out}(z) {\cal L}_{\rm in}^a(z) {\cal L}_{\rm out}(z)\nonumber 
\\&=&\frac{1}{gL^3}
\int_0^L {\rm d}z [z^2+ (L-z)^2]= \frac{2}{3g}, \end{eqnarray}
where double brackets denote cumulants normalized to the average. 
The correlation is thus proportional to $1/g$. 
The inverse power of $g$ counts the
number of four-point vertices, this power
 can be looked upon as the chance that two
intensities interfere. 

One obtains a general expression for $C_2$ by inserting the
appropriate diffusons and the Hikami box \eq{hik2} into \eq{hikc2}. The result
is rather lengthy and was given in \citeasnoun{skin}. 

\subsubsection{Influence of incoming beam profile}
We now study the influence of the beam profile on the correlation. 
In experiments the beam is often focused to a small spot 
 in order to
minimize the dimensionless conductance $g$ and therefore to maximize the
correlations. Physically it is clear that
the correlations increase if the two
incoming channels are closer to each other, i.e., if the beam diameter is
smaller. If the
spot of the incoming beam is finite, amplitudes with different angles,
i.e. different transverse
momenta, are present. They can combine into incoming diffusons with
perpendicular momentum. We suppose that the incoming beam has a Gaussian
profile. We decompose it into plane waves defined in Eq.~(\ref{eqpsiin})
(for convenience we assume the average incidence perpendicular) 
\begin{equation}
\psi_{\rm in}=\frac{2\pi}{W} \sum_a \phi({ \qtl}_a) \psi_{\rm in}^a
,\qquad \phi({ \qtl}) =\frac{\rho_0}{\sqrt{2\pi} }{\rm
e}^{-{ \qtl}^2 \rho_0^2/4} \label{eqpsig},\end{equation} where
$\rho_0$ is the beam diameter. In order to have two diffusons with a momentum
${ \qtl}$ and $-{ \qtl}$, we find that the four incoming
amplitudes combine to a weight function \begin{equation} \int {\rm
d}^2{ P}_1 {\rm d}^2{ P}_3 \; \phi({
P}_{1}) \phi^\ast({ P}_1+{ \qtl}) \; \phi({
P}_{3}) \phi^\ast({ P}_3-{ \qtl}) ={\rm e}^{-\rho_0^2
{ \qtl}^2 /4}. \end{equation} 
We get the correlation function by integrating the $\qtl$ dependent
correlation function over the
momentum with the corresponding weight. 
Neglecting boundary layers we find
for the top of the correlation \cite{deboer} \begin{equation} \langle \langle
T_a^2 \rangle \rangle= \frac{\rho_0^2}{4\pi g} \int {\rm d}^2{ \qtl}
\; {\rm e}^{-\rho_0^2 { \qtl}^2 /4} F_2(\qtl L)
\label{c2} ,\end{equation} with $F_2$ as in Eq.~(\ref{c2dq}). If
the incoming beam is very broad, $\rho_0 \gg L$, only the term
$F_2(\qtl=0)=2/3$ contributes and one recovers the plane wave behavior
$\langle \langle T_a^2 \rangle \rangle= 2/3g $.
The second cumulant decreases as $1/\rho_0^2$ at large $\rho_0$.
 At smaller beam diameters the top of the correlation is
proportional to $1/\rho_0$, and diverge in this
approximation, in which $\qtl
\ell \sim \ell /\rho_0 \ll 1 $.

\subsubsection{Reflection correlations} 

Similar to transmission correlations,
reflection correlations were studied. We have seen that the transmission
corresponds to the diffusons, but in reflection there are two intensity
diagrams: the diffusons bring a constant background,
 the maximally crossed diagrams
yield the enhanced backscatter cone, which is only of importance if incoming
and outgoing angles are the same. This leads to more diagrams and to more peaks
in the reflection correlation than in the transmission correlation. For the
$C_1$ one can pair diffusons and maximally crossed diagrams
\cite{berkovits4}. Indeed in optical experiments two
peaks were seen by \citeasnoun{freund3}.
 The $C_2$ term in reflection was treated
by \citeasnoun {berkovits5}. Measurement of $C_2$ in reflection by
collecting all reflected light is tricky as it is hard not to interfere with
the incoming beam. Nevertheless, measurement may be possible as a long
range component in the angular resolved reflection
 (this however requires quite
small values of $g$). For both $C_1$ and $C_2$ in reflection one does not
expect a very good agreement with theory, the problem is that the precise
behavior of the diffuson near the surface is important. The diffusion
approximation $q \ell \ll 1$ is not valid anymore, in particular if the indices
of refraction match (then the intensity gradient near the surface is the
largest). The calculation of $C_1$ is probably easily extended, yet for
$C_2$ one also would need the Hikami box beyond first order in $q\ell$.

\subsection{Conductance Fluctuations: $C_3$}\label{chc3}

One could be tempted to think
 that the calculation of the $C_3$ or UCF in the above Landauer
approach is also straightforward. However, the calculation in the Landauer
approach is quite cumbersome, as on short length scales 
(of the order of the mean free path)
 divergences show up when one treats the problem on a macroscopic level
using diffusons. As an alternative, the Kubo approach is often used in
mesoscopic electronic systems \cite{altshuler2,lee,lee2}. The results for the
conductance obtained by either Kubo or Landauer formalism should be identical,
see \citeasnoun{fisher} and \citeasnoun{janssen}. Yet the Kubo approach
cannot be applied directly to optical systems when absorption is to be
included. Studies of $C_3$ in the Landauer approach were undertaken by
\citeasnoun{kane}, \citeasnoun{berkovits3}, and by the present
authors \cite{c3pre} where 
it is shown explicitly that all divergences cancel.

The $C_3$ correlation function, defined in Eq.~(\ref{eqdefcb}), involves
incoming
diffusons ${\cal L}_{\rm in}^{a,c}$, and outgoing ones ${\cal L}_{\rm
out}^{b,d}$. Due to the factorization of the external
direction dependence, see Eqs.~(\ref{tfaa}), (\ref{tfbb}), and
(\ref{gtt}), this dependence on external parameters 
 cancels from $C_3$. We can write
\begin{equation}\label{C3-F}
C_3^{abcd}(\kappa,\Omegamark)=C_3(\kappa,\Omegamark)
=\frac{1}{\langle T\rangle^2}\sum_\qt F(\qtl,\kappa,\Omegamark) .
\end{equation}
where $\qt$ is the $(d-1)$-dimensional transverse momentum. The function
$F$ is the main object to be determined in this section.
We thus calculate it at fixed $\qtl$ and with external diffusons for diffuse
illumination $\L_{\rm in}$ and $\L_{\rm out}$,
Eqs.~(\ref{eqdifin}) and (\ref{eqdifout}). Depending on dimension we have
\begin{eqnarray}
\sum_\qt F(\qtl,\kappa,\Omegamark) 
&=& F(0,\kappa,\Omegamark) \qquad \qquad \qquad \quad \mbox{quasi 1-d}, \\
&=& W \int \frac{{\rm d}
\qt }{2\pi} F(\qtl,\kappa,\Omegamark)\qquad\mbox{quasi 2-d},
\\ &=& W^2 \int \frac{{\rm d}^2 \qt}{(2\pi)^2}
F(\qtl,\kappa,\Omegamark)\qquad\mbox{3-d}. \label{eqt2} \end{eqnarray}
In contrast to $C_1$ and $C_2$, 
as a result of the $\qtl$ integral, $C_3$ will depend on the 
dimensionality of the system. 

Let us work out the long range diagrams of $C_3$, roughly depicted in 
\fig{figc123}, in more detail. For all diagrams there are two incoming advanced
fields, which we momentarily term $i$ and $j$, and two retarded ones, $i^\ast$
and $j^\ast$. Consider an optical experiment with integrating spheres at
incoming and outgoing sides. The incoming diffusons cannot have a momentum or
frequency difference, and the pairing must be $ii^\ast$ and $jj^\ast$. In
Fig.~\ref{figreg}(a) the incoming diffusons interfere somewhere in the slab.
In a diagrammatic language the diffusons interchange a propagator so that the
pairing changes into $ij^\ast$ and $ji^\ast$. Propagation continues with these
diffusons, which, due to the different pairings can have nonzero frequency
difference and nonzero momentum.
Only the outgoing diffusons without
a momentum or frequency difference are dominant. Therefore, somewhere else
in the slab a second interference occurs. Again exchanging an amplitude, the
original pairing, $ii^\ast$ and $jj^\ast$, is restored and the two diffusons
propagate out, see Fig.~\ref{figreg}(a)i. Other contributions occur as
well, yet the incoming and outgoing pairings are always $ii^\ast$ and
$jj^\ast$. In Fig.~\ref{figreg}(a)ii the first incoming diffuson meets an
outgoing diffuson and exchanges amplitudes.
 These internal diffusion lines meet
at a second point where the original pairings are restored. Clearly in this
process the intermediate paths are traversed in time reversed order. Due to
time-reversal symmetry they give a contribution similar to
Fig.~\ref{figreg}(a)i. In Fig.~\ref{figreg}(b)i a diffuson breaks up such that
one of its amplitudes makes a large detour, returns to the breaking point and
recombines into an outgoing diffuson. The second incoming diffuson crosses the
long path of the amplitude, and one of its amplitudes follows the same
contour as the first diffuson. 
Finally, Fig.~\ref{figreg}(c) depicts the situation where only one internal
diffuson occurs. Its endpoints must lie within a distance of a few mean free
paths. Because of the local character this class does not show up in the final
result; we need it, however, since it contains terms that cancel divergences
from the other two classes.

\subsection{Calculation of the $C_3$ correlation function}\label{secdifdiv}

The diagrams for the conductance fluctuations contain a loop; the two internal
diffusons have a free momentum, over which one has to integrate. In
Fig.~\ref{figreg}(a) we denote this momentum ${\bf q}$. Physically, one
expects important contributions to the conductance fluctuations if the distance
between the two interference vertices ranges from the mean free path to the
sample size. Yet the ${\bf q}-$integral
diverges for large momentum, i.e. when the two interference processes
are close to each other. The standard picture of diffuse transport with
diffusons and interference described by Hikami vertices, that works so well for
loop-less diagrams such as the $C_2$ correlation function (section
\ref{chc12}), and the third cumulant of the total transmission (section
\ref{cht3}), becomes spoiled by these divergences.

As an example we calculate the diagram presented in Fig.~\ref{figreg}(a)i. This
diagram was first depicted by \citeasnoun{feng} and considered
in detail by \citeasnoun{berkovits3}. 
These authors pointed out that a
short distance divergency appears. They found a cubic divergency in three
dimensions, and in general, a $d$-dimensional divergency in $d$ dimensions. Let
us see how it appears. For simplicity we 
consider a quasi one-dimensional system for which frequency
differences and absorption are
absent, therefore the decay rate vanishes, i.e. $M=0$ for all diffusons. 
From Fig.~\ref{figreg}(a)i one directly reads off its corresponding
expression $F_{a.i}$ \begin{equation}
F_{a.i} = \int \int {\rm d}z {\rm d} z' {\cal L}_{\rm in}^2(z) H_4(z)
 {\cal L}_{\rm int}^2(z,z') H_4 (z') {\cal L}_{\rm out}^2(z). \end{equation}
After performing some partial integrations and using the diffusion
equation Eq.(\ref{eqdifeq}) one finds
\begin{eqnarray} F_{a.i} &=& \frac{\ell^4}{4 k^4}\delta(0)
\int {\rm d}z {\cal L}_{\rm in}^2(z) {\cal L}_{\rm
out}^2(z) \nonumber \\&& + \frac{h_4 \ell^2}{k^2}\int {\rm
d} z {\cal L}_{\rm int}(z,z)\left[{\cal L}_{\rm in}'^2(z)
 {\cal L}_{\rm out}^2(z)
+{\cal L}_{\rm in}^2(z) {\cal L}_{\rm out}'^2(z) 
\right. \nonumber \\ && \left. \qquad 
+ 2 {\cal L}_{\rm in}'(z) {\cal L}_{\rm out}'(z)
{\cal L}_{\rm in}(z) {\cal L}_{\rm out}(z) \right] \nonumber \\
&& + 4 h_4^2\int {\rm d} z' \int {\rm d} z {\cal L}_{\rm int}^2(z,z'){\cal
L}_{\rm in}'^2(z) {\cal L}_{\rm out}'^2(z'), \end{eqnarray}
were the spatial derivative of ${\cal L}(z)$ was denoted ${\cal L}'$.
The divergency is clearly present in the $\delta(0)$,
 it comes about as follows:
For the case of zero external momenta,
the Hikami box yields $H_4({\bf q},0,-{\bf q},0)=2h_4 q^2$,
while the internal diffuson
has the form ${\cal L}_{\rm int}(q)=12\pi/(\ell^3 q^2)$. Omitting the external
lines, the diagram leads to
\begin{equation} \int \frac{d q}{(2\pi)} H_4^2({\bf q},0,-{\bf
q},0) {\cal
L}_{\rm int}^2(q) =\frac{\ell^4}{4k^4} \int \frac{ {\rm d}^3q}{(2\pi)^3} q^0=
\frac{\ell^4}{4k^4}\delta (r=0). \end{equation}

We label the expressions according to the
diagrams in \fig{figreg}, $F_a$, $F_b$ and $F_c$ and sum them. For the 1D case
we find
\begin{equation} F_{a}+F_b+F_{c}= -\frac{2}{15}\delta(0)L+
\frac{2}{15}. \label{fabc} \end{equation}
The second term in Eq.(\ref{fabc}) is 
exactly equal to the well known result for the UCF in one
dimension. 
(Note that the pre-factors of the diffusons and the Hikami boxes have
canceled precisely. This is a manifestation of the universal character of
conductance fluctuations.) But a singular part is
annoyingly present. The term $\delta(0)$ is a linear divergency. In the
three-dimensional case one has to integrate over the transverse momenta 
yielding a cubic divergency.

The cancellation of this divergence is involved and requires a number of
special short range diagrams which resist a general classification or
treatment \cite{c3pre}.
Here we only use that finally the special short range diagrams cancel the
 divergences, yet they give no long range contribution. 
The nondivergent part of the diagrams of Fig.~\ref{figreg} is 
the only remainder. It reads
\begin{eqnarray} F(\qtl,\kappa,\Omegamark)\!&\!=\!&\!4 h_4^2\int \!\int
 \! {\rm d}z \,{\rm d}z' \,
{\cal L}_{\rm int}(z,z';M) {\cal L}_{\rm int}(z,z';M^\ast)\, 
\nonumber \\ &&\qquad
\times \left[ {\cal L}_{\rm in}'^2(z){\cal L}_{\rm out}'^2(z')+
{\cal L}_{\rm in}'(z) {\cal L}_{\rm out} '(z) {\cal L}_{\rm in}'(z')
{\cal L}_{\rm out} '(z') \right]\nonumber\\
\!&\!+\!&\!\frac{h_4^2}{2} \!\int \! \int \! {\rm d}z \,{\rm d}z'
\left[{\cal L}_{\rm int}^2(z,z';M)\!+\!{\cal L}_{\rm int}^2(z,z';M^\ast)\right]
\frac{d^2}{dz^2}
\left[ {\cal L}_{\rm in}(z){\cal L}_{\rm out}(z)\right] \nonumber \\
\!&\!\!&\!\qquad \times \frac{d^2}{dz'^2}
 \left[{\cal L}_{\rm in}(z') {\cal L}_{\rm
out}(z') \right]
\label{eqF}.\end{eqnarray}
in which again $M^2=\qtl^2+\kappa^2+i\Omegamark$.
The upper line of
(\ref{eqF}) corresponds to the diagrams of Fig.~\ref{figreg}(a), whereas the
lower line corresponds to the diagrams of Fig.~\ref{figreg}(b). 
Finally, with Eq.~(\ref{eqt2}) the value at vanishing
transverse momentum gives the variance of the conductance in one dimension,
whereas integration over the transverse momentum yields the 
correlation in two and three dimensions.

Using Eqs.~(\ref{eqF}) and (\ref{eqt2}) and inserting
inserting the appropriate diffusons, various cases can be studied. 
Consider the case of fully transmitting surfaces; 
neglecting absorption and frequency differences one finds
\begin{equation} F(\qtl)=\frac{3}{2} \frac{2+2\qtl^2
L^2-2\cosh 2\qtl L+ \qtl L\sinh 2\qtl L}{\qtl^4 L^4 \sinh^2 \qtl L} ,
\end{equation}
which decays for large momenta as $\qtl^{-3}$, thus converging in three or less
dimensions. The subsequent integration over the loop momentum finally yields
\begin{eqnarray} \langle
 T^2 \rangle_c & =&\frac{2}{15} \approx 0.133 ,\qquad \mbox{quasi 1D} \\
&=&\frac{3}{\pi^3} \zeta(3)\frac{A}{L}
 \approx 0.116 \frac{A}{L}
 , \qquad \mbox{quasi 2D} \\
&=&\frac{1}{2\pi} \frac{A}{L^2}\approx 0.159\frac{A}{L^2} \qquad
\mbox{3D} ,\end{eqnarray} in which $\zeta$ is Riemann's zeta function. These
values are for wide slabs.
For cubic samples \citeasnoun{lee2} find in
 2D that $\langle T^2 \rangle_c 0.186$ and in 3D a value of $0.296$
(the value in 1D is of course the same).
One can also determine the frequency dependency of the correlation using
(\ref{eqF}), which is of importance as it determines the frequency range of the
light needed to see the fluctuations~\cite{c3pre}.
By inserting the appropriate diffusons one can also study
the case of partial reflection at the surfaces of
the sample. It is physically clear that the internal reflections lead to a less
steep diffuse intensity in the sample as a function of the depth. The
fluctuations are proportional to the space derivatives and thus reduce. We
present the results in Fig.~\ref{figc3z03d}
 where we have plotted the correlation
function for various values of the ratio between extrapolation length and
sample thickness. One sees that the correlation is lower than
without internal reflections. We note that neither the variance (the value at
vanishing $\Omegamark$), nor the form of the correlations are fully universal.

\section{Third cumulant of the total transmission}\label{cht3}

So far we considered correlations between intensities defined as
 \begin{equation}
\frac{\expect{T_{ab} T_{cd}} } {\expect{T_{ab} }\expect{ T_{cd}}}=
1+C_1+C_2+C_3 . \end{equation} The $C_1(\qtl,\Omegamark)$,
 $C_2(\qtl,\Omegamark)$, 
and $C_3(\Omegamark)$
correlation functions describe correlations between two intensities. By
definition their value at zero frequency and zero momentum difference, i.e.
the peak value, is the second cumulant of the distribution functions of
the transmission quantities. An obvious question is how higher cumulants
behave. The size of the fluctuations and the shape of
the distribution relates to the ``distance'' from the localization transition.
Far from localization, diffusion channels are almost uncorrelated and
fluctuations are small (except the optical speckle pattern in the angular
resolved transmission). The correlation between the channels increases if 
localization is approached. The relevant parameter is
 the inverse dimensionless
conductance $1/g$, which can be interpreted as the chance that two channels
interfere. Close to Anderson localization $g$ approaches unity, and
fluctuations increase. 

\subsection{Cumulants of the probability distribution} \label{seccum}
We first study the third cumulant of the total transmission.
The total transmission is a constant superposed with fluctuations. To first
order in $g^{-1}$ the fluctuations have a Gaussian distribution
\cite{deboer,kogan}. The relative variance of this distribution, the top of 
$C_2$, is proportional to $g^{-1}$, it is thus a factor $g$ larger than for the
conductance fluctuations. This sensitivity of the total transmission to
interference processes and its simple limiting behavior (as compared to the
angular resolved transmission) make it an ideal quantity to study mesoscopic 
effects in its distribution. 
The third cumulant of the distribution was determined in
very precise experiments by \citeasnoun{t3prl}. In this section we
calculate its theoretical value. Using the diagrammatic
approach we relate the third cumulant normalized to the average total
transmission, $\langle
\langle T_a^3 \rangle \rangle$, to the normalized second cumulant $\langle
\langle T_a^2 \rangle \rangle$.

The moments of the probability distribution can be extracted from the
distribution $P(T_a)$ as \begin{equation} \langle T_a^k \rangle =\int {\rm
d}T_a \, P(T_a) T_a^k. \end{equation} In a diagrammatic approach the $k$-th
moment can be represented by a diagram with $k$ diffusons on both incoming and
outgoing sides. The $k=1$ term is the average total transmission $\langle T_a
\rangle$, as given by the Schwarzschild-Milne equation in Eq.(\ref{eqtaav}),
it corresponds to a single diffuson and is thus independent of
channel-to-channel correlations. The second moment can be decomposed in the
first two cumulants: \begin{equation} \frac{ \langle T_a^2 \rangle }{\langle
T_a \rangle^2} =\frac{\langle T_a\rangle^2}{\langle T_a \rangle^2} +\frac{
\langle T_a^2\rangle_{\rm cum}
}{\langle T_a \rangle^2}=1 +\langle \langle T_a^2 \rangle \rangle ,
\end{equation} where
the double brackets denote cumulants normalized to the average.
Diagrammatically we depict the second moment in Fig.\ \ref{figt2}. The
decomposition in cumulants proves useful as each cumulant corresponds to a
different number of interactions between the diffusons. The first term,
Fig.\ \ref{figt2}(a) contains no interference; it factorizes in the average
transmission squared.
 The second term, Fig.\ \ref{figt2}(b), is the second cumulant
$\langle \langle T_a^2 \rangle \rangle$. It gives the variance of the
fluctuations. This is just a special case of the $C_2$ correlation function,
 $ \langle \langle T_a^2 \rangle \rangle=C_2(0) $.

Likewise, one can distinguish three different
contributions to the third moment, \begin{equation} \frac{ \langle
T_a^3 \rangle}{\langle T_a \rangle^3} =1+3 \langle \langle T_a^2 \rangle
\rangle + \langle \langle T_a^3 \rangle \rangle. \label{t3=} \end{equation} 
We have drawn the
corresponding leading diagrams in Fig.\ \ref{figt3}. The first
term, Fig.\ \ref{figt3}(a), again corresponds to the transmission without
interference. The second term, Fig.\ \ref{figt3}(b), is the product of a single
diffuson and a second cumulant diagram. From the figure it is clear that this
decomposition can be done in three ways which determines the combinatorical
prefactor
of $\langle \langle T_a^2 \rangle \rangle $ in Eq.~(\ref{t3=}). The third
contribution stands for the third cumulant of the distribution and expresses
the leading deviation from the Gaussian distribution. This is the term we are
mainly interested in. It consists of two related diagrams: Fig.\
\ref{figt3}(c+d). The three intensities can interfere twice two by two, or
the intensities can interact all three together, with a Hikami
six-point vertex. Both contributions will prove to be of the same order
of magnitude. The strength of the effect can be easily estimated using the
interpretation of $1/g$ as an interaction probability. By looking at the
diagram, the third cumulant is proportional to the chance of two diffusons
meeting twice, thus of the order $1/g^2$. We have thus found 
the estimate
\begin{equation} \langle \langle T_a^3 \rangle \rangle \propto \langle
\langle T_a^2 \rangle \rangle^2. \end{equation} 

Finally, we note another that there is another contribution to both second and 
third cumulant. Because there is only a finite number of channels in an
experiment, the intensity distribution will always have non-zero width. We have
verified that this effect brings a negligible contribution to the measured
second cumulant and the third cumulant \cite{t3pre}.

\subsection{The calculation of the third cumulant}\label{sect3}

Two processes contribute to the third cumulant. One with two four point
vertices, that we term $ \langle \langle T_a^3 \rangle \rangle_c$ and one with
a six point vertex $ \langle \langle T_a^3 \rangle \rangle_d$, where we have
chosen the subscripts according to Fig.~\ref{figt3}.

\paragraph*{Interference via two four point vertices}

First consider the diagram in Fig.~\ref{figt3}(c). We have
labeled to incoming diffusons with odd numbers, the outgoing ones with
even numbers. Two incoming
diffusons, ${\cal L}^a_1$ and ${\cal L}^a_3$,
 meet at a position $z$, in a Hikami
box they interfere into ${\cal L}_2$
and an internal diffuson ${\cal L}^{\rm int}_{78}$. ${\cal L}_2$
propagates out, whereas ${\cal L}^{\rm int}_{78}$ interferes again at $z'$
with the incoming diffuson ${\cal L}_5$ into two outgoing ones,
 ${\cal L}_4$ and
${\cal L}_6$. Apart from this process, also three other sequences of
interference are possible. This means that the diffusons can also be permuted
as: $({\cal L}^a_1,{\cal L}^a_3, {\cal L}_5,{\cal L}_2,{\cal L}_4,{\cal L}_6)
\rightarrow ({\cal L}^a_3,{\cal L}_5, {\cal L}^a_1,{\cal L}_4,{\cal L}_6,{\cal
L}_2) \rightarrow ({\cal L}_5,{\cal L}^a_1,{\cal L}^a_3,{\cal L}_6,{\cal
L}_2,{\cal L}_4)$. We denote the sum over these permutations as
$\sum_{\rm per}$. As the diagrams can also be complex conjugated, there is
also a combinatorial factor 2 for all diagrams. (Note that
this is different from the $C_2$ calculation;
the {\em second} cumulant diagram is identical to its complex conjugate
and thus there is no such factor.)
The expression for the diagram of Fig.~\ref{figt3}(c) is now
\BA \langle \langle T_a^3\rangle
\rangle_{c}&=&\langle T_a\rangle^{-3} \, 2\sum_{\rm per} A \int_0^L {\rm d}z
\int_0^L {\rm d}z' \; H_4(z) H_4(z') {\cal L}^a_1(z) {\cal L}_2(z) {\cal
L}_3(z)\times \nonumber \\ &&
{\cal L}_4(z') {\cal L}_5(z') {\cal L}_6(z') {\cal L}_{78}^{\rm int}(z,z')
.\EA
In turns out that it is useful to rewrite the form of the Hikami boxes as
introduced in Eq.(\ref{haha}) into an equivalent expression, using momentum
conservation ${\bf q}_a+{\bf q}_{a'}+{\bf q}_{b}+{\bf q}_{b'}=0$. In real
space the use of momentum conservation corresponds to partial integration.
The Hikami box is again simplified using the fact that there are no 
transverse
momentum terms, or ${ \qtl}$ terms, for the outgoing diffusons. 
Using the numbering
in Fig.~\ref{figt3}, we obtain
\begin{equation}
H_4(z)=-h_4[ 2\partial_{z_1}\partial_{z_2}+ 2 \partial_{z_2}\partial_{z_3} ]
, \qquad H_4(z')=h_4[2\partial_{z_4}\partial_{z_6}-
\partial_{z_8}^2 + { \qtl}_8^2]. \label{eqh4keus} \end{equation}
Source terms, i.e., $q_i^2-$terms of the incoming and outgoing diffusons were
again neglected, but the source term of the
diffuson between the vertices is important. As one sees with the diffusion
equation (\ref{eqdifeq}), it brings \begin{equation} 
(-\partial_{z_8}^2 +{ \qtl}_8^2 ){\cal L}^{\rm int}_{78}(z,z') 
=\frac{\ell^3}{12\pi}
\delta(z-z').\end{equation} The contribution
from the
source term, i.e. $H_4(z')\propto -\partial_{z_8}^2 +
{ \qtl}_8^2$,
$H_4(z) \propto \partial_{z_1}\partial_{z_2}+\partial_{z_2}\partial_{z_3}$,
 is \begin{equation}
\frac{-h_4 \ell^2 A}{k^2}\! \int_0^L\!\! {\rm d}z [\partial_{z_1}
\partial_{z_2}\!+\!\partial_{z_2}\partial_{z_3} \!+\ldots + \partial_{z_6} 
\partial_{z_1}] {\cal L}^a_1 \ldots 
{\cal L}_6 \label{h4bron}. \end{equation}
Although this corresponds to a local process
(just one $z$ coordinate is involved), it is of leading order.
Together with the expression coming from $H_4(z')$, proportional to
$\partial_{z_4}\partial_{z_6}$,
we find for the total contribution of the process in
Fig.\ref{figt3}(c)
\begin{eqnarray} \langle \langle T_a^3 \rangle \rangle_{c} \!&\!=\!&\!
\frac{8 h_4^2}{\langle T_a \rangle^3} \sum_{\rm per} A\int_0^L {\rm d}z\;
{\cal L}^a_1(z){\cal L}_2'(z){\cal L}_3(z) \times \nonumber \\ && \qquad 
\int_0^L {\rm d}z'\; {\cal L}_4'(z'){\cal L}_5(z'){\cal L}_6'(z')
\partial_z{\cal L}^{\rm int} (z,z') \nonumber \\
&& -\frac{\ell^7 A}{48\pi k^4 \langle T_a \rangle^3} \int_0^L
\! \!{\rm d}z
[\partial_{z_1} \partial_{z_2}\!+\!\partial_{z_2}\partial_{z_3} \!+\!
\partial_{z_3}
\partial_{z_4}\!+\!\partial_{z_4} \partial_{z_5}\!+\!\partial_{z_5}
\partial_{z_6}\!+\!\partial_{z_6} \partial_{z_1}] \nonumber \\ && 
\times {\cal
L}_1(z) {\cal L}_2(z)
 {\cal L}^a_3(z) {\cal L}_4(z) {\cal L}_5(z) {\cal L}_6(z) \label{eqt3hh},
\end{eqnarray}
where ${\cal L}'(z)$ denotes the $z$ derivative of ${\cal L}(z)$.
Calculated for a plane wave, one finds
\begin{equation} \langle \langle T_a^3 
\rangle
\rangle_{c}=\frac{28}{15g^2},\end{equation} which is indeed proportional to
$g^{-2}$, as predicted.

\paragraph*{Contribution of the six-point vertex} The other diagram
contributing to the third cumulant is depicted in Fig.~\ref{figt3}(d).
 The hexagon
is again the Hikami six point vertex $H_6$. It can be thought of
in the following way: the use of the Hikami box assumes
that the outgoing legs scatter at least once before they propagate out or
interfere again. This is a reasonable assumption for the outgoing diffusons,
but for the internal diffuson ${\cal L}^{\rm int}_{78}$
 it is also possible that
coming from $z$ it directly, i.e., without scattering, 
interferes again at $z'$.
This process was not yet included in the calculation of the previous subsection
but has to be studied separately. The unscattered intensity decays
exponentially over one mean free path, therefore this process is only important
if $z$ and $z'$ are within one mean free path. As the diagrams
can also be complex conjugated, there is also a factor 2 for all diagrams.

After a Fourier transformation in the $z-$direction the six-point vertex
yields a contribution to the third cumulant 
\BA
\langle\langle{T_a^3} \rangle\rangle_{d}\!&=&\! 
\frac{h_4 \ell^2 A}{k^2 \langle T_a
\rangle^3} \int_0^L \!{\rm d}z [\partial_{z_1}
\partial_{z_2}\!+\!\partial_{z_2}\partial_{z_3} \!+\!\partial_{z_3}
\partial_{z_4}\!+\!\partial_{z_4} \partial_{z_5}\!+\!\partial_{z_5}
\partial_{z_6}\!+\!\partial_{z_6} \partial_{z_1}] \nonumber \\ &&
\times {\cal L}_{1}(z)
{\cal L}_{2}(z) {\cal L}_{3}(z) {\cal L}_{4}(z) {\cal L}_{5}(z) {\cal
L}_6(z) ,\EA 
were again the notation of Eq.~(\ref{mcwh4a}) is implied.
We used the fact that all outgoing diffusons
have zero transverse momentum. Therefore all ${ \qtl}_i
{ \qtl}_j$-terms are absent. In case of an incoming plane wave
we find a contribution to the third cumulant $\langle \langle
T_a^3 \rangle \rangle_{d}=-4/ (5g^2) $.
The contribution from the source term, i.e. Eq.(\ref{h4bron}) of the
previous subsection, exactly cancels the
contribution from the six-point vertex. The cancellation
 seems plausible as one does not
expect short distances properties to be important in the total process.
Nevertheless,
this cancellation depends on the precise form of the Hikami four-point
vertex in Eq.(\ref{eqh4keus}). If we use other equivalent forms of the
Hikami box
the contributions of the single and double integral in Eq.(\ref{eqt3hh}) 
shift with respect to each other and a
full cancellation is not present. Of course, neither the result for
Eq.(\ref{eqt3hh}) nor the final result for $\langle \langle T_a^3 \rangle
\rangle$
relies on this choice. 
One obtains for the third cumulant \begin{eqnarray} \langle \langle
T_a^3 \rangle
\rangle&=& \langle \langle T_a^3 \rangle \rangle_{c} +\langle \langle T_a^3
\rangle \rangle_{d}\nonumber \\ &=& 
\frac{8 h_4^2 A k^2}{\langle T_a \rangle^3 }
 \sum_{\rm per} \int_0^L {\rm d}z \;
{\cal L}^a_1(z){\cal L}_2'(z){\cal L}^a_3(z) \times \nonumber \\ &&
\int_0^L {\rm d}z'\; {\cal L}_4'(z'){\cal L}_5(z'){\cal L}_6'(z')
\partial_z{\cal L}^{\rm int}(z,z') \label{eqt3}
.\end{eqnarray} 

\subsection{Influence of incoming beam profile}
\label{secinf} Now that we know the leading interference processes, inserting
the diffusons gives the final value of the third cumulant. We first consider
the simple case of incoming plane waves. As there can be no transverse
momentum difference in the incoming amplitudes, all ${ \qtl}_i$
vanish. As a result
all diffusons are linear functions of $z$. We find from
Eq.~(\ref{eqt3})
\begin{equation} \langle \langle T_a^3 \rangle
\rangle=\frac{16}{15g^2}=\frac{12}{5} \langle \langle T_a^2 \rangle \rangle^2
\qquad \mbox{plane wave;} \; \; \rho_0 \gg L.\label{eqt3pw} 
\end{equation}

In practice, however, one often deals
 with a Gaussian beam with limited spot size,
influencing the cumulants in two ways. First, if the spot size decreases to
values comparable to the sample thickness we have to convolute over a range
of incoming momenta, just like we did when calculating the second cumulant.
Second, the Gaussian profile brings an extra geometrical factor as will be
 shown below. The expression for the diagrams with
 diffusons with arbitrary momentum is calculated as follows. 
 Because of momentum conservation and phase condition on
the outgoing diffusons, the transverse momentum ${ \qt}_7$ of
the diffuson connecting the two four-boxes must equal ${ \qt}_5$.
The integration over the possible momenta results again in a Gaussian weight
function $\exp (-\rho_0^2({ \qtl}_1^2+
{\qtl}_3^2+{ \qtl}_5^2)/8)$. The third cumulant is obtained
by inserting the momentum dependent diffusons into Eq.(\ref{eqt3}). This
gives \begin{equation}\langle \langle T_a^3 \rangle
\rangle=\frac{\rho_0^4}{16 \pi^2 g^2} \int {\rm d}^2{
\qt}_1 {\rm d}^2{ \qt}_3 \; \eexp{-\rho_0^2[{
\qtl}_1^2+{ \qtl}_3^2+({ \qtl}_1+{ \qtl}_3)^2]/8 } 
F_3({\qtl}_1L,\qtl_3 L, |\qt_1+\qt_3 | L) \label{k3}
,\end{equation} with \begin{eqnarray} F_3(x_1,x_3,x_5)& =&
\sum_{\rm per} [
\frac{(x_1\!+\!x_3)^2 x_5 \cosh(x_1\!+\!x_3) }{(x_1\!+\!x_3\!+\!x_5)^2 
(x_1\!+\!x_3-x_5)^2} -
\frac{(x_1-x_3)^2 x_5 \cosh(x_1-x_3) }{(x_1-x_3\!+\!x_5)^2(x_1-x_3-x_5)^2}
\nonumber \\ &&- \frac{(x_1\!+\!x_3) x_5 \sinh(x_1\!+\!x_3)
}{(x_1\!+\!x_3\!+\!x_5)(x_1\!+\!x_3-x_5)} + \frac{(x_1-x_3) x_5 \sinh(x_1-x_3)
}{(x_1-x_3\!+\!x_5)(x_1-x_3-x_5)} \nonumber \\ &&+\frac{(x_1\!+\!x_3)
\cosh(x_1\!+\!x_3\!+\!2x_5) }{4(x_1\!+\!x_3\!+\!x_5)^2} -\frac{(x_1\!+\!x_3)
\cosh(x_1\!+\!x_3-2x_5)
}{4(x_1\!+\!x_3-x_5)^2} \nonumber \\ &&-\frac{(x_1-x_3) \cosh(x_1-x_3\!+\!2x_5)
}{4(x_1-x_3\!+\!x_5)^2} +
\frac{(x_1-x_3) \cosh(x_1-x_3-2x_5) }{4(x_1-x_3-x_5)^2}
]\times \nonumber \\ &&\left[x_5 \sinh(x_1) \sinh(x_3)
\sinh^2(x_5)\right]^{-1}\label{eqdefF} .\end{eqnarray} 
We study again the behavior if the beam diameters are wide. In the limit of
large beam diameter ($\rho_0 \gg L$) one finds $F_3(0,0,0)=\frac{16}{15}$,
this means for the third cumulant $\langle \langle T_a^3 \rangle \rangle= 4
F_3(0,0,0)/3g^2$, or 
\begin{equation} \label{16/5}
\langle \langle T_a^3 \rangle \rangle =
\frac{16}{5} \; \langle \langle T_a^2 \rangle \rangle^2 \label{eqt3g},
\qquad\mbox{(Gaussian profile;} \; \rho_0\gg L) ,
\end{equation} 
which differs by a
factor $\frac{4}{3}$ from the plane wave limit Eq.(\ref{eqt3pw}). This is a
geometrical effect, depending on the profile of the incoming beam. In a real
space picture one understands this effect easiest: 
The correlation depends on the
distance, it is strongest if the incoming beams are close together.
 Therefore it
is not surprising to see the influence of their overlap. In next section we
calculate this geometrical factor also for higher cumulants. For the
experimental relevant case that the beam diameter is roughly equal to the
thickness, it turns out that the behavior of Eq.(\ref{eqt3g}) is found for a
large range of beam diameters. The increase of the correlation for smaller
beams turns out to be roughly the same for both the third cumulant and the
second cumulant squared.

In \citeasnoun{t3pre} theory was extensively compared with the
experiment reported
in \cite{t3prl}. Apart from the correction for the finite beam diameter, as
discussed above, two other experimental corrections were included: Internal
reflections and contributions from disconnected diagrams. All corrections to
(\ref{eqt3g}) turn out to be relatively small. By presenting the results as
the ratio between the second cumulant squared and the third cumulant, errors in
the sample thickness and the mean free path cancel. For the experimental data
it was found that $ \langle \langle T_a^3 \rangle \rangle=(3.3 \pm 0.6) \langle
\langle T_a^2 \rangle \rangle^2 $, in good agreement with Eq.~(\ref{16/5}).

~\citeasnoun{stoytchev} performed microwave scattering on systems
with values of $g$ approaching unity. In that case the plane wave
prediction applies. They found 
$ \langle \langle T_a^3 \rangle \rangle=(2.38 \pm 0.05) \langle
\langle T_a^2 \rangle \rangle^2 $, indeed in good agreement with Eq.
(\ref{eqt3pw}).

The extension of the calculation to higher cumulants is straightforward. The
$n$-th cumulant will contain $(n-1)$ Hikami four-point vertices. So the
contribution is $\langle \langle T_a^n \rangle \rangle \propto g^{1-n}$. But
it is clear that the calculation becomes laborious at large $n$. We follow
another approach to this problem in Sec.~\ref{chprob}.

\subsection{Third cumulant of the conductance}

The reader will not be surprised to learn 
that the third cumulant of the conductance
can also be calculated using a diagrammatic approach \cite{g3}.
 This third cumulant has the particular behavior that
it vanishes in one dimension\cite{macedo}. 
We already saw that in the mesoscopic regime the conductance
shows universal fluctuations (UCF).
The conductance being a random variable showing large fluctuations, one should 
consider its full distribution. It was soon clear that the
first higher cumulants of the conductance are proportional to \cite{altshuler3}
\begin{equation}
\langle g^n\rangle_{\rm cum} \propto \langle g \rangle^{2-n} ,
 \qquad n<g_0, \; 
\langle g \rangle \gg1 . 
\label{eqglown}
\end{equation} 
Here $g_0$ is the mean
conductance at the scale $\ell$. In
the metallic regime far from localization, where $\langle g \rangle
\gg 1$, the higher
cumulants are probably small, and the distribution of the conductance is 
roughly Gaussian. However, for $n\agt g_0$ the decrease in 
magnitude of cumulants as described by Eq.~(\ref{eqglown}) is changed 
into a very rapid increase
($\propto \exp[g_0^{-1}n^2]$). This leads to log-normal
tails of the distribution \cite{altshuler3}. With increasing disorder, the 
log-normal tails become more important. The full
conductance distribution at the threshold of localization ($\langle g \rangle 
\sim1$) is at present unknown,
yet it is quite plausible that the whole distribution crosses
over to a log-normal shape in the strongly localized regime (see Refs.
\cite{shapiro2} and \cite{altshulerboek} for a discussion). Indeed,
it is well known that in the strongly localized regime in one dimension
the conductance
is given by the product of transmission amplitudes, yielding a log-normal 
distribution \cite{anderson2,abrikosov2,melnikov,economou2}.

Although higher order cumulants govern the tails of the distribution,
 they do not
affect it near the center. A deviation from the Gaussian distribution near the
center is revealed, first of all, in the lowest nontrivial cumulants. An
important step in this direction was the calculation of the third cumulant of
the distribution using random matrix theory by \citeasnoun{macedo}. He found
for the orthogonal ($\beta=1$) and symplectic ensemble ($\beta=4$) that the
third cumulant of the conductance is proportional to $1/g^2$, thus the leading
term given by Eq.~(\ref{eqglown}) vanishes (see for instance \cite{metha} for
the definitions of the ensembles). For the unitary ensemble ($\beta=2$) even
this sub-leading term vanishes, and the behavior is $1/g^3$ \cite{macedo}.
 The physical reason behind this is not clear.
(We recently learned that the same result in quasi one dimension was found
 \citeasnoun{tartakovskiprive} by the scaling method described in 
\cite{tartakovski}.) However, random matrix theory is only valid in quasi
one dimensional systems. Therefore, it was not known whether such
 a cancellation
also occurs in higher dimensions. If this is indeed the case,
this might indicate an overlooked symmetry
of the system. That this might be the case is also suggested by the fact
that the leading order contribution to the third cumulant of the density of
states vanishes in two dimensions \cite{altshuler3}. The question 
we thus wish to consider is whether
or not a cancellation occurs for the third cumulant of the
conductance in two and three dimensions.

We rewrite Eq.(\ref{eqglown}) into a relation for the relative cumulants: $
\langle g^n \rangle_c/\langle g \rangle^n \propto \langle
g\rangle^{2-2n}$. The inverse powers of $\langle g \rangle $ on the right
hand side can be interpreted as the number of Hikami four-boxes in the
diagrams. Thus the third cumulant diagram contains four Hikami four-boxes.
Indeed, it proves impossible to draw $\langle g^3 \rangle_c$ diagrams with
three boxes. Diagrams with more boxes are sub-leading as they are of
higher power in $1/ \langle g \rangle$.
The diagrams are
represented in Fig.\ \ref{figtop}. In the figure there are diagrams of the
same order with one six-box and two four-boxes. They can be obtained by
contracting one diffuson in the diagrams with four-boxes. In physical terms the
six-box diagrams correspond to processes where after an interference process,
the amplitudes do not combine into a diffuson. Instead, the amplitudes
interact again directly without being scattered.

The diagrams were evaluated using a Kubo approach, the advantage being that no
divergences emerge; treatment of absorption is difficult though, as indicated
in the discussion on $C_3$.
The evaluation of the diagrams is a painful but straightforward 
exercise. 
The diagrams have two or three free momentum loops, of which one is eliminated 
due to momentum conservation.
In one dimension, where the diffusons are simple linear functions,
the diagrams can be evaluated analytically.
Numerical evaluation of the integrals in $d=2$ and $d=3$ yield
\begin{eqnarray}
\langle g^3
\rangle_c &=&0 \qquad \mbox{(quasi 1D),} \nonumber \\
\langle g^3
\rangle_c&=& -0.0020 \langle g \rangle^{-1} \qquad \mbox{ (quasi
2D, square),} \nonumber \\ \langle g^3 \rangle_c&=& + 0.0076 \langle g
\rangle^{-1} \qquad \mbox{ (3D, cubic).} \end{eqnarray} 
Thus the leading
contribution to the third cumulant in one dimension vanishes. 
This confirms the
random matrix theory result \cite{macedo} diagrammatically.
It is seen that there is no cancellation in higher dimensions.
The results for rectangular samples are
given in Fig.~\ref{figwide3}, where we multiplied
the third cumulant by the average of the dimensionless conductance. The
third cumulant for wide slabs ($L_x \gg L_z , \, L_y \gg L_z$) is proportional
to $(L_x L_y / \langle g \rangle)$. In the figure,
 due to the multiplication by 
$\langle g \rangle$, there is 
proportionality to $(L_x L_y)^2/L_z^4$ for wide slabs. 
Note that the third cumulant passes through zero when
going from 2D to 3D at a sample size of $ 0.46 L_z \times
L_z \times L_z $. 
For very narrow slabs the correct quasi 2D limit
is recovered. 

The random matrix result that the third cumulant vanishes in one dimensions
is thus also found in the diagrammatic approach.
In two and three dimensions, however, the cancellation is not present; 
the leading contribution to the third cumulant is negative in
two dimensions and positive in three dimensions. The fact that the third
cumulant changes sign is surprising. The third cumulant is also known as the
skewness of a distribution. In analogy with the third cumulant of the total
transmission, or if the distribution would be tending to
log-normal, one would have expected a positive third cumulant of the
conductance. Instead, we find that all possible values occur: negative,
positive and zero. Experiments measuring the
conductance distribution could enlighten this point.

\section{Full distribution functions}\label{chprob}

In this section we calculate the full distribution of the total transmission
and the angular resolved transmission; both can be mapped on the eigenvalue
 distribution of the transmission matrix. 
Before we discuss the distribution in the mesoscopic regime, 
let us first discuss the 'classical' situation, which is the limit
 of large $g$. The correlations
between intensities are now 
absent.
Here the angular resolved transmission, or speckle intensity, is distributed
according to the 
Raleigh's law \cite{goodmanboek},
\begin{equation}
 P(T_{ab})= \frac{1}{\langle T_{ab} \rangle}\eexp{-T_{ab}/ \langle
T_{ab} \rangle}. \label{eqrayleigh}\end{equation}
Anyone who ever saw a laser speckle pattern on the wall will remember the
wild pattern with both bright and dark spots.
 This is due to the Rayleigh's law. 
Rayleigh's law can be derived in the context of the ladder diagrams
\cite{shapiro}. The $n-$th moment of the speckle intensities is made up of $n$
amplitudes and $n$ complex conjugated amplitudes. As stated before, there is no
restriction on the pairing of two amplitudes into a diffuson for this type of
measurement. Therefore the amplitudes $\{a_1^\ast, a_2^\ast,
a_3^\ast,\ldots,a_n^\ast\}$ have $n!$ possibilities to pair with the amplitudes
$\{a_1, a_2, a_3,\ldots,a_n \}$. Thus
 \begin{equation} \langle T_{ab}^n \rangle = n! \langle
T_{ab} \rangle^n \end{equation} which corresponds with Rayleigh's law. 
Yet, as we know from the example of a laser beam reflecting from a rough wall,
 {\em multiple} scattering is not essential for Rayleigh's law.
Indeed, Rayleigh's original derivation goes as follows \cite{goodmanboek}:
The field $\psi$ at a given
position on the outgoing side is the sum of many fields, and therefore the real
and imaginary part each have a independent Gaussian distribution. The
intensity $I$ is the amplitude squared and has thus an exponential distribution
\begin{equation}
 P \sim\eexp{-[({\rm Re} \psi)^2+ ({\rm Im} \psi)^2 ]/2\sigma^2 }
\sim \eexp{-I/\expect{I}} .\end{equation} 
Thus reflection from a rough wall will also do.

\citeasnoun{kogan2} addressed the question of the speckle statistics of
(optically) very thin samples. In optically very thin samples a
considerable part of the light does not scatter at all, but is transmitted 
coherently. In
the limit of zero thickness, i.e. only coherent light, 
the distribution becomes a
delta-function. In the intermediate regime a simple counting argument yields
the distribution, which interpolates between the delta distribution and
Rayleigh's law. In the opposite case of strong scattering, the mesoscopic
regime, interferences modify the speckle distribution. The leading correction
was derived by \citeasnoun{shnerb}. Genack and Garcia observed a
deviation from Rayleigh's law at large intensities \cite{genack2,garcia2}. A
crossover to stretched exponential behavior was derived by 
\citeasnoun{kogan}.

For the total transmission and the conductance the distributions are similar in
the large $g$ limit. For both quantities the outgoing diffusons are
frequency and momentum independent as we explained in Sec.~\ref{chc12}. As
there is no interaction between the diffusons in this limit, only one type of
pairing is possible for both quantities. This yields in principle a
delta-distribution if one probes an infinite number of channels. In practice of
course only a limited number of channels is probed, and the law of large
numbers predicts a narrow Gaussian distribution.

\subsection{Eigenvalues of the transmission matrix} 

In mesoscopic systems the observables are random
quantities and are therefore not always characterized by their mean values, but
their entire distribution function is of interest. This is particularly
prominent in the distribution of eigenvalues of the transmission matrix. 
Assuming that all eigenvalues contribute equally to the conductance, one 
expects a Gaussian distribution. As there are $N$ eigenmodes, the eigenvalue
distribution should be peaked around $g/N={\ell}/{L}$. But this picture proves
wrong, as was first put forward by \citeasnoun{dorokhov} and
later also by \citeasnoun{imry}.
Not the eigenvalues but the inverse localization lengths $1/\xi_n$ are
uniformly distributed, see \citeasnoun{pendry} and 
\citeasnoun{stoneboek}. As a result the eigenvalues have a ``bimodal'' 
distribution peaked
around 0 and 1. The eigenvalues $T_n$ of the transmission
matrix $t^\dagger t$ can be expressed as 
\begin{equation} T_n=\cosh ^{-2} (L/ \xi_n),\end{equation}
implying that \begin{equation} \langle {\rm Tr} (t
t^\dagger)^j\rangle= \langle \sum_{n=1}^N T_n^j\rangle=g\int_0^1\frac{{\rm
d}T}{2T\sqrt{1-T}} T^j \label{pt}.\end{equation}

This distribution is plotted in Fig.\ref{figptn}. The first derivations of the
distribution used random matrix theory 
and were therefore valid in quasi-1D only 
\cite{mello2}. However, \citeasnoun{nazarov} showed that it is
not only true in quasi-1D, but under very general conditions. Note that the
normalization of the distribution is ill-defined, the distribution should be
understood in the sense that all its moments are defined. This problem can
be avoided by adjusting the lower boundary of the integral instead of 0.
The minimal value corresponds to the decay of unscattered intensity, therefore
$T_n \geq \cosh^{-2} (L/\ell) \approx \exp(-2L/\ell)$, normalizing the
distribution \eq{pt}. Although this cutoff is important for the normalization
of the distribution, its influence on the momenta is very small and we refrain
from it. The distribution Eq.~(\ref{pt}) is valid only in the regime
where $g$ is not too small. Loop effects near the Anderson transition, where
$g\sim 1$, will change it, as we will discuss below.

The concept of open and closed channels explains the
physical meaning of this curious distribution function.
An eigenmode of the transmission matrix is, according
to this distribution, either essentially blocked or, with a much smaller
probability, essentially conducting. This picture was confirmed in various
computer simulations \cite{pendry,pendry2,oakeshott}. 
The eigenvalue
distribution also provides us with a nice picture explaining why the
correlations in mesoscopic systems are so large. If all the channels are
equally conducting, fluctuations of the channels average out by the law of
large numbers. Yet if only a few channels are conducting, fluctuations in one
channel will be clearly seen. 
Using these concepts the probability distribution of the
total transmission and deviations from Rayleigh's law will
 be explained below.

We should point out that for the UCF the same argument goes wrong. 
The variance in the conductance
$N \ell/ L$ is caused by the finiteness of the number of channels $N$. In the
picture of either closed or open channels the number of open channels $N_{\rm
eff}$ equal roughly $N\ell/L \ll N$. The conductance
would thus fluctuate according the law of large numbers with a variance:
$\expect{\sum_n T_n^2}=\frac{2}{3}g$. However, this is incorrect as we know
that $\expect{\sum_n T_n^2} \sim O(1)$, see Sec.~\ref{chc3}. For the UCF
the interaction between the eigenmodes is essential.

In practice it is difficult, if not impossible, to measure the eigenmodes and
eigenvalues directly. As they are the eigenmodes of the very large random
transmission matrix, they have a very complex structure. Nevertheless, the
eigenvalue distribution has
observable consequences. It was shown by \citeasnoun{beenakker}
that the shot noise of electronic conductors universally reduces by a factor 3
because of this distribution. Shot noise only occurs in electronic systems and
is thus not of much interest for us. Instead, we will show that the total
transmission and speckle intensity distribution function are related to the
eigenvalue distribution function.

\subsection{Distribution of total transmission} \label{secdisTa}

We
first calculate the probability distribution of the total transmission. From
the distribution without interference effects we saw that it has a simpler
distribution (in the sense of cumulants) than the speckle distribution. 
We neglect 
corrections as absorption, skin layers, and disconnected diagrams. Consider
again an incoming plane wave in direction $\mu_a$ ($\mu$ denotes again the
cosine with respect to the $z$ axis). The wave is transmitted into outgoing
channel $b$ with transmission amplitude $t_{ab}$ and transmission probability
$T_{ab} \equiv |t_{ab}|^2$. The speckle and total transmission factorize as
$\langle T_a\rangle =\epsilon_a g$ and $\langle T_{ab}\rangle= \epsilon_a
\epsilon_b g$.

We consider the $j^{\rm th}$ cumulant of $T_a$. In a diagrammatic approach
this object has $j$ transmission amplitudes $t_{ab}$ and an equal number of
Hermitian conjugates $t_{ba}^\dagger=t_{ab}^*$. The explicit calculation of
the second cumulant $C_2$ (Sec.~\ref{chc12}), and third cumulant (section
\ref{cht3}) showed that the leading diagrams are connected and have no loops.

Let us fix the external diffusons in the term $t_{a b_1}^{
}t^\dagger_{b_1a}t_{ab_2}^{ }\cdots t_{ab_j}^{ }t^\dagger_{b_ja}$.
Contributions to the sum over $b_i$ only come from diagrams with outgoing
diffusons that have no transverse momentum. These are the diagrams where the
lines with equal $b_i$ pair into diffusons. We indicate this pairing of the
outgoing diffusons with brackets 
\begin{equation} ( t_{a b_1}^{ } t^\dagger_{b_1a})
(t_{ab_2}^{ } t^\dagger_{b_2a} ) \cdots ( t_{ab_j}t^\dagger_{b_ja}). 
\end{equation} The
outgoing diffusons are now fixed. At the incoming side it is convenient to
keep for instance the $t_{ab}$'s fixed, while the $t_{ab}^\dagger$'s are
permuted among them. This leaves $j!$ possibilities for the incoming side
in total, but of these only $(j-1)!$ permutations correspond to connected 
diagrams (these diagrams contain $j-1$ Hikami boxes. 
The $C_2$ diagram contained one box, the third cumulant diagram two boxes).
Next we factor out the incoming and outgoing diffusons and
group the remainder of the diagrams into a skeleton $K$. Making use of Eq.\
(\ref{tfaa}) we obtain:
\begin{eqnarray} \label{inte} \langle T_a^j\rangle _{\rm 
con}\!&\!= \!&\!\epsilon_a^j
(j\!-\!1)! \!
\int \! {\rm d}{\bf r}_1 {\rm d}{\bf r}_{1}'\cdots {\rm d}{\bf r}_j {\rm d}
{\bf r}_{j}'
{\cal L}_{\rm in}({\bf r}_1) {\cal L}_{\rm out} ({\bf r}_{1}') \nonumber \\
&& \cdots {\cal L}_{\rm in}({\bf r}_j) {\cal L}_{\rm out}({\bf r}_{j}')
K({\bf r}_1, {\bf r}_{1}', \cdots ,
{\bf r}_j,{\bf r}_{j}').\end{eqnarray}
The integral just
describes \BA \langle{\rm Tr} (t t^\dagger)^j\rangle &\equiv&
{\rm Tr} \sum _{b_1,a_2,b_2,\cdots,a_j,b_j}\langle t^{ }_{a_1b_1}
t^\dagger_{b_1a_2}
t_{a_2b_2}^{ }\cdots t_{a_jb_j}^{ }t^\dagger_{b_ja_1}\rangle \nonumber \\
&=&
\sum _{a_1,b_1,\cdots,a_j,b_j}\langle t^{ }_{a_1b_1}t^\dagger_{b_1a_2}
t_{a_2b_2}^{ }\cdots t_{a_jb_j}^{ }t^\dagger_{b_ja_1}\rangle 
.\EA
There is only one way to attach incoming and outgoing diffusons to $K$. The
sums over the indices lead exactly to the total flux diffusons
in Eq.~(\ref{inte}). One finds the following important relation between the
moments of the eigenvalue distribution and the connected total transmission
diagrams \begin{equation} \langle
T_a^j\rangle_{\rm
con}=(j-1)! \epsilon_a^j \langle {\rm Tr} (t t^\dagger)^j\rangle.
\end{equation} 
Normalizing with respect to the average,
we introduce
$ s_a=T_a / \langle T_a\rangle$.
The generating function of the connected diagrams
is easily calculated
\begin{eqnarray} \Phi_{\rm con}(x)&\equiv&\sum_{j=1}^\infty
\frac{(-1)^{j+1} x^j}{j!}\; \langle s_a^j\rangle_{\rm con} \nonumber \\
&=&g\log^2\left(\sqrt{1+x/g}+\sqrt{x/g}\right).
\label{phiconpw}\end{eqnarray}
Since the cumulants are solely given by connected diagrams, the
distribution of $s_a$ follows as \cite{probprl}
\begin{equation} P(s_a)
=\int_{-i\infty}^{i\infty}\frac{{\rm d}x}{2\pi i} \exp
 \left[xs_a- \Phi_{\rm con}(x) \right]
\label{PsaPW}.\end{equation}

Let us examine some properties of the distribution.
For limiting values of $s_a$ we use a saddle point analysis.
The saddle point is found by the condition $\frac{d}{dx} \left[ xs_a-
\Phi_{\rm con}(x) \right] =0$. Thus we find 
\begin{equation} s_a=\frac{\log\left(\sqrt{1+x/g}
+\sqrt{x/g}\right)}{\sqrt{x/g(1+x/g)}}\label{eqsaddle}.
 \end{equation} The r.h.s. of
Eq.(\ref{eqsaddle}) diverges if $x$
approaches $-g$ (from above) and decreases monotonically for larger $x$.
Thus for large $s_a\gg 1$, we find the saddle point near $x=-g$.
By inserting the saddle point one finds a simple exponential tail
\begin{equation} P(s_a)\approx \exp(-gs_a+g\frac{\pi^2}{4}), \qquad s_a \gg
1.\end{equation}

The saddle point also dominates the shape for small $s_a\ll 1$ (and large $g$).
One finds essentially a log-normal growth: \begin{equation}
P(s_a) \sim \exp\left[ \frac{g}{4}-\frac{g}{4} \left( \log \frac{2}{s_a}
+\log \log \frac{2}{s_a} -1 \right)^2 \right]. \end{equation}

In Fig.\ \ref{figPtaPW} the full curves depict
the distribution Eq.~(\ref{PsaPW}) for
some values of $g$. At moderate $g$ we clearly see the deviation from a
Gaussian. In interesting, recent experiments 
on quasi 1D microwave scattering, \citeasnoun{stoytchev} 
measured the total transmission distribution for conductances as low
as $g \sim 3$. Although one would expect important
loop contributions for such low values of $g$,
excellent agreement with the above prediction is found.
To obtain the fit, the value of $g$ had to be extracted 
from the variance of the
distribution rather than from the measured conductance. This adjustment is
probably necessary because of the strong absorption occurring in
waveguides. The effect of absorption was calculated by \citeasnoun{brouwer97}
for a waveguide.
For strong absorption the total transmission has a log-normal distribution,
while the 
ratio of $\langle  \langle T_a^3 \rangle \rangle /
\langle  \langle T_a^2 \rangle \rangle^2$ becomes 3.0 instead of 2.4.

\subsection{Influence of beam profile}
Above we have considered the case of an incoming plane wave. Again in optical
systems a Gaussian intensity profile is more realistic. 
For the third cumulant we saw a nontrivial dependence on the incoming beam
profile, suggesting that also higher cumulants are sensitive to this effect. 
For perpendicular incidence
the incoming amplitude is $ \psi_{\rm in}({\bf r})= W^{-1}
\sum_a \phi(q_a) \psi^a_{\rm in} ({\bf r})$,
where $\psi^a_{\rm in} ({\bf r})$ is the plane wave of Eq.\
(\ref{eqpsiin}), and where \begin{equation}
\phi(q_a)=\sqrt{2\pi}\rho_0\exp(-\frac{1}{4}\rho_0^2q_a^2).\end{equation}
We consider the limit where the beam is much broader than the
sample thickness ($\rho_0\gg L$) but still much smaller than the transverse
size of the slab ($\rho_0\ll W$). 
With a smaller beam diameter, the
 incoming transverse momenta, which are
of order $1/\rho_0$, are of the order of $1/L$.
The diffusons will then become the well-known $\cosh$ functions, 
Sec.~\ref{rmpintro}. 
Here the momentum dependence of the diffusons
can be neglected; apart from a geometrical factor, the diffusons are identical
to the plane wave case.
Due to the integration
over the center of gravity, each diagram involves a factor
$A\delta_{\Sigma q,\Sigma q'}$.
In the $j^{\rm th}$ order term there occurs a factor
\begin{eqnarray} F_j&=&\frac{A}{A^{2j}}\sum_{q_1 q_{1}'\cdots
q_{j}q_{j}'} \phi(q_1)\phi^{\ast}(q_{1}')\cdots\phi(q_{j})\phi^\ast(q_{j}')
\delta_{\Sigma q,\Sigma q'}\nonumber \\&=&\int {\rm d}^2\rho \;
|\phi(\rho)|^{2j}.\end{eqnarray} For a plane
wave we have $|\phi(\rho)|=\sqrt{A}$,
and $F_j=A^{1-j}$. For our Gaussian beam we
obtain \begin{equation}
F_j=\frac{1}{j}\left(\frac{\pi\rho_0^2}{2}\right)^{1-j}.\end{equation} It is
thus convenient to identify $A_G=\frac{1}{2}\pi\rho_0^2$ with the
effective area of a Gaussian beam. (This definition is different from
the one used in previous sections, were $A_G= \pi \rho_0^2$). 
As compared to the plane
wave case, the $j^{\rm th}$ order term
is smaller by a factor $1/j$ for a Gaussian profile.
This implies for the generating function of the connected diagrams
\begin{equation} \Phi_{\rm con}(x) = g\int_0^1 \frac{{\rm
d}y}{y}\log^2\left(\sqrt{1+\frac{xy}{g}}+\sqrt{\frac{xy}{g}}\right)
\label{phicong}
.\end{equation} For small $s_a$ (and large $g$), there is again a log-normal
saddle point. For large $s_a$, the dominant shape of the decay is given by the
singularity at $x=-g$ and again yields $P(s_a)\sim\exp(-gs_a)$.
The shape of the distribution is quite similar to the plane wave case. 
Expanding the generating function we recover the results for the second
and third cumulant obtained in Secs.~\ref{chc12} and \ref{cht3}.

\subsection{Speckle intensity distribution} \label{secdisTab}

We apply the same method to obtain the distribution of the angular
transmission coefficient. 
The angular and total transmission distributions can be related to each other
in a fairly 
simple way, because the interference processes are dominated by the same
type of diagrams: the loopless connected diagrams. 
The different distribution for total transmission and speckle is the
consequence of a counting argument only.
In the plane wave situation the average angular transmission reads
$\langle T_{ab} \rangle = \epsilon_a \epsilon_b g$. Let us count the
number of connected loopless diagrams that contribute to
$T_{ab}^j=t_{ab}^{ }t^\dagger_{ba}t_{ab}^{ }\cdots t_{ab}^{ }t^\dagger_{ba}$.
Now not only massless outgoing diffusons, but all
 pairings into outgoing diffusons contribute.
This yields an extra combinatorial factor $j!$ in the
$j^{\rm th}$ moment: \begin{equation} \langle T_{ab}^j \rangle_{\rm con}=
j! (j-1)! \epsilon_a^j \epsilon_b^j \langle {\rm Tr} (t^\dagger t)^j\rangle.
\end{equation}
For the normalized angular transmission coefficient
$s_{ab}=T_{ab}/\langle T_{ab}\rangle$, we define the generating function
of the connected diagrams as \cite{probprl}
\begin{equation} \Phi_{\rm
con}(x)\equiv \sum_{j=1}^\infty \frac{(-1)^{j-1}x^j}{j!^2} \langle
s_{ab}^j\rangle_{\rm con}, 
\end{equation}
with $\Phi_{\rm con}$ given by Eq.~(\ref{phiconpw}) for
plane wave incidence and by Eq.~(\ref{phicong}) for a broad
Gaussian beam, respectively.
In contrast to the total transmission distribution, the cumulants are not
only given by the connected diagrams. \citeasnoun{kogan}
showed that the extra summation of the disconnected diagrams corresponds
to an additional integral, which finally yields a Bessel function:
\begin{equation}
P(s_{ab})=\int_{-i\infty}^{i\infty}
\frac{{\rm d}x}{\pi i}K_0(2\sqrt{-s_{ab}x}) \exp\left(-\Phi_{\rm con}(x)
\right) \label{PtabPW}. \end{equation}
The integral can be evaluated numerically.
The speckle intensity distribution is shown in Fig.\ \ref{figPtabP}
for an incoming plane wave.

For large $g$ and moderate
$s_{ab}$, one has $\Phi_{\rm con}(x) \approx x $ and we recover Rayleigh's
law: $P(s_{ab})=\exp(-s_{ab})$.
The leading correction is found by expanding in $1/g$
\begin{equation}
P(s_{ab})= {\rm e}^{-s_{ab}} \left[ 1+\frac{1}{3g} (s_{ab}^2 -4 s_{ab}+
2)\right]. \end{equation}
A similar equation was derived previously by \citeasnoun{shnerb}. 

For obtaining
the tail of the distribution, one can again apply steepest descent, which
yields \begin{equation} P(s_{ab})\sim \exp(-2\sqrt{g s_{ab} }\;).
\end{equation} 
This stretched exponential tail of the distribution
 was also seen in experiments. Already
in 1989 Garcia and Genack observed it in microwave experiments. But
unfortunately their dynamic range was rather small. The fit of stretched
exponential is therefore rather unprecise. The maximum intensity was 5 times
the average where one does not expect to see the tail behavior, rather one
observes a cross-over behavior.
Using \eq{phicong} one easily derives the speckle distribution 
due to an incoming beam with 
Gaussian profile; it leads to a different distribution
with the same asymptotic behavior.

\subsection{Joint distribution} 
It turns out that the speckle distribution and
the total transmission distribution of a certain incoming direction are
related. This is not surprising as a large total transmission for
 a given incoming
direction will also be reflected in the individual speckles. The moments of
the joint distribution are 
\begin{equation} \expect{s_{ab}^k s_a^l}_{\rm con} = (k+l-1)!
\;k!\; \expect{{\rm Tr} (t t^\dagger)^{k+l}}. \end{equation}
 The combinatorial factor 
$k!$ is for the
speckle pattern, the factor $(k+l-1)!$ is the number of possible pairings of
incoming beam. Defining $\sigma_{ab}$ as $s_{ab}=\sigma_{ab} s_a $,
 one obtains
$\expect{ \sigma_{ab}^k s_a^m}=k! \; (m-1)! \; \expect{T_n^{m}}$ with
$m=k+l$. Thus
 $P(\sigma_{ab},s_a) =\exp(-\sigma_{ab}) P(s_a)$. 
This shows 
that $\sigma_{ab}$ and $s_a$ are independent variables.
Nevertheless, $s_{ab}$ and
$s_a$ are dependent. Their joint distribution is 
\begin{equation} P(s_{ab},s_a)
=\frac{\exp(-s_{ab}/s_a)}{s_a} \int \frac{\d x}{2\pi i} \exp(xs_a
-\Phi_{con}(x)) \label{Pjoint}. \end{equation} 
Integration over $s_a$ or $s_{ab}$ yields again the distribution $P(s_{ab})$,
or $P(s_a)$, respectively. This is somewhat surprising. Indeed, as the right
hand side of \eq{Pjoint} contains only connected diagrams, the same holds thus
for the left hand side. This seems to contradict our earlier remark that
$P(s_{ab})$ contained both connected and disconnected diagrams. However, in
deriving \eq{Pjoint} we have tacitly performed analytic continuation for
negative $l$ values. 
By performing the integration over $s_{a}$ we find the
correct expression for $P(s_{ab})$. 

In conclusion we derived the full
distribution function of both total transmission and angular transmission
coefficient by mapping it to the distribution of eigenvalues of $T_{ab}$. But
at least for the lowest cumulants one can use this mapping the other way
around. First, experiments on the distribution functions thus confirm the
eigenvalue distribution. Secondly, from our detailed evaluation
of the low order diagrams, i.e. the second and third cumulant, insight in the
eigenvalue distribution can be obtained. It is clear from those calculations
that also the eigenvalue distribution is based on loopless connected
diagrams.

The effect of absorption has been neglected here. Our explicit calculation of
second and third cumulants and the experimental data show that the effects of
absorption and skin layers can be incorporated to a large extent 
by considering the
normalized cumulants, such as $\langle T_a^3 \rangle_c / \langle T_a^2
\rangle_c^2 $.  Indeed, also in the presence of strong absorption it was seen
that these dimensionless ratios are similar to the values calculated without
absorption \cite{stoytchev}.  A theoretical analysis of absorption effects was
carried out by ~\citeasnoun{brouwer97}. 

The distribution of the conductance cannot be calculated using the above
technique. The correlations in the conductance correspond to correlations in
the eigenvalues. These correlations were not included, as we only used the
distribution of a single eigenvalue, and not the joint distribution of multiple
eigenvalues. 

\acknowledgements{
We have been stimulated by the continuous interest in our work of
A. Z. Genack,
H. C. van de Hulst,
I. V. Lerner,
J. M. Luck, and
G. Maret.
We gained many insights from discussion and collaboration
with our co-workers on this subject: 
E. Amic, 
M. P. van Albada,
B. L. Altshuler,
J.F. de Boer, 
A. L. Burin,
Zh.S. Gevorkian,
E. Hofstetter, 
Yu. M. Kagan,
A. Lagendijk,
D. J. Lancaster,
I. V. Lerner,
J. M. Luck,
R. Maynard,
A. Mosk,
P. N. den Outer, 
M. Schreiber,
G. V. Shlyapnikov,
B. A. van Tiggelen, and
R. Vlaming.

Further we like to thank: 
C. W. J. Beenakker,
P.W. Brouwer,
the late S. Feng,
J. J. M. Franse, 
N. N. Garcia,
E. Kogan,
Yu. Nazarov,
M. Nieto-Vesperinas,
I. Ya. Polishchuk,
J. Ricka,
J. J. Saens,
R. Sprik,
M. Stoytchev,
A. Tip,
P. de Vries, 
and G. H. Wegdam.

This research was supported by the EEC HCM-Network ERBCHRXCT930373
and by NATO \ (grant nr.\ CRG 921399).
}

\newpage

\appendix

\section{Loop integrals}
\label{seciint} 
The same loop momentum integrals recur
often in the diagram calculations:
\begin{equation} I_{k,l} \equiv
\int \frac{\d^3 \p}{(2\pi)^3} G^k(\p)G^{\ast\,l}(\p), \end{equation}
 where $ G(\p) =
[\p^2-k^2-nt]^{-1}$. For the simplest integral, $I_{1,1}$
 we find \begin{equation}
I_{1,1}=\int\frac{\d^3\p}{ (2\pi)^3 } G(\p)G^\ast(\p) = \int_{-\infty}^\infty
\frac{ \d p }{ (2\pi)^2 } \,\frac{p^2}{(p^2 +\mu^2)(p^2 +\overline{\mu}^2)},
\end{equation} where $\mu^2=-k^2-nt$. The sum of the residues yields
$1/[4\pi(\mu+\overline{\mu})]$, and without absorption the optical theorem
gives \begin{equation} 
\mu + \overline{\mu} = \frac{n\,{\rm Im}t}{k} = \frac{1}{\ell}.
\label{opttheorema} \end{equation} 
Therefore \begin{equation} I_{1,1}=\frac{\ell}{4\pi} . \end{equation} The
general $I_{k,l}$ integral 
can be found from the $I_{1,1}$ integral using \begin{equation} I_{k+1,l}=
\frac{-1}{2k\mu}\frac{\d}{\d\mu}I_{k,l} \mbox{ ;} \qquad I_{k,l+1}=
\frac{-1}{2l\overline{\mu}}\frac{\d}{\d\overline{ \mu} }I_{k,l} . 
\end{equation}
For example, $I_{1,2}=-I_{2,1}=-i\ell^2/(8\pi k)$ and $I_{2,2}=\ell^3/(8\pi
k^2)$. We shall also use 
\begin{equation} \int \frac{\d^3 \p}{(2\pi)^3} (\p \cdot \q)^2 G^k(\p)
G^{\ast\,l} (\p) = \frac{1}{3} k^2 q^2 I_{k,l}, \end{equation}
where  the factor $1/3$
comes from the angular integral. 
If absorption is present $ \mu + \overline{\mu} =
[\ell ( 1-\kappa^2 \ell^2/3 )]^{-1}$, thus 
\begin{equation} I_{1,1} =
\frac{1}{4\pi(\mu+\overline{\mu})} = \frac{\ell}{4\pi}\left(1-\frac{1}{3}
\kappa^2\ell^2\right). \end{equation} 
In general there is a prefactor \begin{equation} I_{n,m} \propto (1-\kappa^2
\ell^2 /3)^{n+m-1}, \end{equation}
 where we have assumed that $\kappa \ell \ll 1$.
However, this prefactor needs only to be included for terms not proportional to
$q$ or $\Omegamark$, as we work to first order of $(M\ell)$.

\newpage

\begin{table}
\caption
{Numerical values of several fundamental quantities 
in radiative transfer theory
for typical values of the index ratio $m$.}
\[
\begin{array}{|r|rrrr|}
\hline \hline
 m & \tau_0 & \tau_1(1) & \gamma(1,1) & \Delta Q \ell  \\ 
\hline 2 &6.08 &21.7 &21.5 &0.136 \\
3/2 &2.50 &10.8 &10.6 &0.269 \\ 
4/3 &1.69 &8.34 &7.94 &0.343 \\ 
1 &0.710446 &5.03648 &4.22768 &1/2 \\ 
3/4 &0.815 &5.39 &4.63 &0.479 \\ 
2/3 &0.881 &5.60 &4.85 &0.465 \\ 
1/2 &1.09 &6.25 &5.55 &0.427 
\\ \hline \hline
\end{array}
\]
\label{tabel1} \label{table1}
\end{table}

\begin{table}
\caption{Anisotropic versus isotropic scattering.
$\tau_0\ell_{tr}$ is the thickness of a skin layer;
$\tau_1(1)$ and $\gamma(1,1)$, respectively, yield the
transmitted and reflected intensities in the normal direction;
$B(0)$ is the peak value of the enhancement factor
at the top of the backscatter cone;
$\tau_{tr}\Delta Q=k_1\ell_{tr}\Delta\theta$ is the dimensionless
width of this backscatter cone.
The third row gives the relative difference in the two cases.}
\[
\begin{array}{|r|rrrrr|}
\hline \hline 
&\tau_0 & \tau_1(1)& \gamma(1,1)& B(0)& \tau_{tr}\Delta Q \\
\hline
\mbox{Isotropic} (\tau_{tr}=1) &
0.710446 &5.036475 &4.227681 &1.881732 &1/2 \\
\mbox{Very anisotropic} (\tau_{tr}\gg 1) 
&0.718211 &5.138580 &4.889703 &2 &0.555543 \\
\hline 
\mbox{Difference (\%)} & 1.1 &2.0 &15.7 &6.3 &11.1 \\ \hline \hline
\end{array}
\]
\label{table2}
\end{table}

EDITOR: table 3 can be set narrower, I couldn't figure out how to make Latex 
do so.

\begin{table}
\begin{tabular}{|l|l|l|}
Object &  Small radius behavior  &  Large radius behavior \\
& $(R\ll\ell)$ & $(R\gg\ell)$ \\
\hline
Absorbing sphere & $ Q\approx\frac{3R^2}{4\ell}$ & 
$Q\approx R \left(1-\frac{\tau_0\ell}{R}\right), \quad \tau_0\approx 0.710446 $ \\
Transparent sphere  & $P\approx-\frac{R^3}{3}\left(1+\frac{3R}{4\ell}\right) $
& $P\approx -R^3
\left(1-\frac{3\tau_{01}\ell}{R}\right) , \quad \tau_{01}\approx 0.4675  $\\
Reflecting sphere & $P\approx\frac{R^2\ell}{4} $
& $P\approx \frac{R^3}{2}
\left(1+\frac{6\ell^2}{5R^2}\right)$ \\
Absorbing cylinder & 
$R_{\rm eff}\approx a\ell\left(\frac{R}{\ell}\right)^{\frac{16}{3\pi^2}}
e^{-\frac{4\ell}{3R}}, \quad a\approx1.20 $ &
$R_{\rm eff}\approx R \left(1-\frac{\tau_0\ell}{R}\right), \quad \tau_0\approx0.710446 $ \\
Transparent cylinder & $P\approx-\frac{R^2}{2}\left(1+\frac{R}{\ell}\right)$&
$P\approx -R^2 \left(1-\frac{2\tau_{01}\ell}{R}\right), \quad \tau_{01}\approx0.2137$ \\
Reflecting cylinder & $P\approx\frac{3R\ell}{8} $ & 
$P\approx R^2 \left(1+\frac{4\ell^2}{5R^2}\right) $ 
\end{tabular}
\caption{
Capacitance $Q$ and polarizability $P$ of small and large spheres and cylinders 
immersed in an opaque medium. In column for large radius the diffusion approximation is
given by the term outside the brackets. 
For narrow and wide cylinders the effective
radius and the polarizability are presented.} \label{table3}
\end{table}

\newpage

\begin{figure}
\epsfxsize=12cm
\centerline{\epsffile{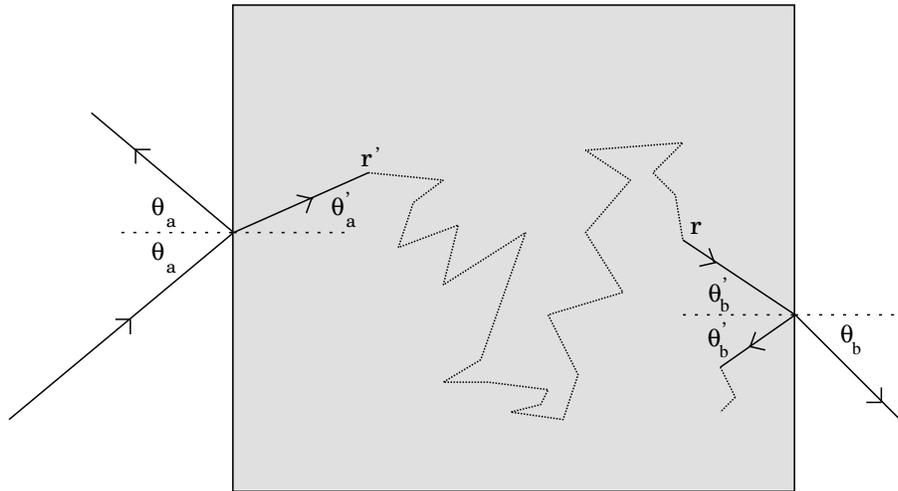}}
\caption{Schematic representation of multiple scattering in a slab.
A plane wave, impinging at an angle $\theta$ with respect to the
$z$ axis, is refracted to an angle $\theta'$ inside the multiple
scattering medium, while it is partly specularly reflected. 
After the diffusive transport in the bulk, a wave arriving under
angle $\theta'_b$ at one of the boundaries, partially 
leaves the medium under
angle $\theta_b$, and is partially reflected internally.}
\label{skin.eps}
\end{figure}

\begin{figure}
\epsfxsize=12cm
\epsfysize=8cm
\centerline{\epsfbox{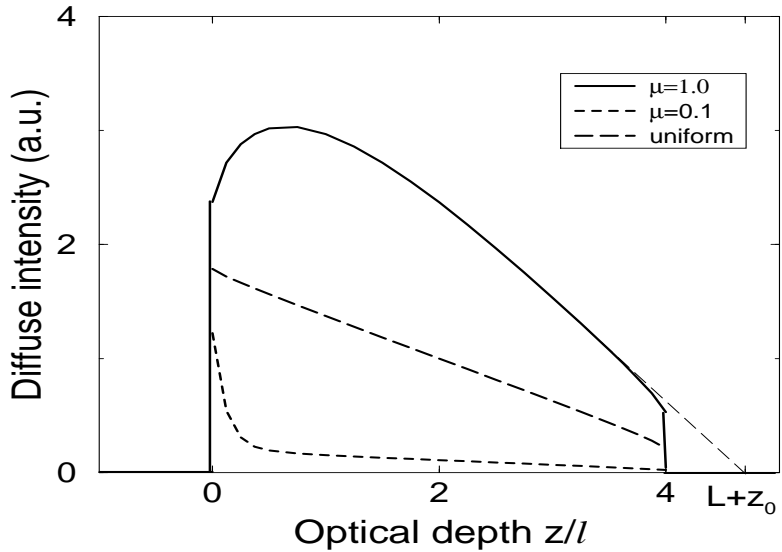}}
\caption{Solution of the Schwarzschild Milne equation as taken from 
table 17 in Van der Hulst (1980) for a slab of thickness $L=4\ell$ with
no internal reflection, $a=1$. The precise solution near the incoming plane
strongly depends on $\mu=\cos \theta'$, where $\theta'$
is the angle of the incoming beam. 
Apart from a multiplication factor, the bulk behavior
is the same; all lines extrapolate to
$L+z_0$ (thin dashed line).} \label{figdifsm}
\end{figure}

\begin{figure}[htbp]
\epsfxsize=12cm
\centerline{\epsffile{gvg.eps}}
\caption{The dressed Green's function.} \label{gvg}
\end{figure}

\begin{figure}[htbp]
\epsfxsize=12cm
\centerline{\epsffile{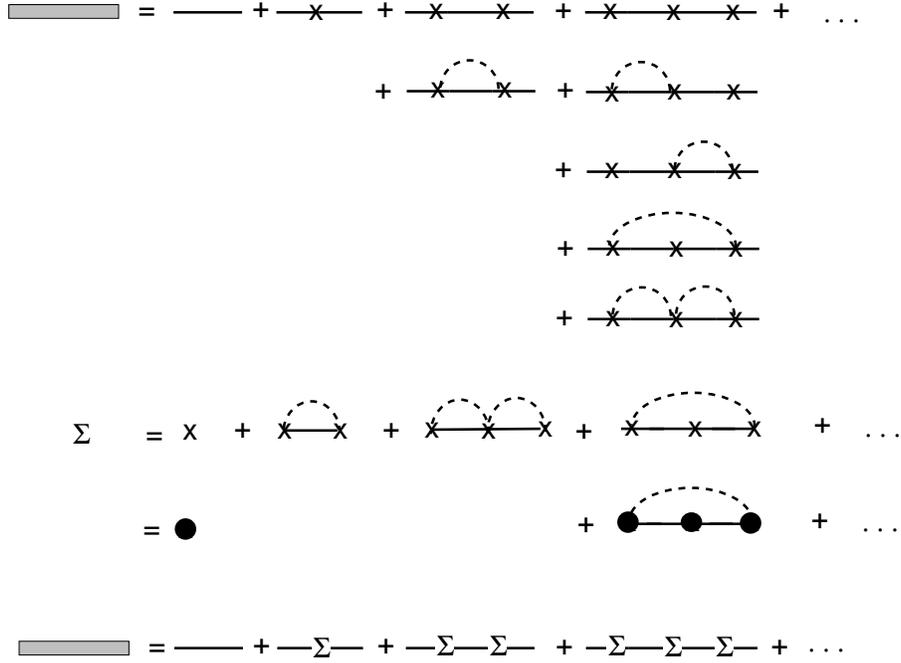}}
\caption{Average Green's function (heavy line):
Thin lines, the bare Green's function; dashed lines
indicate that scattering takes place from the same potential. 
The self-energy $\Sigma$
only contains irreducible diagrams. 
It generates the amplitude Green's function according to the
lowest line.} \label{sigma}
\end{figure}

\begin{figure}[htbp]
\epsfxsize=10cm
\centerline{\epsffile{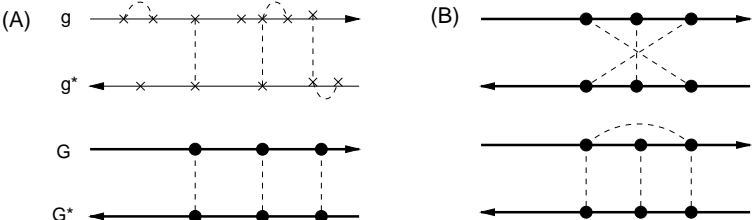}}
\caption{Various intensity diagrams. (A)
The ladder diagrams. Upper line: A typical ladder diagram before
averaging. The crosses are scattering potentials, the Green's functions are
bare. Lower line: Averaging over the possible scattering diagrams
leads to the ladder diagrams, here a diagram with three common
scatterers,  where
the circles represent $t$ matrices and the Green's functions are dressed.
The ladder sum contains diagrams with an arbitrary number of common scatterers.
(B) Some scattering diagrams that are not elements of the ladder diagrams.
The upper diagram is a maximally crossed diagram, responsible for the enhanced
backscatter cone. The lower diagram is of second-order in the density.}
\label{ladderfig}
\end{figure}

\begin{figure}
\epsfxsize=8cm
\centerline{\epsfbox{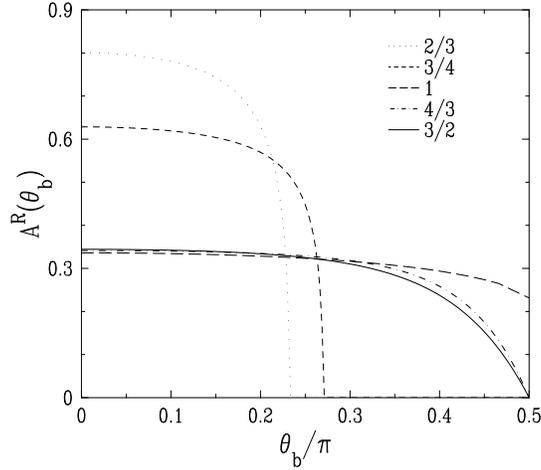}}
\caption{Angular resolved reflection coefficient for perpendicular
incidence on a semi-infinite medium, for different values
of the index ratio $m$. For $m<1$ no radiation can exit the
random medium at an angle exceeding the Brewster angle.
}\label{jmlar}
\end{figure}

\begin{figure}
\epsfxsize=10cm
\centerline{\epsfbox{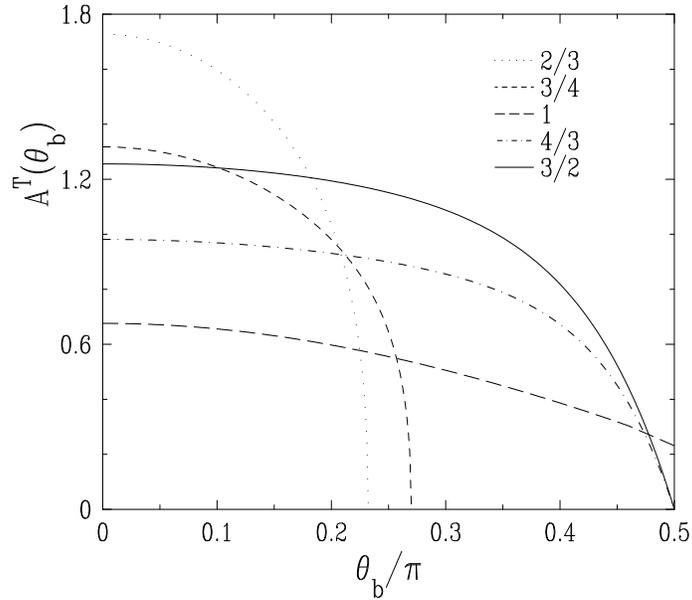}}
\caption{
Angular resolved transmission coefficient for
perpendicular incidence on a thick slab. The ratio of refractive
indices, $m$, has been indicated.}
\label{jmlat}
\end{figure}

\begin{figure}[htpb]
\centerline{\epsffile{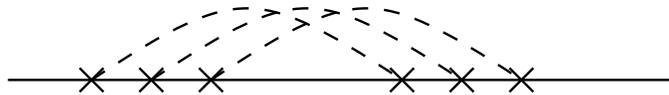}}
\vspace{3mm}
\caption{Handle diagram with three scatterers. 
The set of handle diagrams with an arbitrary number of scatterers
generates the minimal set of self-energy diagrams necessary for flux
conservation with the backscatter cone taken into account.}
\epsfxsize=8cm
\label{handlediagrams}
\end{figure}

\begin{figure}
\epsfxsize=10cm
\centerline{\epsfbox{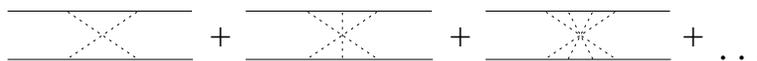}}
\caption{Maximally crossed diagrams}\label{figcoop}
\end{figure}

\begin{figure}
\epsfxsize=10cm
\centerline{\epsfbox{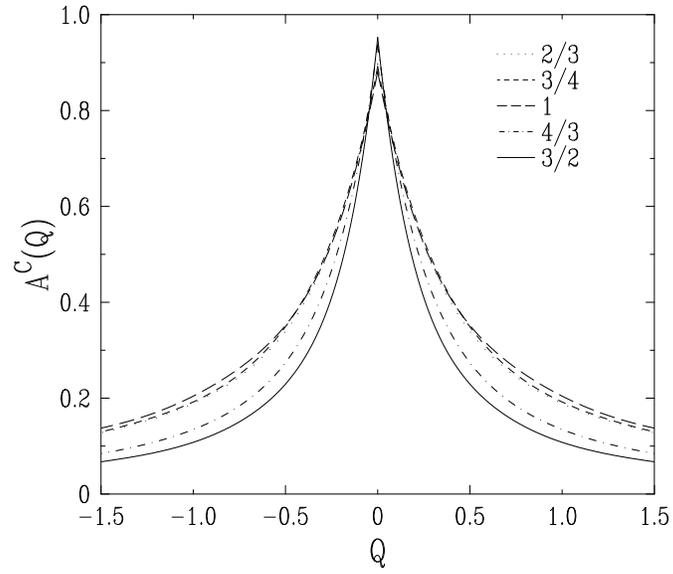}}
\caption{Backscatter cone of a semi-infinite medium at
normal incidence 
for several values of $m$. The diffusive background was substracted.
}
\label{jmlac} \end{figure}

\begin{figure}[htb]
\epsfxsize=5cm
\centerline{\epsfbox{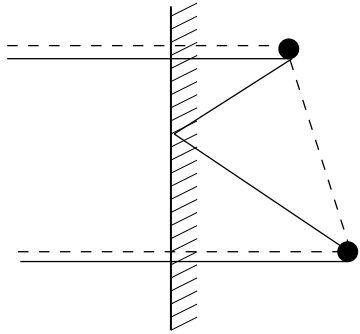}} 
\caption{At large angles the backscatter cone is mainly determined
by second-order scattering. The term with internal reflection
has a larger optical path and therefore decays faster.} 
\label{back2} \end{figure}

\begin{figure}
\epsfxsize=10cm
\epsfysize=10cm
\centerline{\epsffile{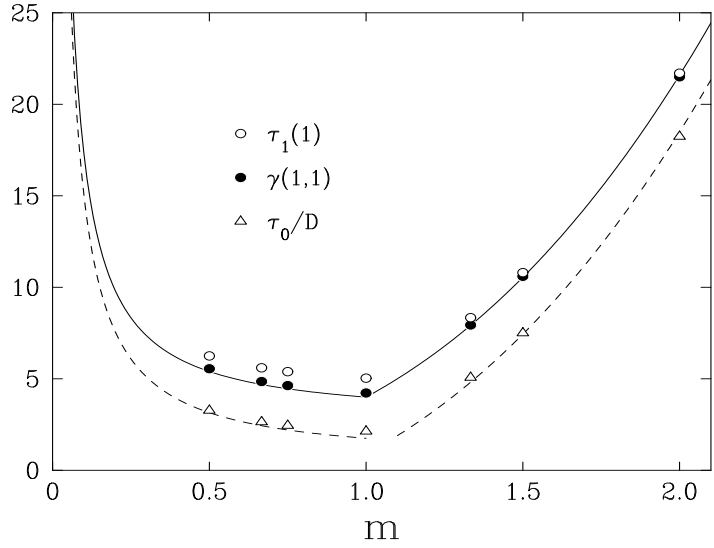}}
\caption{Comparison of exact values and diffusion approximation
for $ \tau_1(1) $, $ \gamma(1,1)$
and $ 3\tau_0 $. The solid line is $ 4/\T $,
the dashed line is Eqs.~(\protect{\ref{mcwtn8}})+(\protect{\ref{mcwtn13}}).
Even for $m$ close to unity the asymptotic results are quite good.}
\vspace{3mm}
\label{jmltau}
\end{figure}

\begin{figure}
\epsfxsize=11cm 
\epsfysize=8cm 
\centerline{\epsfbox{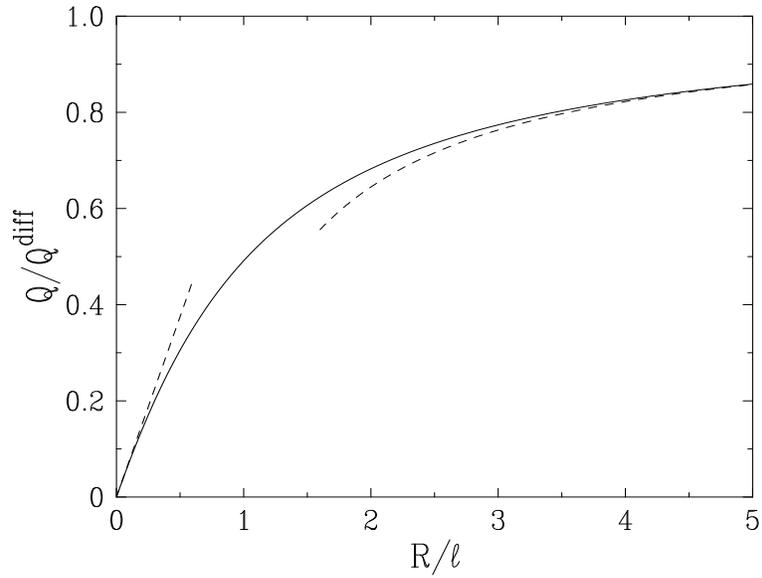}} 
\caption{
Plot of the size factor $Q/Q_{\rm diff}$ of 
the capacitance of an absorbing sphere, against the ratio $R/\ell$.
Full line: outcome of the numerical analysis;
Dashed lines: small-radius and large-radius behaviors of Table~3.}
\vspace{3mm}\label{jmlfig1}
\end{figure}

\begin{figure} 
\epsfxsize=11cm
\epsfysize=8cm
 \centerline{\epsfbox{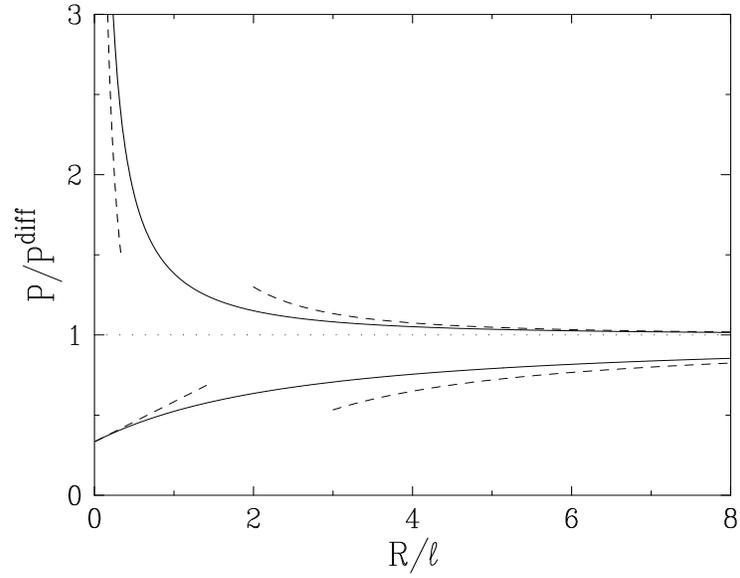}} 
\caption{
Plot of the size factor $P/P^{\rm diff}$ of the polarizability
of a transparent sphere (lower curves) and a reflecting sphere
(upper curves).
Full lines: outcome of the numerical analysis;
Dashed lines: small-radius and large-radius behaviors of Table~3.}
\vspace{3mm}\label{jmlfig23}\end{figure}

\begin{figure} \epsfxsize=11cm \centerline{\epsfbox{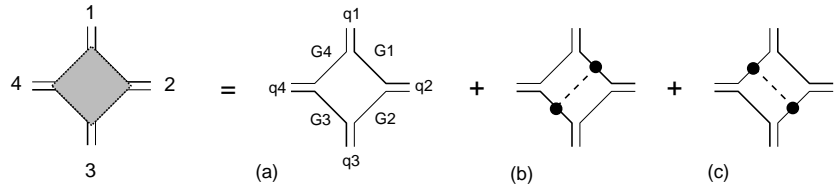}} 
\caption{Hikami four-point vertex. It describes the exchange of
amplitudes of two incoming diffusons 1 and 3 into two outgoing diffusons 2 and
4. The dots linked with the dashed line denote an extra scatterer on which both
amplitudes scatter. The solid lines are dressed amplitude propagators. }
\label{figh4sob}
\end{figure}

\begin{figure}
\epsfxsize=11cm
\centerline{\epsfbox{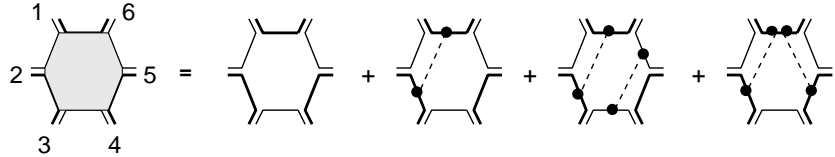}}
\caption{Six-point vertex $H_6$, describing the interaction of six
diffusons. We did not draw possible rotations of the three right diagrams. In
total there are sixteen diagrams.} \label{figh6}
\end{figure}

\begin{figure} \epsfxsize=11cm \centerline{\epsfbox{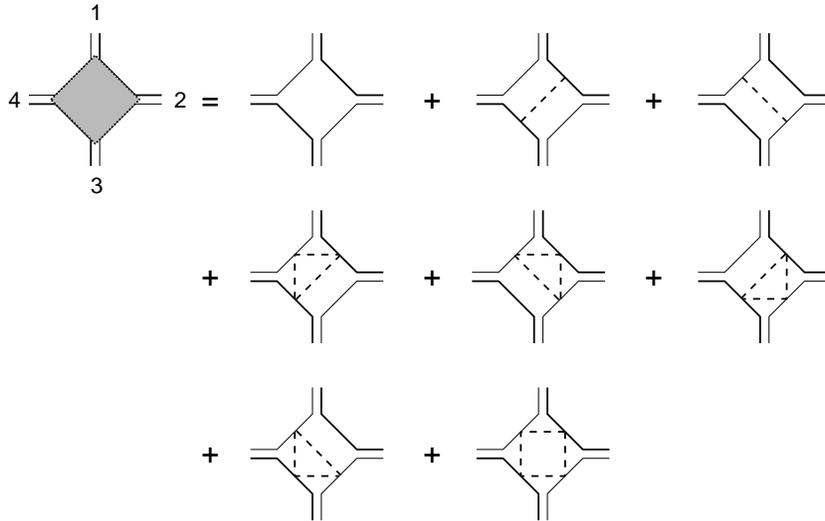}}
\caption{Four-point vertex, beyond the second-order 
Born approximation. Instead of three, there are now eight diagrams to be
calculated. The resulting expression, 
however, is apart from a renormalization of the mean
free path the same as in second-order Born approximation.} \label{hikbox}
\end{figure}

\begin{figure}
\epsfxsize=12cm
\centerline{\epsfbox{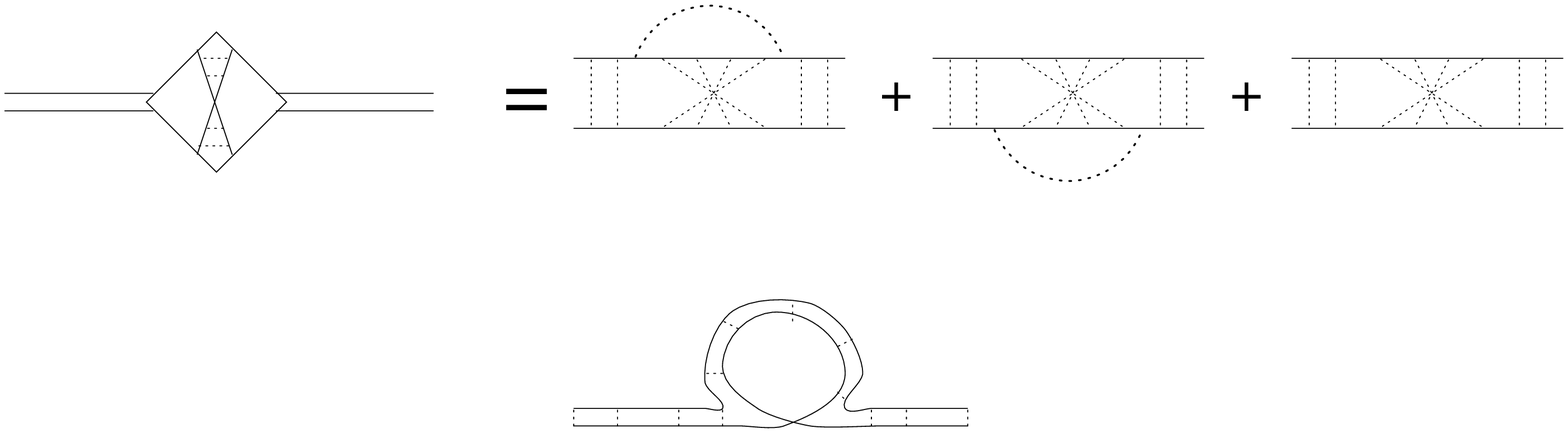}}
\caption{First-order correction to the conductivity
in the Hikami formalism. At the right we have written out 
the vertices in the second-order Born approximation.
On the second line the first diagram in the r.h.s. of the upper line
is drawn in a different way,
showing it is a loop effect.} \label{figghik}
\end{figure}

\begin{figure}
\epsfxsize=11cm
\centerline{\epsfbox{c123.eps}}
\caption{Schematic picture of the different correlation functions
present in the transmission of two intensities. The arrows distinguish
between advanced and retarded propagators. Propagators with equal
transverse momentum or frequency have the same style of line (dashed
or solid). The black box contains in principle all contributions to
$\expect{T_{ab}T_{cd}}$. The main part just factorizes, but
correlations are present. The $C_1$ correlation involves a simple interchange
of amplitudes. The $C_2$ and $C_3$ correlation involve a Hikami box (circled
H); they are a perturbative effect for weak scattering. }\label{figc123}
\end{figure}

\begin{figure}
\epsfxsize=5cm
\centerline{\epsfbox{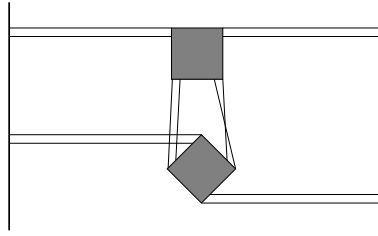}}
\caption{A contribution of order $1/g^2$ to the $C_2$ correlation. 
The boxes are Hikami boxes, the parallel lines are the diffusons.
By following the amplitudes, one can check that the incoming pairings
$ij^*$ and $ji^*$ change on the outgoing side into pairings $ii^*$
and $jj^*$. 
This diagram is thus not contributing to the $C_3$, which is also of
order $1/g^2$ but requires diagrams without amplitude exchange.
The exchange in pairing of incoming and outgoing amplitudes means that the 
diagram is a higher order $C_2$ diagram.
}
\label{figc2g2}\end{figure}

\begin{figure}
\epsfxsize=10cm
\epsfysize=7cm
\centerline{\epsfbox{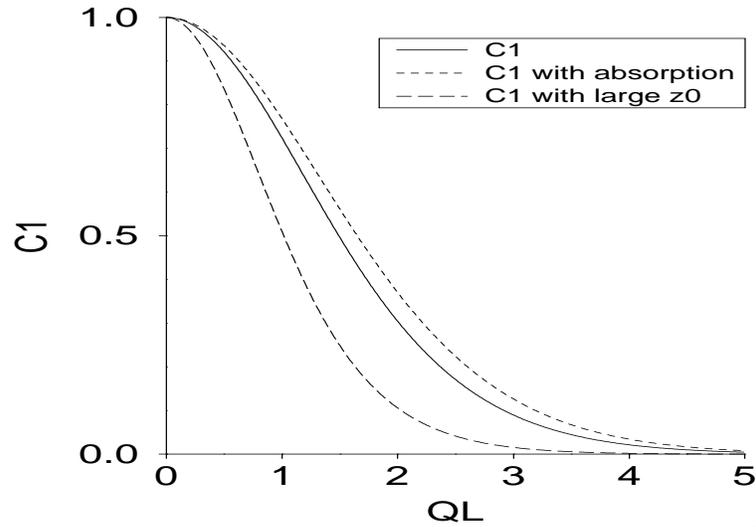}}
\caption{$C_1$ angular correlation function plotted against the
scaled perpendicular momentum difference: Solid line, small $z_0$,
no absorption;
short dashed line, absorption ($\kappa=2/L$), small $z_0$;
long dashed line, large skin layers ($z_0=L/3$), no absorption.}
\label{figc1}
\end{figure}

\begin{figure}
\epsfxsize=6cm
\centerline{\epsfbox{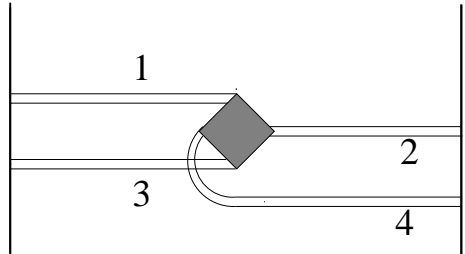}}
\caption{Diagram of the long-range $C_2$ correlation function. The
shaded box denotes the Hikami four-point vertex depicted in
Figs.~\protect{\ref{figh4sob}} and \protect{\ref{hikbox}}.}
\label{figc2diagram}
\end{figure}

\begin{figure}
\epsfxsize=10cm
\epsfysize=7cm
\centerline{\epsfbox{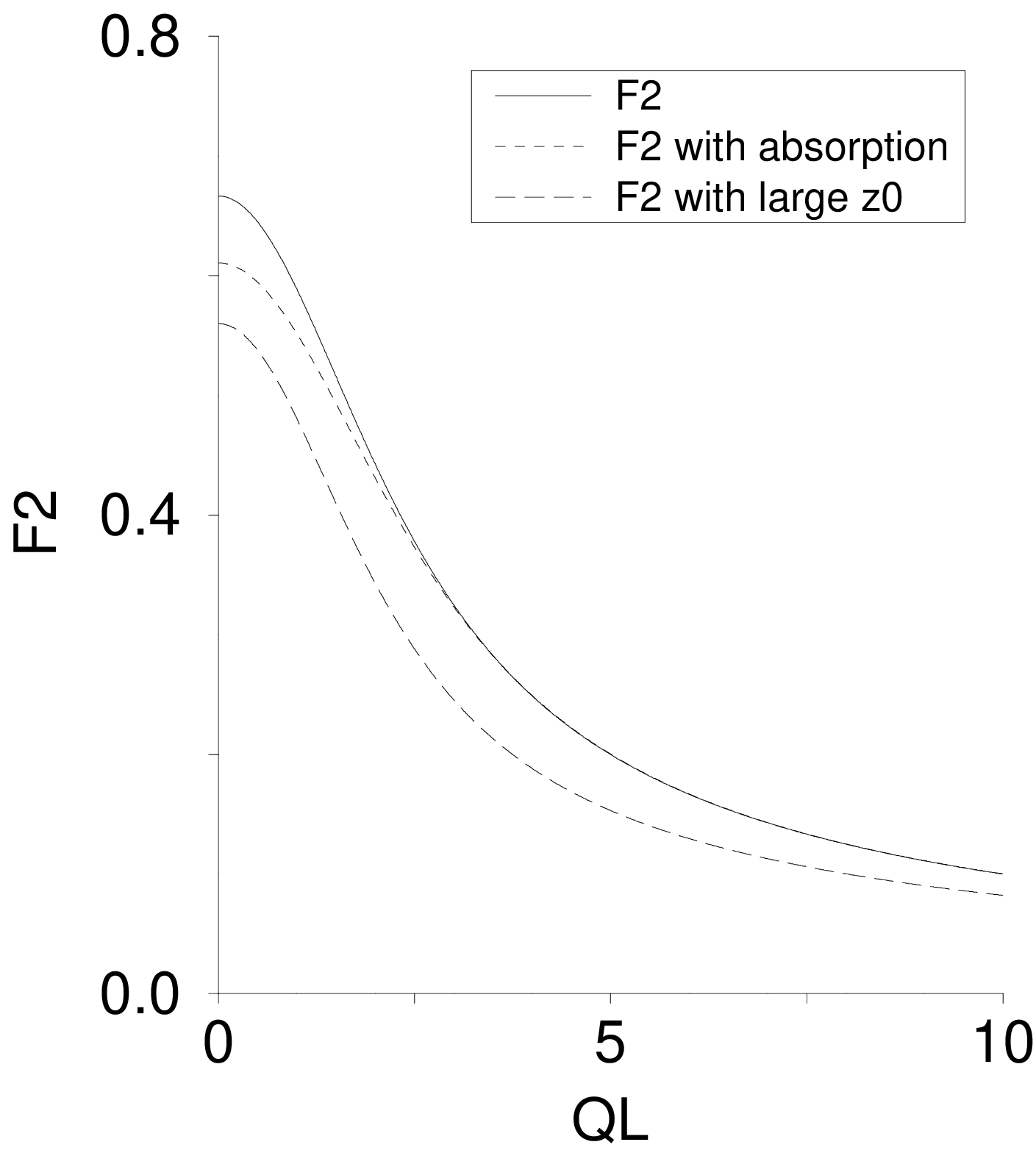}}
\caption{$F_2$, the normalized form of the angular 
correlation function $C_2$, plotted against the scaled
perpendicular momentum difference: Solid line, small $z_0$,
no absorption Eq.~(\protect{\ref{c2dq}});
short dashed line, absorption ($\kappa=2/L$), small
$z_0$; long dashed line, large skin layers ($z_0=L/3$), no absorption.}
\label{figc2}
\end{figure}

\begin{figure}
\epsfxsize=14cm
\centerline{\epsfbox{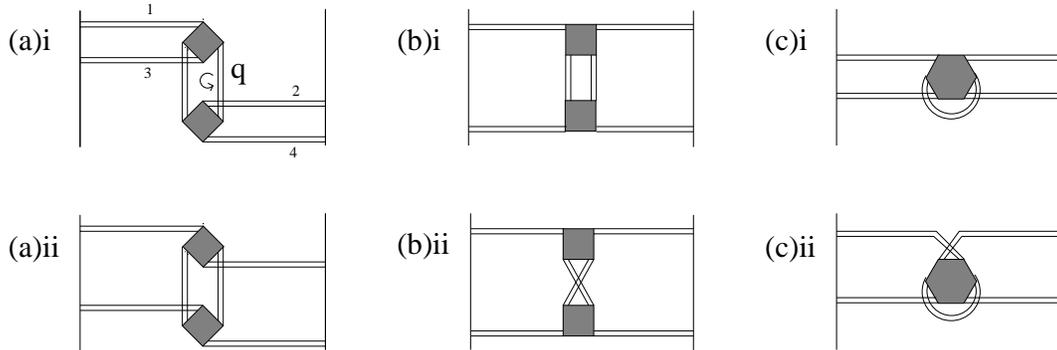}}
\caption{Leading contributions to the conductance fluctuations apart
from short-distance processes.
 The incoming diffusons from the left interfere
twice before they go out on the right. The close parallel lines correspond to
diffusons; the shaded boxes are Hikami vertices; ${\bf q}$ denotes the free
momentum which is to be integrated over.} \label{figreg} \end{figure}

\begin{figure}
\epsfxsize=10cm
\centerline{\epsfbox{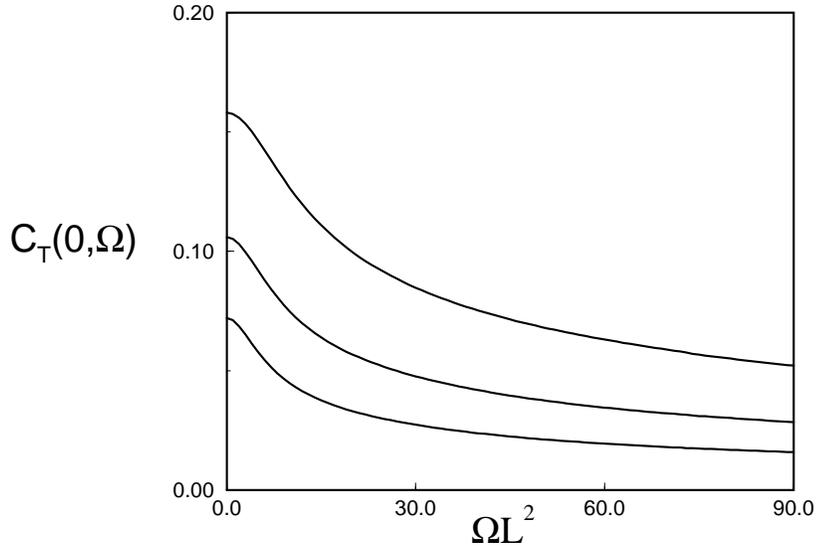}}
\caption{The $C_3$ frequency correlation in
3D as a function of frequency difference; different internal reflection
values: upper curve, $z_0=0$; middle curve, $z_0=L/10$; lower curve 
$z_0=L/5 $; no absorption.}
\label{figc3z03d} \end{figure}

\begin{figure}
\epsfxsize=11cm
\centerline{\epsfbox{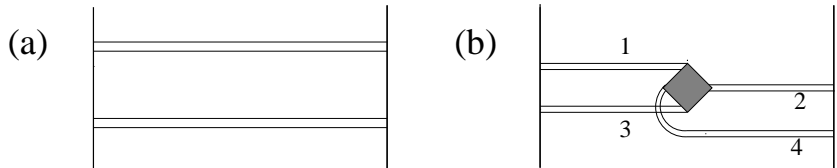}}
\caption{The two contributions to the second moment of the total
transmission. (a) Independent  transmission channels. This
process is of order unity and is almost completely reducible to the mean
value squared. (b) Two interfering channels; this is
the second cumulant or $C_2$ diagram of order $1/g$. The close
parallel lines are diffusons.}\label{figt2} \end{figure}

\begin{figure}
\epsfxsize=12cm
\centerline{\epsfbox{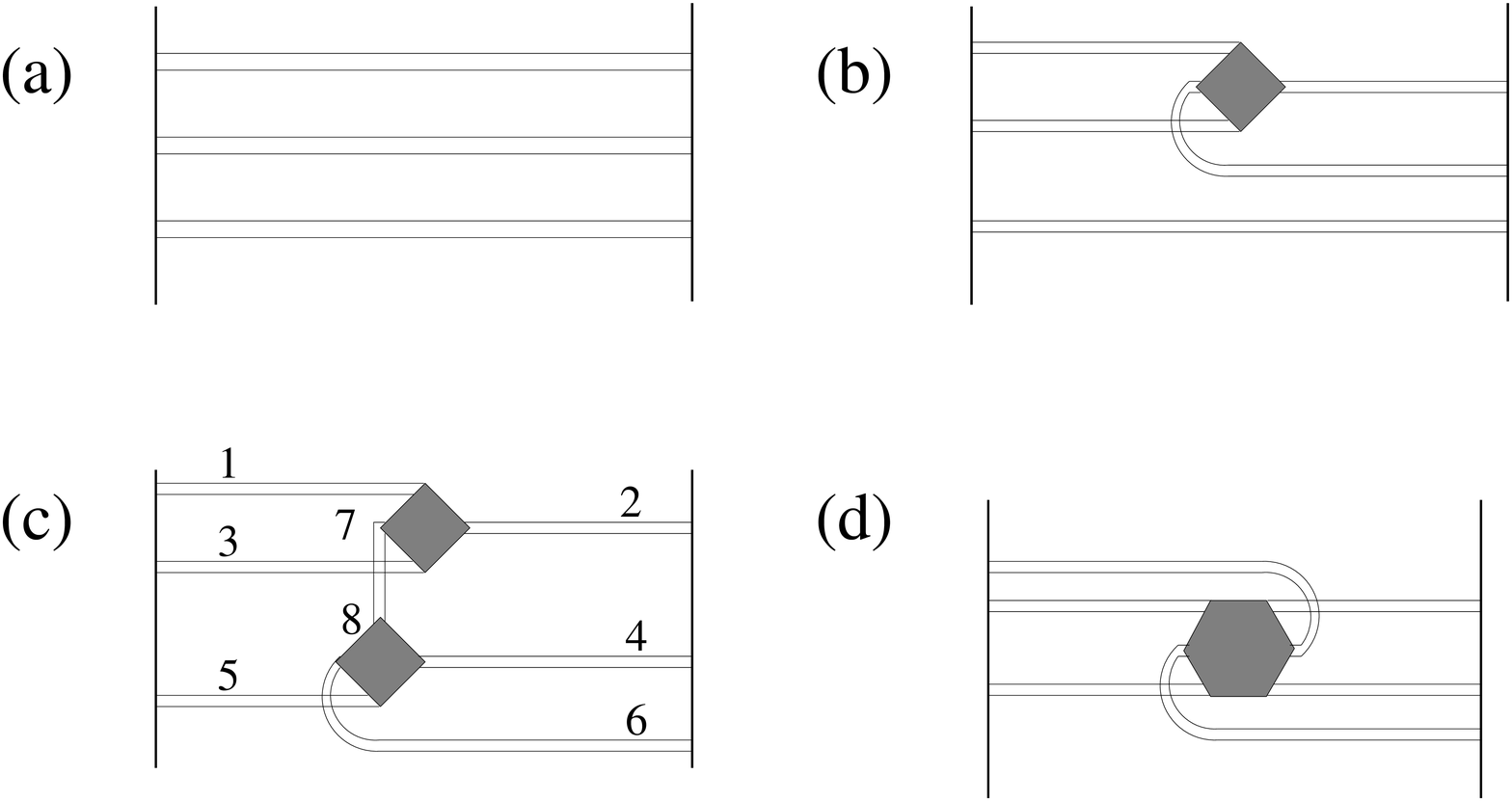}}
\caption{The three contributions to the third moment of the total
transmission. (a) independent transmission channels; it
is of order 1. (b) correlation which can be decomposed
into the second cumulant; this is of order $g^{-1}$. 
(c) and (d)
contributions to the third cumulant, $O(g^{-2})$. }\label{figt3}
\end{figure}

\begin{figure} \epsfxsize=12cm 
\centerline{\epsffile{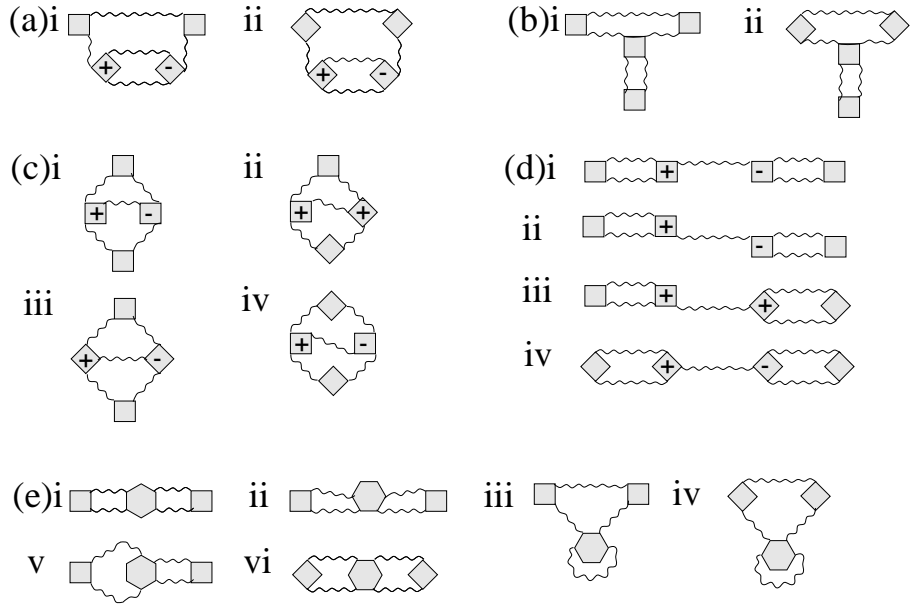}}
\caption{Set of diagrams for the third cumulant of 
the conductance. The wavy lines denote diffusons (in the Kubo
approach). \label{figtop}} \end{figure}

\begin{figure}
\epsfxsize=8cm
\centerline{\epsffile{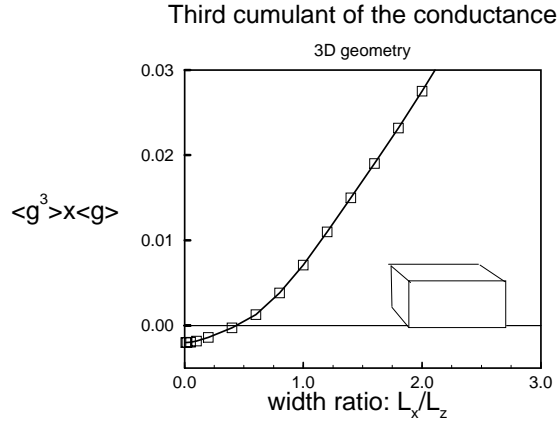}}
\caption{Third cumulant multiplied by the average conductance
in a 3D sample, plotted against the transverse size $L_x$. 
The geometry is $L_x \times L_z \times L_z$ for the solid line.
Note that the sign changes in going from a quasi-2D to a 3D sample.
The dashed line corresponds to the geometry $L_x \times L_x \times
L_z $. \label{figwide3}}
\end{figure}

\begin{figure} \epsfxsize=10cm 
 \centerline{\epsfbox{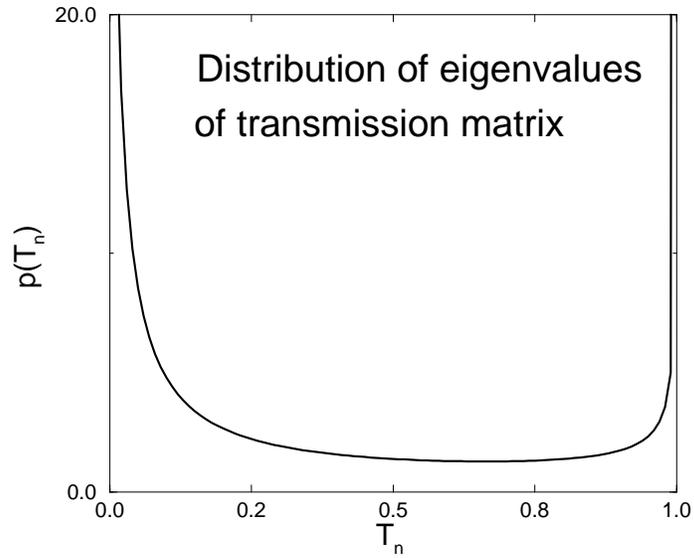}} \caption{The bimodal
eigenvalue distribution. The average value occurs only with 
a small probability,
Instead the eigenvalues are almost all either 0 or 1.} \label{figptn}
\end{figure}

\begin{figure}
\epsfxsize=11cm
\epsfysize=7cm
\centerline{\epsfbox{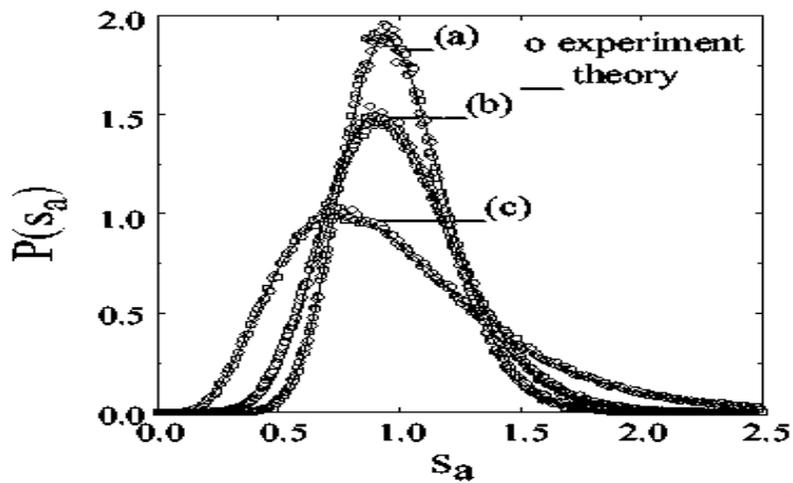}}
\caption{Distribution of the total transmission
vs normalized intensity $s_a$, theory and microwave experiments 
for three different sample sizes (a,b,c) (Stoychev and Genack, 1997).
For curve (c), the conductance is the lowest at 3.06 (absorption corrected).
Reproduction with kind permission of the authors.
}
\label{figPtaPW}
\end{figure}

\begin{figure}
\epsfxsize=11cm
\epsfysize=7cm
\centerline{\epsfbox{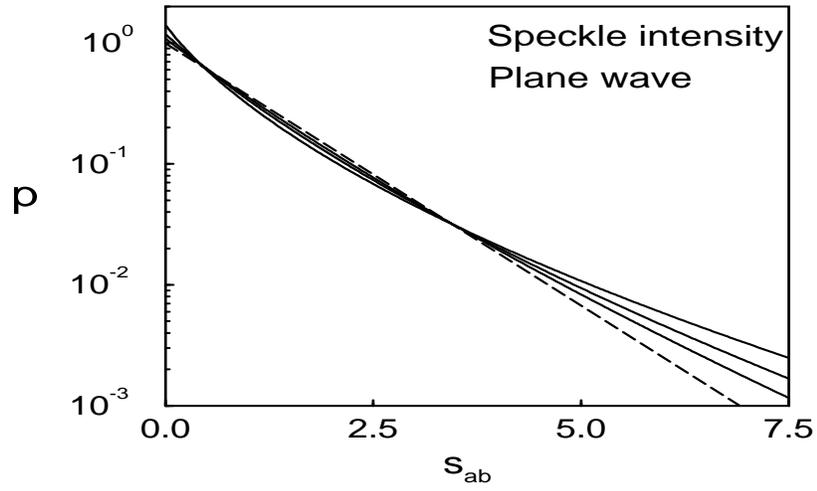}}
\caption{
Distribution of speckles versus the normalized intensity $s_{ab}$ for an
incoming plane wave, $g=2,4$, and 8 (upper to lower curve at
$s_{ab}=5$). The dashed line corresponds to
Rayleigh's law ($g= \infty$).} \label{figPtabP}
\end{figure}

\end{document}